\documentclass[aps,prd,preprintnumbers,showpacs,showkeys,nofootinbib,
superscriptaddress,fleqn,floatfix,tightenlines,10pt]{revtex4-1}
\usepackage{amsmath,amsfonts,amssymb,amscd,amsxtra,amsthm}
\usepackage{graphicx}  
\usepackage{epstopdf}
\usepackage{dcolumn}  
\usepackage{bm}          
\usepackage{slashed}
\usepackage{cancel}
\usepackage{float} 
\usepackage{mathtools}
\usepackage{amsbsy}
\usepackage{amstext}

\usepackage[utf8]{inputenc} 
\usepackage{booktabs}
\usepackage[normalem]{ulem} 
\usepackage[dvipsnames]{xcolor} 
\usepackage{tabularx}
\usepackage{enumitem}  
\usepackage{array} 
\usepackage{slashed}
\usepackage{tikz}
\usepackage{float}
\usepackage{multirow}
\renewcommand\sout{\bgroup \color{red} \ULdepth=-.5ex \ULset}

\makeatletter

\begin{document}  
\preprint{INHA-NTG-06/2019}
\title{Electromagnetic form factors of the baryon decuplet \\ with flavor
  SU(3) symmetry breaking}
\author{June-Young Kim}
\email[E-mail: ]{junyoung.kim@inha.edu}
\affiliation{Department of Physics, Inha University, Incheon 22212,
Republic of Korea}

\author{Hyun-Chul Kim}
\email[E-mail: ]{hchkim@inha.ac.kr}
\affiliation{Department of Physics, Inha University, Incheon 22212,
Republic of Korea}
\affiliation{School of Physics, Korea Institute for Advanced Study 
  (KIAS), Seoul 02455, Republic of Korea}
\date{\today}
\begin{abstract}
We investigate the electromagnetic form factors of the baryon decuplet
within the framework of the $\mathrm{SU(3)}$ self-consistent chiral
quark-soliton model, taking into account the $1/N_c$ rotational
corrections and the effects of flavor $\mathrm{SU(3)}$ symmetry
breaking. We first examine the valence- and sea-quark contributions to
each electromagnetic form factor of the baryon decuplet and then the
effects of the flavor SU(3) symmetry breaking. 
The present results are in good agreement with the recent lattice
data. We also compute the charge radii, the magnetic radii, the
magnetic dipole moments and the electric quadrupole moments, comparing
their results with those from other theoretical works. 
We also make a chiral extrapolation to compare the present results
with the lattice data in a more quantitative manner. The results show
in general similar tendency to the lattice results. In particular, the
results of the $M1$ and $E2$ form factors are in good agreement with
those of lattice QCD.
\end{abstract}
\maketitle
\section{Introduction}
The $\Delta$ isobars are the first excited baryons with spin
3/2. Its electromagnetic (EM) structure and properties have
been experimentally studied at a rather gradual pace over decades
because of its ephemeral nature, the $\gamma^* N\to \Delta$
transitions being mainly focused~\cite{Bartel:1968tw,
  Stein:1975yy}. By the turn of the new century, however, being
equipped with a new generation of electron beam accelerators and
detectors with high precision, one was 
able to take a better grasp of the EM transitions of $\Delta$
isobars~\cite{Krusche:2003ik, Pascalutsa:2006up}. For example, various
experimental groups have announced the results on the $\gamma^* N\to
\Delta$ excitation: the Laser Electron Gamms Source (LEGS)
Collaboration at the National Synchrotron Light Source (NSLS) of
Brookhaven National Laboratory~\cite{LEGS:1992, Blanpied:1997zz,
  Blanpied:2001ae}, the CLAS Collaboration~\cite{Joo:2001tw,
  Biselli:2003ym, Aznauryan:2009mx} at Jefferson Laboratory, and the
A1 and A2 Collaborations~\cite{Beck:1997ew, Beck:1999ge, 
  Kotulla:2002cg, Ahrens:2004pf, Stave:2008aa, Sparveris:2013ena} at
MAMI. The EM structure of the strangeness members of the baryon
decuplet has been also experimentally investigated at these 
experimental facilities. So far the static magnetic dipole moments
have been only measured~\cite{Wallace:1995pf,LopezCastro:2000cv,
Tanabashi:2018oca}. Future experiments will even provide the EM
data on the $\Xi^*$ and $\Omega^-$ baryons~\cite{Afanasev:2012fh} 
in detail. The EM structure of the baryon decuplet is more complicated
than that of the  baryon octet, since all the decuplet baryons have
spin 3/2. Thus, a decuplet baryon has two more terms of the EM form  
factors than the baryon octet, i.e. the electric quadrupole ($E2$)
form factor and the magnetic octupole ($M3$) form factors in addition
to the electric monopole ($E0$) and magnetic dipole ($M1$) form
factors.   

The EM form factors of the $\Delta$ isobars and the
other members of the baryon decuplet are still experimentally
unknown. However, recently, a series of calculations in lattice
quantum chromodynamics (QCD) was carried out. Alexandrou et
al. reported the lattice results on the EM form factors of the
$\Delta$ and $\Omega$~\cite{Alexandrou:2007we,
  Alexandrou:2008bn,Alexandrou:2009hs, Alexandrou:2010jv}, which
improved greatly the old lattice work~\cite{Leinweber:1992hy}. The 
magnetic dipole moments of the $\Delta$ and $\Omega$ were examined
also in lattice QCD~\cite{Lee:2005ds, Aubin:2008qp,
  Boinepalli:2009sq}. Theoretically, the EM properties of the baryon
decuplet have been studied within various different models and
theoretical frameworks: for example, effective field theory and baryon
chiral perturbation 
theories~\cite{Butler:1993ej, Geng:2009ys, Ledwig:2011cx,
  Li:2016ezv}, $1/N_c$ expansions~\cite{Luty:1994ub, Buchmann:2002et,   
  Flores-Mendieta:2015wir}, various versions of the quark  
models~\cite{Krivoruchenko:1991pm, Schlumpf:1993rm, Wagner:2000ii,
Berger:2004yi}, the Skyrme models~\cite{Kroll:1994um, Oh:1995hn}, 
QCD sum rules~\cite{Lee:1997jk, Azizi:2008tx, Aliev:2009pd}, and 
Bethe-Salpeter approaches~\cite{Nicmorus:2010sd,
  Sanchis-Alepuz:2013iia, Sanchis-Alepuz:2015fcg}, and holographic
QCD~\cite{Druks:2018hif}. 

In the present work, we will investigate the EM form factors of all
the members in the baryon decuplet within the chiral quark-soliton
model ($\chi$QSM), taking into account the effects
of explicit flavor SU(3) symmetry breaking. The $\chi$QSM has been
developed as a pion mean-field approach, based on the seminal papers
by Witten~\cite{Witten:1979kh, Witten:1983tx}. In the limit of the
large number of colors ($N_c$), the nucleon mass is proportional to
$N_c$, whereas its decay width is of order $\mathcal{O}(1)$. It
implies that the meson-loop corrections can be suppressed in this
limit. Then the nucleon arises as a state consisting of the $N_c$
valence quarks, which is bound by the pion mean
fields~\cite{Diakonov:1987ty, Wakamatsu:1990ud, 
  Christov:1995vm, Diakonov:1997sj}. The presence of the $N_c$ valence
quarks brings about the vacuum polarization that produces effectively
the pion mean fields. Then, they are again affected by these pion mean
fields in a self-consistent manner. The $\chi$QSM has been successfully 
applied to the description of the SU(3) baryons. For example, the
model explains very well various form factors of the lowest-lying
SU(3) baryons such as electromagnetic structures~\cite{Kim:1995ha,
  Wakamatsu:1996xm, Kim:1997ip, Silva:2013laa}, strange form
factors~\cite{Kim:1995hu, Silva:2001st, Silva:2005fa, Silva:2005qm}, 
tensor charges and form factors~\cite{Kim:1995bq, Kim:1996vk,
  Pobylitsa:1996rs, Ledwig:2010tu, Ledwig:2010zq, Goeke:2007fp},
semileptonic decays~\cite{Kim:1997ts, Ledwig:2008ku, Yang:2015era},
radiative decay~\cite{Watabe:1995xy, Silva:1999nz, Ledwig:1900ri} and
the parton distributions~\cite{Diakonov:1996sr, 
  Diakonov:1997vc, Goeke:2000wv, Schweitzer:2001sr, Pobylitsa:1998tk},    
etc. Moreover it was shown that the $\chi$QSM can be extended to
describing the singly heavy baryons by changing the pion mean
fields arising from the $N_c-1$ valence quarks. With this extension,
the model reproduces very well the observables of the singly heavy
baryons such as mass splittings ~\cite{Yang:2016qdz, Kim:2018xlc},
strong decays~\cite{Kim:2017jpx, Kim:2017khv,Kim:2018cxv} and
electromagnetic properties ~\cite{Yang:2018uoj, Kim:2018nqf}. 

In fact, The EM form factors of the  $\Delta$ baryon have been already
studied by Ref.~\cite{Ledwig:2008es} within the same framework, with
flavor SU(3) symmetry assumed~\footnote{Note that
  Ref.~\cite{Ledwig:2008es} considered the effects of flavor SU(3)
  symmetry breaking on the magnetic dipole moments of the baryon
  decuplet}. In this work, we want to investigate the EM form
factors of all the members in the baryon decuplet, taking into account
the $1/N_c$ rotational corrections and the effects of flavor SU(3)
symmetry breaking. Since there exist in particular the lattice results
of the $\Omega$ EM form factors~\cite{Alexandrou:2010jv}, it is of
great interest to compute and compare them with those from 
lattice QCD. Thus, we will present and discuss the results of the
$E0$, $M1$, and $E2$ form factors of the baryon decuplet, emphasizing
the comparison with the lattice data. Concerning the $M3$ form
factors, any chiral soliton approach yields null results 
of the $M3$ form factors. It is consistent with the fact that the
magnitudes of the $M3$ form factors are very small. 

The structure of the present work is sketched as follows: In
Section~\ref{sec:2}, we define the EM form factors of the baryon
decuplet in terms of the matrix elements of the EM current. In
Section~\ref{sec:3}, we briefly review the formalism of the $\chi$QSM
in the context of the derivation of the EM form
factors of the baryon decuplet. In Section~\ref{sec:4} we present the
numerical results of them. We discuss first the valence and sea
contributions to the EM form factors, and then examine the effects of
flavor SU(3) symmetry breaking. We also show the numerical results of
the charge and magnetic radii, the magnetic dipole moments, and the
electric quadrupole moments. In the last subsection of
Section~\ref{sec:4}, we make a chiral extrapolation of the present
work so that we can compare the numerical results with those from
lattice QCD in a quantitative manner. In the final Section, we
summarize the present work and draw conclusions. 

\section{Electromagnetic multipole form factors 
  of the baryon decuplet}
\label{sec:2}
The EM current is defined as 
\begin{align}
  \label{eq:LHcurrent}
J_\mu (x) = \bar{\psi} (x) \gamma_\mu \hat{\mathcal{Q}} \psi(x),  
\end{align}
where $\psi(x)$ denotes the quark field $\psi=(u,\,d,\,s)$ in flavor
space. The charge operator of the quarks $\hat{\mathcal{Q}}$ is
written in terms of the flavor SU(3) Gell-Mann matrices $\lambda_3$
and $\lambda_8$ 
\begin{align}
 \label{eq:chargeOp}
\hat{\mathcal{Q}} =
  \begin{pmatrix}
   \frac23 & 0 & 0 \\ 0 & -\frac13 & 0 \\ 0 & 0 & -\frac13
  \end{pmatrix} = \frac12\left(\lambda_3 + \frac1{\sqrt{3}}
                                                  \lambda_8\right).
\end{align}
The matrix elements of the EM current between baryons with spin 3/2
can be parametrized by four real form factors $F_i^{B}$
($i=1,\cdots,4$) as follows: 
\begin{align}
\langle B(p',s) | e_B J^{\mu}(0) | B(p,s) \rangle 
&= - e_B \overline{u}^{\alpha}(p',s) \left[ \gamma^{\mu} \left \{
  F^{B}_{1}(q^2) \eta_{\alpha \beta} + F^{B}_{3}(q^2) \frac{ q_{\alpha} q_{\beta}
  }{4M_{B}^{2}}  \right \}\right. \cr
&\hspace{2cm} \left.  + \,i\frac{\sigma^{\mu \nu} q_{\nu}}{2M_{B}}
  \left \{ F^{B}_{2}(q^2) \eta_{\alpha \beta} + F^{B}_{4} (q^2)\frac{q_{\alpha}
  q_{\beta}}{4 M_{B}^2}  \right \}  \right ]{u}^{\beta}(p,s), 
\label{eq:MatrixEl1}
\end{align}
where $M_B$ denotes the mass of the corresponding baryon in the
decuplet, and $e_B$ stands for the corresponding electric charge of 
the baryon $B$. The metric tensor $\eta_{\alpha\beta}$ of 
Minkowski space is expressed as $\eta_{\alpha\beta}
=\mathrm{diag}(1,\,-1,\,-1,\,-1)$. $q_\alpha$ designates the momentum
transfer $q_\alpha=p'_\alpha-p_\alpha$ and its square is given as
$q^2=-Q^2$ with $Q^2 >0$. $u^\alpha (p,\,s)$ represents the
Rarita-Schwinger spinor that describes the decuplet baryon with spin
3/2, carrying the momentum $p$ and the spin component $s$ projected
along the direction of the momentum. $\sigma^{\mu\nu}$ is the
well-known antisymmetric tensor
$\sigma^{\mu\nu}=i[\gamma^\mu,\,\gamma^\nu]/2$.   

It is often more convenient to introduce the multipole EM form factors,
which can be expressed in terms of the $F_i^B$ given in
Eq.~\eqref{eq:MatrixEl1}
\begin{align}
&G_{E0}^B(Q^{2}) = \left(1 + \frac{2}{3} \tau\right)[F_{1}^{B}-\tau F_{2}^{B}] -
  \frac{1}{3} \tau (1+ \tau) [F_{3}^{B} - \tau F_{4}^{B}], \cr 
&G_{E2}^B(Q^{2}) = [F_{1}^{B}-\tau F_{2}^{B}]- \frac{1}{2} (1+ \tau)
  [F_{3}^{B} - \tau F_{4}^{B}] , \cr 
&G_{M1}^B (Q^{2}) = \left(1+\frac{4}{5} \tau\right)[F^{B}_{1}+F^{B}_{2}]  -
  \frac{2}{5}  \tau (1+ \tau)[F^{B}_{3}+F^{B}_{4}], \cr 
&G_{M3}^B(Q^{2}) =  [F^{B}_{1}+F^{B}_{2}] -\frac{1}{2}(1+\tau)
  [F^{B}_{3}+F^{B}_{4}], 
\end{align}
where $\tau=Q^2/4M_B^2$. These four form factors are called,
respectively, the electric or Coulomb monopole ($E0$), magnetic dipole
($M1$), electric quadrupole ($E2$), and magnetic octupole ($M3$) form
factors. Each form factor at $Q^2=0$ has a clear physical meaning:
$G_{E0}^B(0)$ is the normalized charge, i.e. the corresponding
decuplet baryon $B$ with positive or negative charge will have the
value of $G_{E0}^B(0)=1$ or $G_{E0}^B=-1$, whereas the neutron one
will have $G_{E0}^B(0)=0$. $G_{M1}^B(0)$ means the magnetic dipole
moment $\mu_B$ for the corresponding baryon, whereas  $G_{E2}^B(0)$
yields the corresponding electric quadrupole moment $Q_B$.
$G_{M3}^B(0)$ will give us the magnetic octupole ($O_B$) moment of the
corresponding decuplet baryon $B$~\cite{Pascalutsa:2006up}. Thus,
these physical quantities can be written in terms of the charge and
the form factors $F_i^{B}$
\begin{align}
e_B & = e G_{E0}^B(0) = e F_1^{B}(0),\cr
\mu_B &= \frac{e}{2M_B} G_{M1}^B = \frac{e}{2M_B}
        \left[e_B+F_2^{B}(0)\right],\cr 
Q_B &= \frac{e}{M_B^2} G_{E2}^B(0) = \frac{e}{M_B^2} \left[e_B
      -\frac12 
      F_3^{B}(0)\right],\cr 
O_B &= \frac{e}{M_B^3} G_{M3}^B(0) =  \frac{e}{M_B^3}
      \left[e_B+F_2^{B}(0) 
      - \frac12 (F_3^{B}(0) + F_4^{B}(0))\right].
  \label{eq:q2z}
\end{align}
 
If one chooses the Breit frame, i.e. $p'=-p=q/2$, then the EM
multipole form factors can be explicitly expressed as 
\begin{align}
G_{E0}^B(Q^{2}) &= \int \frac{d \Omega_{q}}{4 \pi}\langle B(p',3/2) |
  J^{0}(0) | B (p,3/2) \rangle,  \cr 
G_{E2}^B(Q^{2}) &= -\int {d \Omega_{q}} \sqrt{\frac{5}{4\pi}
  }\frac{3}{2} \frac{1}{\tau}\langle B(p',3/2) | Y^{*}_{20}
  (\Omega_{q}) J^{0}(0) | B (p,3/2) \rangle, \cr 
G_{M1}^B(Q^{2})&= \frac{3 M_B}{4\pi}\int \frac{d\Omega_{q}}{ i
  |\bm{q}|^{2}}q^{i}\epsilon^{ik3}\langle B (p',3/2) | J^{k}(0) | B
  (p,3/2) \rangle,  \cr
G_{M3}^B(Q^{2}) &=-\frac{35M_{B}}{8}\sqrt{\frac{5}{\pi}} \int 
  \frac{d\Omega_{q}}{ i  |\bm{q}|^{2} \tau}q^{i}\epsilon^{ik3}\langle
  B (p',3/2) |
  \left(Y^{*}_{20}(\Omega_{q})+\sqrt{\frac{1}{5}}Y^{*}_{00}(\Omega_{q})
                  \right) J^{k}(0) | B (p,3/2) \rangle,
                  \label{eq:matel6}
\end{align}
where $J^0$ and $J^i$ denote the temporal and spatial components of
the EM current, respectively. Thus, we can straightforwardly compute
the matrix elements of the EM current in Eq.~\eqref{eq:matel6} within
the $\chi$QSM. 
\section{Electromagnetic form factors 
  in the chiral quark-soliton model}
\label{sec:3}
The $\chi$QSM is characterized by the effective chiral partition
function 
\begin{align}
\label{eq:partftn}
\mathcal{Z}_{\chi\mathrm{QSM}} = \int \mathcal{D}\psi \mathcal{D}
  \psi^\dagger \mathcal{D} \pi^a \exp\left[-\int d^4 x \psi^\dagger i D(\pi^a)
  \psi\right]  = \int \mathcal{D} \pi^a \exp
  (-S_{\mathrm{eff}}[\pi^a]),   
\end{align}
where $\psi$ and $\pi^a$ represent the quark and
pseudo-Nambu-Goldstone (NG) boson fields, respectively. The
$S_{\mathrm{eff}}$ denotes the effective chiral action given as a
functional of $\pi^a$
\begin{align}
S_{\mathrm{eff}}[\pi^a] \;=\; -N_{c}\mathrm{Tr}\ln D\,,
\label{eq:echl}
\end{align}
where $\mathrm{Tr}$ stands for the trace running over space-time and
all relevant internal spaces. The $N_c$ is the number 
of colors, and $D(\pi^a)$ is the one-body Dirac differential operator
defined by  
\begin{align}
 D(\pi^a) \;=\; 
 (i\slashed{\partial} + i MU^{\gamma_5} + i \hat{m})\,.
 \label{eq:Dirac}  
\end{align}
We assume isospin symmetry, so the up and down
current quark masses are set equal to each other,
i.e. $m_{\mathrm{u}}=m_{\mathrm{d}}$. We define the average mass of
the up and down current quarks by $\overline{m}=(m_{\mathrm{u}} +
m_{\mathrm{d}})/2$. Then, the mass matrix 
of the current quark is written as $\hat{m}
=\mathrm{diag}(\overline{m},\, \overline{m},\,  m_{\mathrm{s}}) =
\overline{m} +\delta m$.  $\delta m$ contains the mass of the strange
current quark, which can be decomposed in to the flavor singlet and
octet parts 
\begin{align}
\delta m  \;=\; \frac{-\overline{m} + m_{s}}{3}\gamma_{4}\bm{1} +
\frac{\overline{m} - m_{s}}{\sqrt{3}} \gamma_{4} \lambda^{8} =
m_{1}\gamma_{4} \bm{1} + m_{8} \gamma_{4} \lambda^{8}\,,
\label{eq:deltam}
\end{align}
where $m_1$ and $m_8$ designate respectively the singlet and octet
components of the  current quark masses. They are expressed as 
$m_1=(-\overline{m} +m_{\mathrm{s}})/3 $ and $m_8=(\overline{m}
-m_{\mathrm{s}})/\sqrt{3}$. The one-body Dirac operator multiplied by
$\gamma_4$ can be written as 
\begin{align}
\gamma_4 D = -i\partial_{4} + h(U(\pi^a)) - \delta m,
\end{align}
where $\partial_4$ denotes the derivative with respect to the
Euclidean time. The SU(3) single-quark Hamiltonian $h(U)$ is defined as 
\begin{align}
h(U) \;=\;
i\gamma_{4}\gamma_{i}\partial_{i}-\gamma_{4}MU^{\gamma_{5}} -
\gamma_{4} \overline{m}\, ,
\label{eq:diracham}  
\end{align}
where $U^{\gamma_5}$ represents the SU(3) chiral field. Since it is
well known that the isovector charge radii ~\cite{Beg:1973sc} diverge
in the chiral limit, it is essential to include $\overline{m}$ in the
Dirac Hamiltonian~\eqref{eq:diracham}. Moreover, $\overline{m}$ in
Eq.~\eqref{eq:diracham} produces the correct Yukawa tail of the pion
field.

Since the pseudo-NG fields carry the flavor indices, the
hedgehog ansatz is usually imposed, in which each pion field with
$a=1,\,2,\,3$ is taken to be aligned along each three-dimensional
spatial component 
\begin{align}
  \label{eq:hedgehog}
\pi^a (\bm{x}) = {n}^a P(r), 
\end{align}
where $n^a=x^a/r$ with $r=|\bm{x}|$. $P(r)$ is called the profile
function of the chiral soliton. Then the flavor SU(2) chiral field is
written as     
\begin{align}
U_{\mathrm{SU(2)}}^{\gamma_5} \;=\; \exp(i\gamma^{5}\hat{\bm{n}}\cdot 
\bm{\tau} P(r))
\;=\; \frac{1+\gamma^{5}}{2}U_{\mathrm{SU(2)}} +
  \frac{1-\gamma^{5}}{2}U_{\mathrm{SU(2)}}^{\dagger}
\label{eq:embed}
\end{align}
with $U_{\mathrm{SU(2)}}=\exp(i\hat{\bm{n}}\cdot \bm{\tau} P(r))$. 
The flavor SU(3) chiral field is constructed by embedding the SU(2)
soliton into SU(3)~\cite{Witten:1983tx}  
\begin{align}
U^{\gamma_{5}}(x) \;=\; \left(\begin{array}{lr}
U_{\mathrm{SU(2)}}^{\gamma_{5}}(x) & 0\\
0 & 1
\end{array}\right).
\end{align}
Since we employ the pion mean-field approximation, we perform 
the integration over $U$ in Eq.~\eqref{eq:partftn} around the saddle
point, which yields $\delta S_{\mathrm{eff}}/\delta P(r) =0$. This
saddle point approximation furnishes the classical equation of motion 
that can be solved self-consistently. The solution provides the pion
mean field and consequently the self-consistent profile function
$P(r)$. For the detailed formalism of constructing the collective
Hamiltonian, we refer to reviews~\cite{Christov:1995vm, Diakonov:1997sj}.

The matrix elements of the EM current~\eqref{eq:MatrixEl1} can be
computed by considering the following functional integral 
\begin{align}
\langle B,\,p'| J_\mu(0) |B,\,p\rangle  &= \frac1{\mathcal{Z}}
  \lim_{T\to\infty} \exp\left(i p_4\frac{T}{2} - i p_4'
  \frac{T}{2}\right) \int d^3x d^3y \exp(-i \bm{p}'\cdot \bm{y} + i
  \bm{p}\cdot \bm{x}) \cr
& \hspace{-1cm} \times \int \mathcal{D} \pi^a \int \mathcal{D} \psi
                            \int   \mathcal{D} \psi^\dagger
                            J_{B}(\bm{y},\,T/2) \psi^\dagger(0) 
  \gamma_4\gamma_\mu \hat{Q} \psi(0) J_B^\dagger (\bm{x},\,-T/2)
  \exp\left[-\int d^4 z   \psi^\dagger iD(\pi^a) \psi\right],  
\label{eq:correlftn}
\end{align}
where the baryon states $|B,\,p\rangle$ and $\langle B,\,p'|$ are 
respectively defined by 
\begin{align}
|B,\,p\rangle &= \lim_{x_4\to-\infty}   \exp(i p_4 x_4)
                \frac1{\sqrt{\mathcal{Z}}} \int d^3 x
                \exp(i\bm{p}\cdot \bm{x}) J_B^\dagger
                (\bm{x},\,x_4)|0\rangle,\cr
\langle B,\,p'| &= \lim_{y_4\to\infty}   \exp(-i p_4' y_4)
                \frac1{\sqrt{\mathcal{Z}}} \int d^3 y
                \exp(-i\bm{p}'\cdot \bm{y}) \langle 0| J_B^\dagger 
                (\bm{y},\,y_4).
\end{align}
The baryon current $J_B$ can be constructed as an Ioffe-type current
in terms of $N_c$ valence quarks
\begin{align}
J_B(x) = \frac1{N_c!} \epsilon_{i_1\cdots i_{N_c}} \Gamma_{JJ_3
  TT_3 Y}^{\alpha_1\cdots \alpha_{N_c}} \psi_{\alpha_1 i_1} (x)
  \cdots \psi_{\alpha_{N_c} i_{N_c}}(x),  
\end{align}
where $\alpha_1\cdots \alpha_{N_c}$ and $i_1\cdots i_{N_c}$ denote
spin-flavor indices and color ones, respectively. The matrices
$\Gamma_{JJ_3 TT_3 Y}^{\alpha_1\cdots \alpha_{N_c}}$ will project out
the baryon state with quantum numbers $JJ_3TT_3Y$. The creation
current operator $J_B^\dagger$ can be expressed in a similar manner.  
As for the detailed formalism of the zero-mode quantization and
the techniques of computing the baryonic correlation function given in
Eq.~\eqref{eq:correlftn}, we refer to Refs.~\cite{Christov:1995vm,
  Kim:1995mr}. See also Ref.~\cite{Ledwig:2008es}.

Having performed the zero-mode quantization, we derive the collective
Hamiltonian as     
\begin{align}
H_{\mathrm{coll}} = H_{\mathrm{sym}} + H_{\mathrm{sb}},   
\end{align}
where
\begin{align}
  \label{eq:Hamiltonian}
H_{\mathrm{sym}} &= M_{\mathrm{cl}} + \frac1{2I_1} \sum_{i=1}^3
                   J_i^2 + \frac1{2I_2} \sum_{p=4}^7 J_p^2,\cr
H_{\mathrm{sb}} &= \alpha D_{88}^{(8)} + \beta \hat{Y} +
  \frac{\gamma}{\sqrt{3}} \sum_{i=1}^3 D_{8i}^{(8)} \hat{J}_i.
\end{align}
$I_1$ and $I_2$ denote the soliton moments of inertia. The parameters
$\alpha$, $\beta$, and $\gamma$ for the part of explicit flavor SU(3)
symmetry breaking are defined by  
\begin{align}
\alpha=\left (-\frac{{\Sigma}_{\pi N}}{3m_0}+\frac{
  K_{2}}{I_{2}}{Y'}  
\right )m_{\mathrm{s}},
 \;\;\;  \beta=-\frac{ K_{2}}{I_{2}}m_{\mathrm{s}}, 
\;\;\;  \gamma=2\left ( \frac{K_{1}}{I_{1}}-\frac{K_{2}}{I_{2}} 
 \right ) m_{\mathrm{s}},
\label{eq:alphaetc}  
\end{align}
where that the three parameters $\alpha$, $\beta$, and $\gamma$ are 
expressed in terms of the moments of inertia $I_{1,\,2}$ and
$K_{1,\,2}$. 

The symmetry-breaking part of the collective Hamiltonian
$H_{\mathrm{sb}}$ being considered as a perturbation, 
the states of the baryon decuplet are blended with other SU(3) 
representations as follows :   
\begin{align}
|B_{{\bm10}_{3/2}}\rangle = |{\bm{10}}_{3/2},B\rangle + 
a^{B}_{{27}}|{{\bm{27}}}_{3/2},B\rangle + 
a^{B}_{{35}}|{{\bm{35}}}_{3/2},B\rangle,
\label{eq:mixedWF1}
\end{align}
with the mixing coefficients
\begin{eqnarray}
a_{{27}}^{B}
\;=\;
a_{{27}}\left[\begin{array}{c}
\sqrt{15/2}\\
2 \\
\sqrt{3/2} \\
0
\end{array}\right], 
& 
a_{35}^{B}
\;=\; 
a_{35}\left[\begin{array}{c}
5/\sqrt{14}\\
2\sqrt{{5}/{7}} \\
3\sqrt{{5}/{14}} \\
2\sqrt{{5}/{7}}
\end{array}\right], 
\label{eq:pqmix}
\end{eqnarray}
respectively, in the basis
$\left[\Delta,\;\Sigma^{*},\;\Xi^{*},\;\Omega\right]$. The  
parameters $a_{{27}}$ and $a_{35}$ are given by 
\begin{eqnarray}
a_{27} \;=\;
{\displaystyle -\frac{{I}_{2}}{8} \left ( \alpha + \frac{5}{6}\gamma
  \right)}, & 
a_{35} \;=\; {\displaystyle -\frac{{I}_{2}}{24} \left( \alpha -
  \frac{1}{2}\gamma \right)}. 
\label{eq:pqmix2}
\end{eqnarray}
The $a_{27}$ and the $a_{35}$ were derived in Ref.~\cite{Kim:2018xlc}:
$a_{27}=0.126$ and $a_{35}=0.035$. The collective wavefunction of a
baryon with flavor $F=(Y,T,T_3)$ and spin $S=(Y'=-N_{c}/3,J,J_3)$ in
the representation $\nu$ is expressed in terms of a tensor with two
indices, i.e. $\psi_{(\nu;\, F),(\overline{\nu};\,\overline{S})}$, one running over
the states $F$ in the representation $\nu$ and the other one over the
states $\overline{S}$ in the representation $\overline{\nu}$. Here,  
$\overline{\nu}$ denotes the complex conjugate of the
$\nu$, and the complex conjugate of $S$ is written by
$\overline{S}=(N_{c}/3,\,J,\,J_3)$. Thus, the collective wavefunction
is expressed as  
\begin{align}
  \label{eq:SolitonWF1}
\psi_{(\nu;\, F),(\overline{\nu};\,\overline{S})}(R) =
  \sqrt{\mathrm{dim}(\nu)} (-1)^{Q_S} [D_{F\,S}^{(\nu)}(R)]^*,
\end{align}
where $\mathrm{dim}(\nu)$ stands for the dimension of the
representation $\nu$ and $Q_S$ a charge corresponding to the baryon
state $S$, i.e. $Q_S=J_3+Y'/2$.   

Having taken into account the rotational $1/N_c$ and linear
$m_{\mathrm{s}}$ corrections, we obtain the final expression of the
$E0$ form factors for the baryon $B$
\begin{align}
{{G}}^{B}_{E0}  (q^{2})= \int d^{3} z j_{0}(|\bm{q}| |\bm{z}|)
  {\mathcal{G}}_{E0}^B (\bm{z}),
\label{eq:app1}
\end{align}
where $ {\mathcal{G}}^{B}_{E0} (\bm{z})$ stands for the corresponding
electric charge distribution 
\begin{align}
{\cal{G}}^{B}_{E0} (\bm{z})=& \frac{1}{\sqrt{3}} \langle
  D^{(8)}_{Q8}\rangle_B \mathcal{B}(\bm{z}) -
  \frac{2}{I_{1}} \langle
  D^{(8)}_{Qi} \hat{J}_{i} \rangle_B
  {\cal{I}}_{1}(\bm{z}) -
  \frac{2}{I_{2}} \langle
  D^{(8)}_{Qp} \hat{J}_{p} \rangle_B
  {\cal{I}}_{2}(\bm{z}) \cr 
    & -\frac{4 m_{8}}{I_{1}} \langle D^{(8)}_{8i} D^{(8)}_{Qi}
      \rangle_B (I_{1}{\cal{K}}_{1}(z) - K_{1}{\cal{I}}_{1}(z)) 
 -\frac{4 m_{8}}{I_{2}} \langle D^{(8)}_{8p} D^{(8)}_{Qp}
   \rangle_B (I_{2}{\cal{K}}_{2}(z) - K_{2}{\cal{I}}_{2}(z)) \cr 
 &-2\left (  \frac{m_1}{\sqrt{3}}\langle D^{(8)}_{Q8} \rangle_B +
   \frac{m_8}{3}\langle D^{(8)}_{8 8}D^{(8)}_{Q8} \rangle_B  \right)
   {\cal{C}}(\bm{z}), 
\label{eq:elecfinal}
\end{align}
where the explicit expressions for the densities
$\mathcal{B}(\bm{z})$, $\mathcal{I}_{1}(\bm{z})$,
$\mathcal{I}_{2}(\bm{z})$, $\mathcal{K}_{1}(\bm{z})$,
$\mathcal{K}_{2}(\bm{z})$ and $\mathcal{C}(\bm{z})$ can be found in   
Refs.~\cite{Kim:1995mr,Ledwig:1900ri}. The indices $i$ and $p$  
denote dummy ones running over $i=1,\cdots, 3$ and $p=4,\cdots 7$,
respectively.  Since the integrations of the densities in
Eq.~\eqref{eq:elecfinal} are given as 
\begin{align}
\int d^3 z \,\mathcal{B}(\bm{z}) ={N_{c}},\;\;\;
\frac1{I_i}\int d^3 z \,\mathcal{I}_i (\bm{z}) = 1,\;\;\;  
\frac1{K_i}\int d^3 z \,\mathcal{K}_i (\bm{z}) = 1,\;\;\;  
\int d^3 z \,\mathcal{C}(\bm{z}) = 0,
\end{align}
we find that the electric charge is preserved by the relation
\begin{align}
  \label{eq:e_charge28}
\mathcal{Q}_B = \left \langle \hat{T}_3 +
  \frac{\hat{Y}}{2}\right\rangle_B =   \frac{N_c}{\sqrt{3}} \langle
  D^{(8)}_{Q8}\rangle_B  - 2 \langle   D^{(8)}_{Qi} \hat{J}_{i}
  \rangle_B - 2 \langle D^{(8)}_{Qp}   \hat{J}_{p} \rangle_B,  
\end{align}
where $\hat{T}_3$ and $\hat{Y}$ denote the operators for the third
component of the isospin and the hypercharge, respectively. Thus,  
the electric form factor $G_{E0}^B$ at $Q^2=0$ is just the charge of
the corresponding baryon. Note that we also have considered the
symmetry-conserving quantization~\cite{Praszalowicz:1998jm}. 

The expression for the magnetic dipole moment form factor of a baryon
$B$ is written as  
\begin{align}
{{G}}^{B}_{M1}  (q^{2})=\frac{ M_{B}}{|\bm{q}|} \int d^{3} z
  \frac{j_{1}(|\bm{q}| |\bm{z}|)}{ |\bm{z}|}
 \mathcal{G}^{B}_{M1}  (\bm{z}),
\label{eq:magfinal}
\end{align}
where the magnetic distribution $\mathcal{G}_{M1}^B$ is given
by 
\begin{align}
\mathcal{G}^{B}_{M1}(\bm{z}) &=  \langle D^{(8)}_{Q3 }
  \rangle_B  \left(  {\cal{Q}}_{0} (\bm{z})  + \frac{1}{I_{1}}
 {\cal{Q}}_{1} (\bm{z}) \right) -  \frac{1}{\sqrt{3}} \langle
 D^{(8)}_{Q 8}J_{3} \rangle_B \frac{1}{I_{1}}
 {\cal{X}}_{1} (\bm{z}) - \langle
 d_{pq3} D^{(8)}_{Qp} J_{q} \rangle_B
 \frac{1}{I_{2}}  {\cal{X}}_{2} (\bm{z})   \cr 
& + \frac{2}{\sqrt{3}} m_{8} \langle D^{(8)}_{83}
  D^{(8)}_{Q8} \rangle_B \left(\frac{K_{1}}{I_{1}}{\cal{X}}_{1}
  (\bm{z}) -   {\cal{M}}_{1} (\bm{z})\right)  
  +2 m_{8} \langle  d_{pq3}  D^{(8)}_{8p}
  D^{(8)}_{Qq} \rangle_B \left(\frac{K_{2}}{I_{2}}{\cal{X}}_{2} (\bm{z})
     -    {\cal{M}}_{2} (\bm{z})\right)  \cr 
& - 2  \left( m_{1} \langle D^{(8)}_{Q3} \rangle_B +
  \frac{1}{\sqrt{3}} m_{8} \langle D^{(8)}_{88} D^{(8)}_{Q3}
  \rangle_B  \right) {\cal{M}}_{0} (\bm{z}). 
\label{eq:magden1}
\end{align}
Here the indices $p$ and $q$ are the dummy indices running over
$4\cdots 7$. The explicit forms of the magnetic densities can be
found in Ref.~\cite{Kim:1995mr, Ledwig:1900ri}. The matrix elements of 
the collective operators are explicitly presented in
Appendix~\ref{app:A}.  

The expression for the electric quadrupole form factor of a
baryon $B$ is obtained as  
\begin{align}
G^{B}_{E{2}}(Q^{2}) &=  6 \sqrt{{5} } M^{2}_{B} \int d^{3} z
                      \frac{j_{2}(|\bm{q}||\bm{z}|)}{|\bm{q}|^{2}}
                      \mathcal{G}^{B}_{E2}(\bm{z}) ,  
\label{eq:magfinal2}
\end{align}
where $\mathcal{G}_{E2}^B$ is given by 
\begin{align}
\mathcal{G}^{B}_{E2}(\bm{z}) =& -2 \left( \frac{3}{I_{1}} \langle
\ D^{(8)}_{Q 3} J_{3} \rangle_{B} - \frac1{I_1} \langle D^{(8)}_{Q i}
J_{i} \rangle_{B} \right) \mathcal{I}_{1E2}  (\bm{z}) \cr   
 & + 4 m_{8}\left( \frac{K_{1}}{I_{1}} \mathcal{I}_{1E2}(\bm{z}) -
    \mathcal{K}_{1E2}(\bm{z})\right) \left( 3\langle D^{(8)}_{8 3}
    D^{(8)}_{Q 3}\rangle_{B} -\langle D^{(8)}_{8 i} D^{(8)}_{Q i}
    \rangle_{B} \right). 
\label{eq:magden}
\end{align}
The expressions for the densities $\mathcal{I}_{1E2}$ and $
\mathcal{K}_{1E2}$ can be found in Appendix~\ref{app:B}. Though we
could also express the magnetic octupole form factors, we will not
present them here, because they all vanish within the present model.  

It is convenient to decompose the densities into the flavor SU(3)
symmetric and explicit symmetry-breaking terms. In fact, there are two
different contributions from the explicit symmetry-breaking terms: The 
one arises from the linear $m_{\mathrm{s}}$ term in the effective
chiral action~\eqref{eq:echl} and the other comes from the mixed
collective wavefunctions~\eqref{eq:mixedWF1}. Thus, we can split the
densities in terms of these three different terms 
\begin{align}
\mathcal{G}^{B}_{E0}(\bm{z}) &= \mathcal{G}^{B(0)}_{E0}(\bm{z}) +
                             \mathcal{G}^{B(\text{op})}_{E0}(\bm{z})
                             +\mathcal{G}^{B(\text{wf})}_{E0}(\bm{z}),
                             \cr 
\mathcal{G}^{B}_{M1}(\bm{z}) &= \mathcal{G}^{B(0)}_{M1}(\bm{z})
                             +\mathcal{G}^{B(\text{op})}_{M1}(\bm{z})
                             +\mathcal{G}^{B(\text{wf})}_{M1}(\bm{z}),
                             \cr 
\mathcal{G}^{B}_{E2}(\bm{z}) &= \mathcal{G}^{B(0)}_{E2}(\bm{z})
                             + \mathcal{G}^{B(\text{op})}_{E2}(\bm{z})
                             + \mathcal{G}^{B(\text{wf})}_{E2}(\bm{z}),   
\end{align}
where $\mathcal{G}_{E0,M1,E2}^{B(0)}$,
$\mathcal{G}_{E0,M1,E2}^{B(\mathrm{op})}$, and
$\mathcal{G}_{E0,M1,E2}^{B(\mathrm{wf})}$ represent the symmetric
terms, the flavor SU(3) symmetry-breaking ones from the
effective chiral action, and those from the mixed collective
wavefunctions, respectively. They are written explicitly as  
\begin{align}
{\cal{G}}^{B(0)}_{E0}(\bm{z}) =&
  \left(\frac{1}{24}{\mathcal{B}}(\bm{z}) + \frac{5}{8I_{1}}
  {\mathcal{I}}_{1}(\bm{z})+\frac{1}{4I_{2}} {\mathcal{I}}_{2}(\bm{z})
  \right) \mathcal{Q}_B, 
\label{eq:E0leading} \\
{\cal{G}}^{B(\text{op})}_{E0}(\bm{z}) =& -\frac{1}{168\sqrt{3}}\left(
\begin{array}{c} 13\mathcal{Q}_{\Delta} + 19 \\ 12
  \mathcal{Q}_{\Sigma^*} + 31 \\ 11 
  \mathcal{Q}_{\Xi^*} + 43 \\ -45 \mathrm{Q}_{\Omega^-} \end{array}
  \right) \frac{4m_{8}}{I_{1}}\bigg{(} I_{1} {\mathcal{K}}_{1}(\bm{z})
  - K_{1} {\mathcal{I}}_{1}(\bm{z})\bigg{)} \cr 
    & \hspace{-1cm}
-\frac{1}{168\sqrt{3}}\left( \begin{array}{c}
                                    -10\mathcal{Q}_{\Delta} + 50 
\\ -6 \mathcal{Q}_{\Sigma^*}+44 \\ -2 \mathcal{Q}_{\Xi^*} + 38 \\ -30
                                    \mathrm{Q}_{\Omega^-}  \end{array} \right)
  \frac{4m_{8}}{I_{2}} 
  \bigg{(} I_{2} {\mathcal{K}}_{2}(\bm{z})-
  K_{2} {\mathcal{I}}_{2}(\bm{z})\bigg{)} 
-2 \bigg{(} \frac{m_{1}}{24}+
 \frac{m_{8}}{168\sqrt{3}}\left( \begin{array}{c} -
                                   \mathcal{Q}_{\Delta} + 5
\\ -2 \mathcal{Q}_{\Sigma^*} + 3 \\ -3 \mathcal{Q}_{\Xi^*} + 1 \\ -3
                                   \mathcal{Q}_{\Omega^-}  \end{array}
  \right) \bigg{)} 
  {\mathcal{C}} (\bm{z}), \\ 
  {\cal{G}}^{B(\text{wf})}_{E0}(\bm{z}) =&+ a_{27}\left(
\begin{array}{c c c} {\frac{5}{24}}(- \mathcal{Q}_{\Delta} + 2)
  \\ {\frac{1}{12}}(-3 \mathcal{Q}_{\Sigma^*} + 2) \\  {\frac{1
  }{24}}(-7 \mathcal{Q}_{\Xi^*} - 2) \\ 0
\end{array} \right) \left({\mathcal{B}}(\bm{z})-\frac{5}{I_{1}}
  {\mathcal{I}}_{1}(\bm{z})+\frac{2}{I_{2}}{\mathcal{I}}_{2}(\bm{z})
  \right) \cr
& +a_{35}\left( \begin{array}{c c c}
                  {\frac{1}{56}}(-\mathcal{Q}_{\Delta} - 2)
                  \\ {\frac{2}{56}}(-\mathcal{Q}_{\Sigma^*} - 2)
                  \\ {\frac{3}{56}}(-\mathcal{Q}_{\Xi^*} - 2)
                  \\ {\frac{4}{56}}(-\mathcal{Q}_{\Omega^-} -
                  2)  \end{array} \right) 
  \left(-5 {\mathcal{B}}(\bm{z})-\frac{5}{I_{1}}
  {\mathcal{I}}_{1}(\bm{z})+\frac{10}{I_{2}}{\mathcal{I}}_{2}(\bm{z})\right),
\label{eq:E0lmscorr}
\end{align}
for the electric monopole form factors,
\begin{align}
  {\cal{G}}^{B_{10}(0)}_{M1}(\bm{z}) &= -\frac{1}{8} \mathcal{Q}_B
\left[ \left(  {\cal{Q}}_{0} (\bm{z})  + \frac{1}{I_{1}}
 {\cal{Q}}_{1} (\bm{z}) \right)  +  \frac{1}{2}  \frac{1}{I_{1}}
 {\cal{X}}_{1} (\bm{z}) + \frac{1}{2}
 \frac{1}{I_{2}}  {\cal{X}}_{2} (\bm{z}) \right ],
\label{eq:M1leadingcon} \\
{\cal{G}}^{B_{10}(\text{op})}_{M1}(\bm{z}) &=
-\frac{m_{8}}{84\sqrt{3}}\left (
\begin{array}{c} 5 \mathcal{Q}_{\Delta} - 4 \\ 3
  \mathcal{Q}_{\Sigma^*} - 1 \\ \mathcal{Q}_{\Xi^{*}} + 2 \\ -6
  \mathcal{Q}_{\Omega^-}  \end{array} \right)
  \left(\frac{K_{1}}{I_{1}}{\cal{X}}_{1}
  (\bm{z}) -   {\cal{M}}_{1} (\bm{z})\right) \cr 
 & \hspace{-1cm} + \frac{m_1}{4} \mathcal{Q}_B {\mathcal{M}}_{0} (\bm{z})
   -\frac{m_{8}}{84\sqrt{3}} \left ( 
            \begin{array}{c} 11 \mathcal{Q}_{\Delta} - 13 \\
     15 \mathcal{Q}_{\Sigma^*} + 2 \\  19 \mathcal{Q}_{\Xi^*} + 17 \\
              -9 \mathcal{Q}_{\Omega^-}  \end{array}
  \right)  \left(\frac{K_{2}}{I_{2}}{\cal{X}}_{2} (\bm{z})
     -    {\cal{M}}_{2} (\bm{z})\right)  
+  \frac{m_{8}}{84\sqrt{3}}
\left ( \begin{array}{c}
5\mathcal{Q}_{\Delta} - 4 \\ 3 \mathcal{Q}_{\Sigma^*} - 1 \\
          \mathcal{Q}_{\Xi^*} +2 \\ -6 \mathcal{Q}_{\Omega^-} \end{array}
  \right) \mathcal{M}_{0} (\bm{z}), 
\label{eq:M1mslinear1} \\
  {\cal{G}}^{B_{10}(\text{wf})}_{M1}(\bm{z}) &=
a_{27}\left( \begin{array}{ccc}
               {\frac{5}{24}}(- \mathcal{Q}_{\Delta} + 2) \\
               {\frac{1}{12}}(-3 \mathcal{Q}_{\Sigma^*} + 2)
               \\ {\frac{1 }{24}}(-7 \mathcal{Q}_{\Xi^*} - 2) \\
               0  \end{array} \right) 
  \left[  \left(  {\cal{Q}}_{0} (\bm{z})  + \frac{1}{I_{1}}
 {\cal{Q}}_{1} (\bm{z}) \right)  -  \frac{3}{2}  \frac{1}{I_{1}}
 {\cal{X}}_{1} (\bm{z}) - \frac{1}{2}
 \frac{1}{I_{2}}  {\cal{X}}_{2} (\bm{z}) \right ] \cr
&+ a_{35}\left( \begin{array}{c c c} {\frac{1}{56}}(-
                  \mathcal{Q}_{\Delta} - 2)
                  \\ {\frac{2}{56}}(-\mathcal{Q}_{\Sigma^*} - 2) \\
                  {\frac{3}{56}}(- \mathcal{Q}_{\Xi^*} - 2)
                  \\ {\frac{4}{56}}(-\mathcal{Q}_{\Omega^-} -
                  2)  \end{array} \right) 
  \left[  \left(  {\cal{Q}}_{0} (\bm{z})  + \frac{1}{I_{1}}
 {\cal{Q}}_{1} (\bm{z}) \right)  +  \frac{5}{2}  \frac{1}{I_{1}}
 {\cal{X}}_{1} (\bm{z}) - \frac{5}{2}
 \frac{1}{I_{2}}  {\cal{X}}_{2} (\bm{z}) \right ],
\label{eq:M1final}
\end{align}
for the magnetic dipole form factors, and 
\begin{align}
  {\cal{G}}^{B_{10}(0)}_{E2} &= \frac{1}{2I_1} \mathcal{Q}_B
{\cal {I}}_{1E2} (\bm{z}), \;\;\; {\cal{G}}^{B_{10}(\text{op})}_{E2} =
                               \frac{4}{63\sqrt{3}}m_{8}
                               \left(\frac{K_{1}}{I_{1}}
                               \mathcal{I}_{1E2}(\bm{z}) -
                               \mathcal{K}_{1E2}(\bm{z}) \right) 
\left( \begin{array}{ccc}
 4Q_{\Delta}+1 \\
 -6Q_{\Sigma^{*}}-5 \\
-16Q_{\Xi^{*}}-11 \\
 9  \end{array} \right), \cr
{\cal{G}}^{B_{10}(\text{wf})}_{E2} &= -\frac{4}{I_{1}} \bigg{
(}a_{27}\left( \begin{array}{ccc}{\frac{5}{24}}(-
  \mathcal{Q}_{\Delta}+2) \\ {\frac{1}{12}}(-3 \mathcal{Q}_{\Sigma^*}
                 + 2) \\ {\frac{1}{24}}(-7 \mathcal{Q}_{\Xi^*} - 2) \\
0  \end{array} \right)+ a_{35}\left( \begin{array}{c c c}
{\frac{1}{56}}(- \mathcal{Q}_{\Delta}-2) \\
{\frac{2}{56}}(-\mathcal{Q}_{\Sigma^*}-2) \\
{\frac{3}{56}}(-\mathcal{Q}_{\Xi^*}-2)
\\ {\frac{4}{56}}(-\mathcal{Q}_{\Omega^-} - 2)
                                     \end{array} \right)
  \bigg{)}{\cal {I}}_{1E2} (\bm{z}) 
\label{eq:E2final}
\end{align}
for the electric quadrupole form factors. Note that all the symmetric 
contributions to the EM form factors of a decuplet baryon are 
proportional to the corresponding charge. It indicates that all the
EM form factors of the neutral baryons in the decuplet vanish in
the chiral limit~\cite{Kim:1997ip}. Moreover, the $m_{\mathrm{s}}$
corrections on the $E2$ form factors of the neutral decuplet baryons
come from the collective operators and collective wavefunctions.
It implies that the 
magnitudes of the $E2$ form factors of the neutral baryons in the
decuplet are in particular small, compared to those of other members
of the baryon decuplet.

In the $\chi$QSM, various sum rules for the magnetic dipole moments of the
baryon decuplet can be derived, as already shown in
Ref.~\cite{Kim:1997ip}. If we examine the expressions for the $E2$
form factors of the baryon decuplet given in 
Eq.~\eqref{eq:E2final}, then we can find that they have the same group 
structure as those of the $M1$ form factors~\eqref{eq:M1final}.
Thus, we can obtain the similar sum rules for the electric quadrupole
moments. In the chiral limit, one can find the following
relations~\cite{Kim:1997ip}: 
\begin{align}
Q_{\Sigma^{*0}} &= \frac{1}{2} (Q_{\Sigma^{*+}} + Q_{\Sigma^{*-}}), \cr
Q_{\Delta^{-}} + Q_{\Delta^{++}}&= Q_{\Delta^{0}} + Q_{\Delta^{+}}, \cr
\sum_{B\in\mathrm{decuplet}} Q_{B} &= 0.
\end{align}
Even though the flavor SU(3) symmetry is broken, we still can find the
following sum rules
\begin{align}
-4 Q_{\Delta^{++}} + 6 Q_{\Delta^{+}} + 3 Q_{\Sigma^{*+}} 
- 6 Q_{\Sigma^{*0}}  + Q_{\Omega^{-}} &= 0, \cr
-2 Q_{\Delta^{++}} + 3 Q_{\Delta^{+}} + 2 Q_{\Sigma^{*+}}
 - 4 Q_{\Sigma^{*0}}  + Q_{\Xi^{*-}} &= 0, \cr
-Q_{\Delta^{++}} + 2 Q_{\Delta^{+}}  - 2 Q_{\Sigma^{*0}}
  + Q_{\Xi^{*0}} &= 0, \cr
 Q_{\Sigma^{*+}} - 2 Q_{\Sigma^{*0}}  + Q_{\Sigma^{*-}} &= 0, \cr
 2Q_{\Delta^{++}} -3 Q_{\Delta^{+}} + Q_{\Delta^{-}} &= 0, \cr
 Q_{\Delta^{++}} -2 Q_{\Delta^{+}} + Q_{\Delta^{0}} &= 0.
\end{align}
We also get the differences between the electric quadrupole moments of
the baryon decuplet. 
\begin{align}
\frac{1}{3}(Q_{\Omega^{-}}-Q_{\Delta^{++}}) = 
\frac{1}{2}(Q_{\Xi^{*-}}-Q_{\Delta^{+})} =
 Q_{\Xi^{*0}}-Q_{\Sigma^{*+}} = Q_{\Sigma^{*-}}-Q_{\Delta^{0}}.
\end{align}

\section{Results and discussion}
\label{sec:4}
Before we present the numerical results of the present work, we first
explain how to fix the parameters of the model. In the $\chi$QSM, the
dynamical quark mass $M$ is the only free parameter, which was already
determined by computing various form factors of the nucleon in the
previous works~\cite{Kim:1995mr,Christov:1995vm}. The most 
preferable value is $M = 420$ MeV. We have already examined that the
numerical results of all the EM form factors in this work are
insensitive to the value of $M$. Thus, we adopt the same value of $M =
420$ MeV in the present work. The average mass of the current up and
down quarks $\overline{m}$ is fixed by reproducing the pion mass in
the mesonic sector. The strange current quark mass $m_{\mathrm{s}}$ is
taken to be 180 MeV, which produces very well the mass splittings of
the SU(3) baryons and singly heavy baryons~\cite{Christov:1995vm,
  Kim:2018xlc}. Since the divergences from the sea quarks or from the
vacuum polarizations are tamed by introducing the regularization, we
fix the cutoff parameter $\Lambda$ by reproducing the pion decay
constant $f_\pi = 93$ MeV. 

\subsection{Contributions of the valence quarks and sea-quark
  polarization} 
 \begin{figure}[htp]
 \includegraphics[scale=0.2]{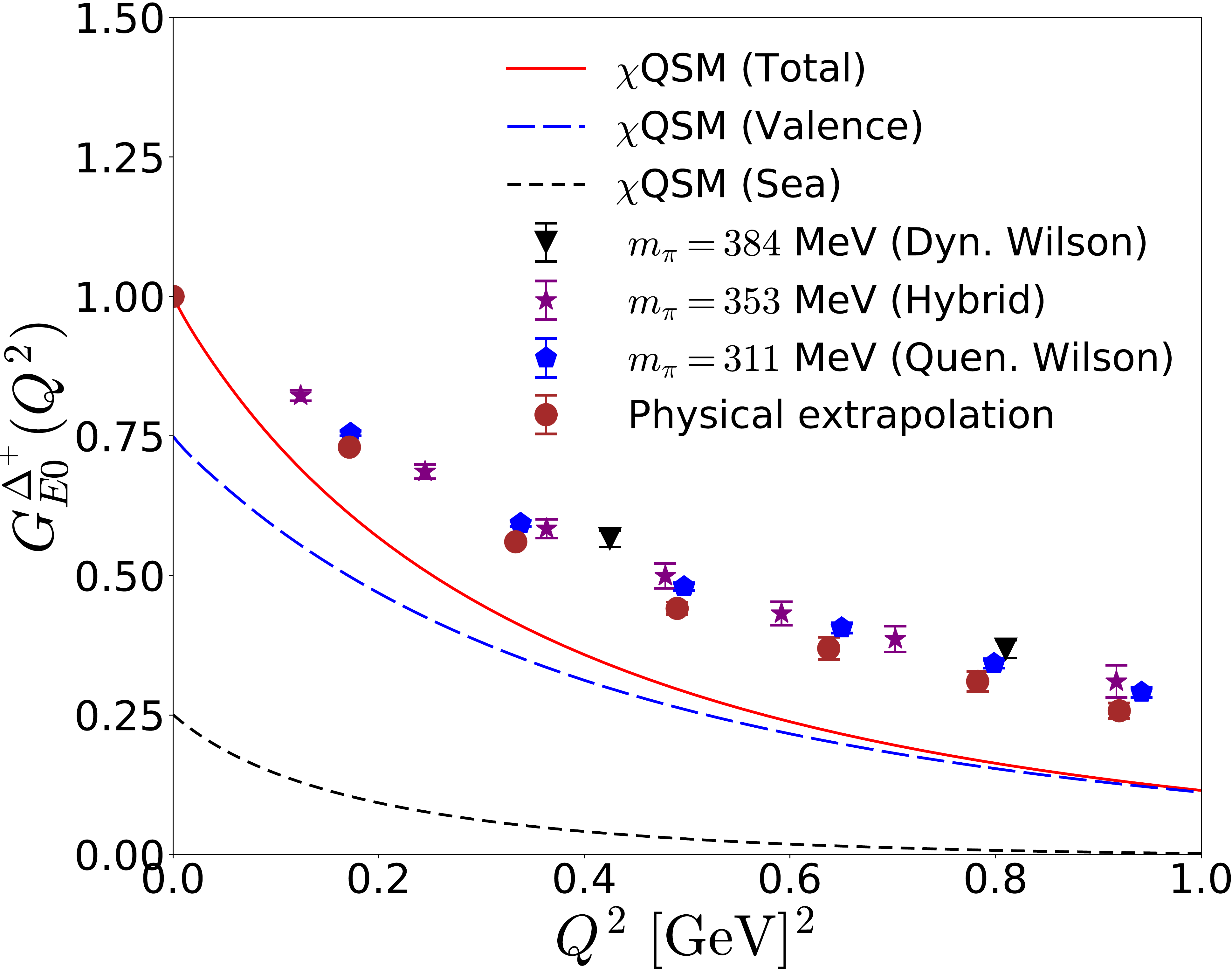}\hspace{0.5cm}
  \includegraphics[scale=0.2]{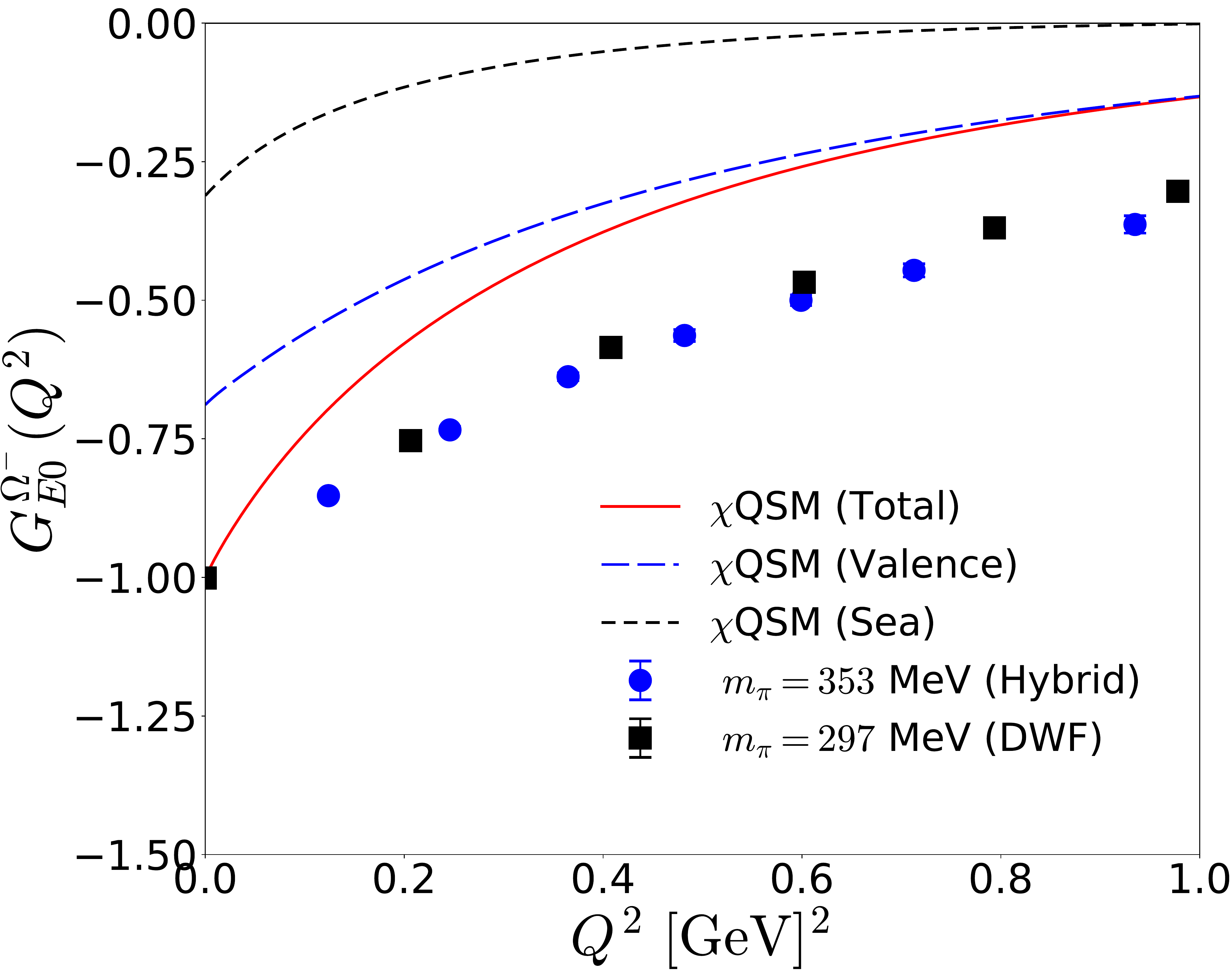}
\caption{Contributions of the valence and sea quarks to the electric
  monopole form factors of the $\Delta^+$ isobar and $\Omega^-$ in the
  left and right panels, respectively. The dashed curves
  depict the valence-quark contributions, the short-dashed ones draw
  those of the sea quarks, and the solid ones show the total
  results. The lattice data are taken from
  Refs.~\cite{Alexandrou:2007we,Alexandrou:2009hs,Alexandrou:2010jv}.  
The data of the EM from factors of the $\Delta^{+}$ can be found in
Refs.~\cite{Alexandrou:2009hs}, and especailly, the values of the
  physical extrapolation are taken from Ref.~\cite{Alexandrou:2007we},
   whereas those of the EM form factors of $\Omega^{-}$ can be found
   in Ref.~\cite{Alexandrou:2010jv}.}    
\label{fig:1}
\end{figure}
Since there exist the lattice data on the $\Delta^+$ isobar and
$\Omega^-$ hyperon EM form factors, we will first focus on those of
the $\Delta^+$  and $\Omega^-$ and then will discuss the EM form
factors of all the other members in the baryon decuplet.
In Fig.~\ref{fig:1}, we draw the numerical results for the $E0$ form
factors of the $\Delta^+$ and $\Omega^-$ in the left and right panels
respectively, examining the valence-quark and sea-quark 
contributions. As shown in Fig.~\ref{fig:1} explicitly, the
valence-quark contributions are the dominant ones in general. 
The valence quarks contribute to the $\Delta^+$ electric monopole form
factor by about 75~\% whereas the sea quarks provide approximately
25~\% contribution to it.  
We also see that the general feature of each contribution 
seems very similar in the case of the $\Omega^-$, except that the
effect of the sea-quark polarization becomes slightly larger than that
on the $\Delta^+$ $E0$ form factor. Note that the sea-quark
contribution to the proton electric monopole form factor is smaller
than those to the $\Delta^+$ and $\Omega^-$ form factors. 

Compared with the lattice data of the $\Delta^+$ and $\Omega^-$
electric monopole form factors, we find that the present results fall off
faster than those of the lattice calculation as $Q^2$
increases. However, the lattice calculations tend to provide the
results of the electric monopole form factor of the proton, which
decrease more slowly than the experimental data, in particular, when    
the unphysical value of the pion mass is used~\cite{Capitani:2015sba,
  Abdel-Rehim:2015jna, Djukanovic:2015hnh, Chambers:2017tuf}. Even
though the physical pion mass is employed, the lattice results fall
off still more slowly than the experimental
data~\cite{Alexandrou:2017ypw}. One can find the similar tendency 
in the case of the tensor and anomalous form factors of the
nucleon. Compared with the lattice data on these form
factors~\cite{Gockeler:2005cj, Gockeler:2006zu} , the results of the
$\chi$QSM again decrease more slowly~\cite{Ledwig:2010tu,
  Ledwig:2010zq}. In this respect, it is natural for the present
results to fall off faster than the lattice ones. 
We will discuss the comparison with the lattice data in detail in
Subsection~\ref{subsec:IV-D},
considering the chiral extraponlation.

\begin{figure}[htp]
\includegraphics[scale=0.2]{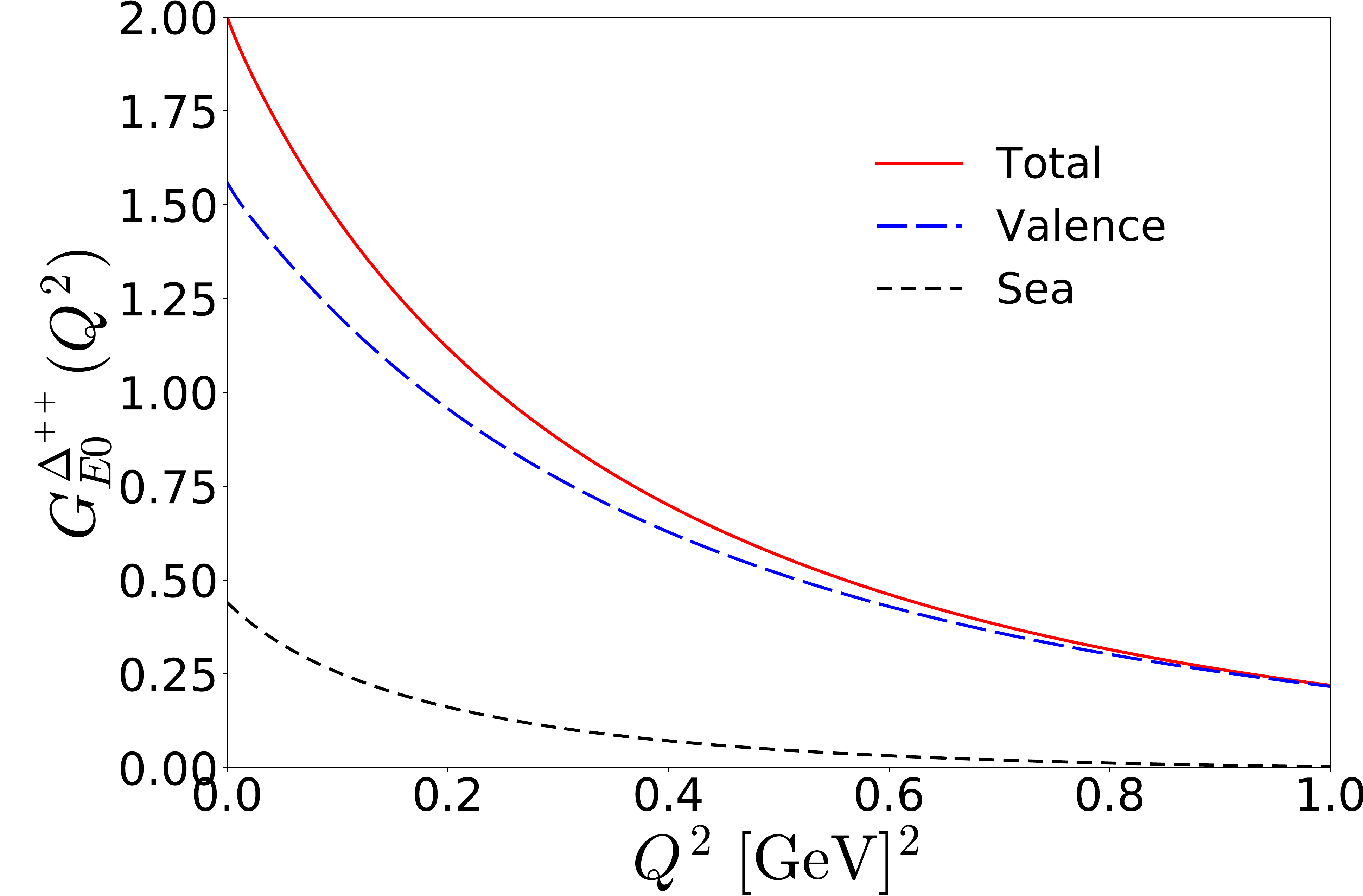} \hspace{0.5cm}
\includegraphics[scale=0.2]{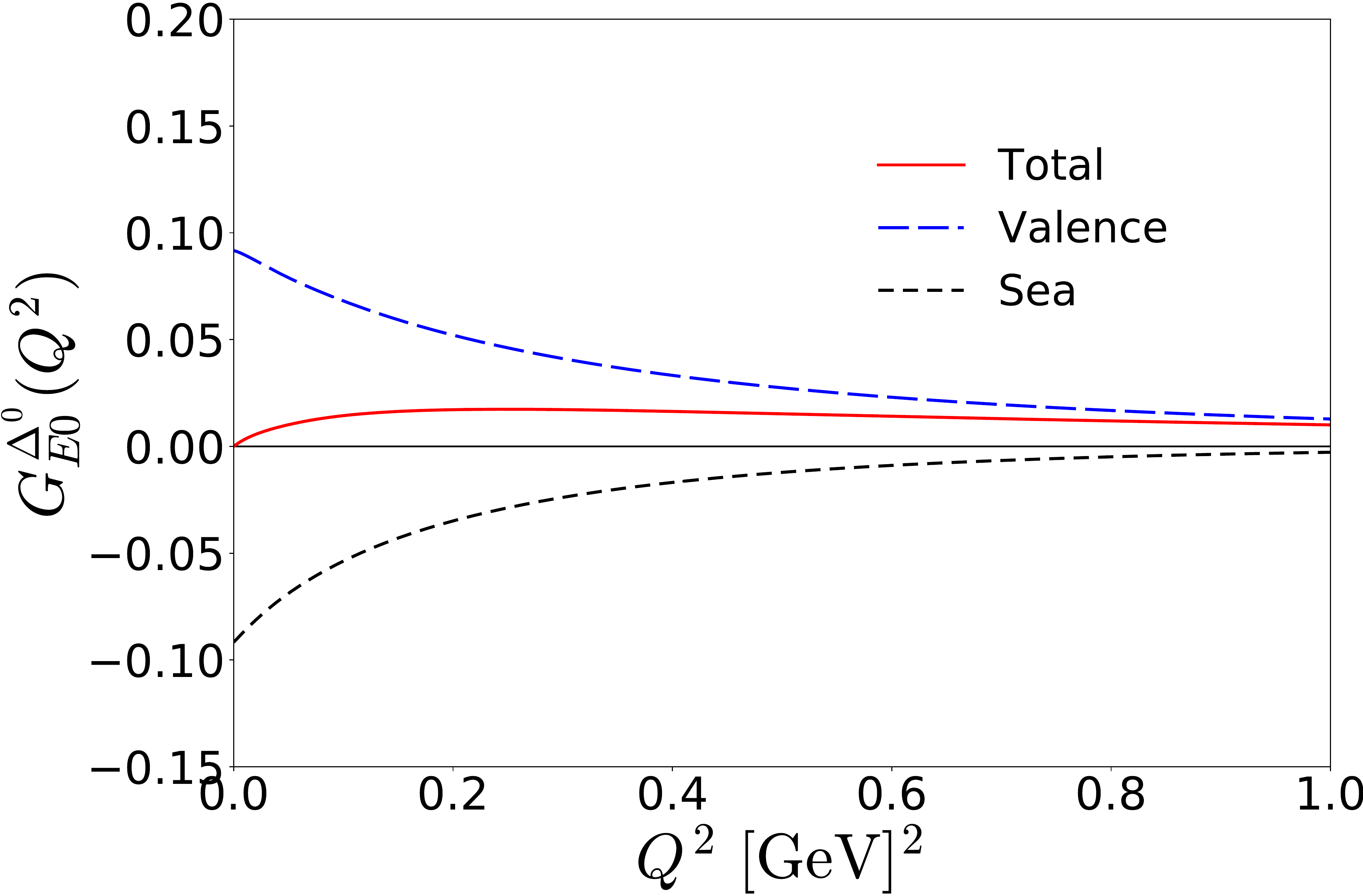}
\includegraphics[scale=0.2]{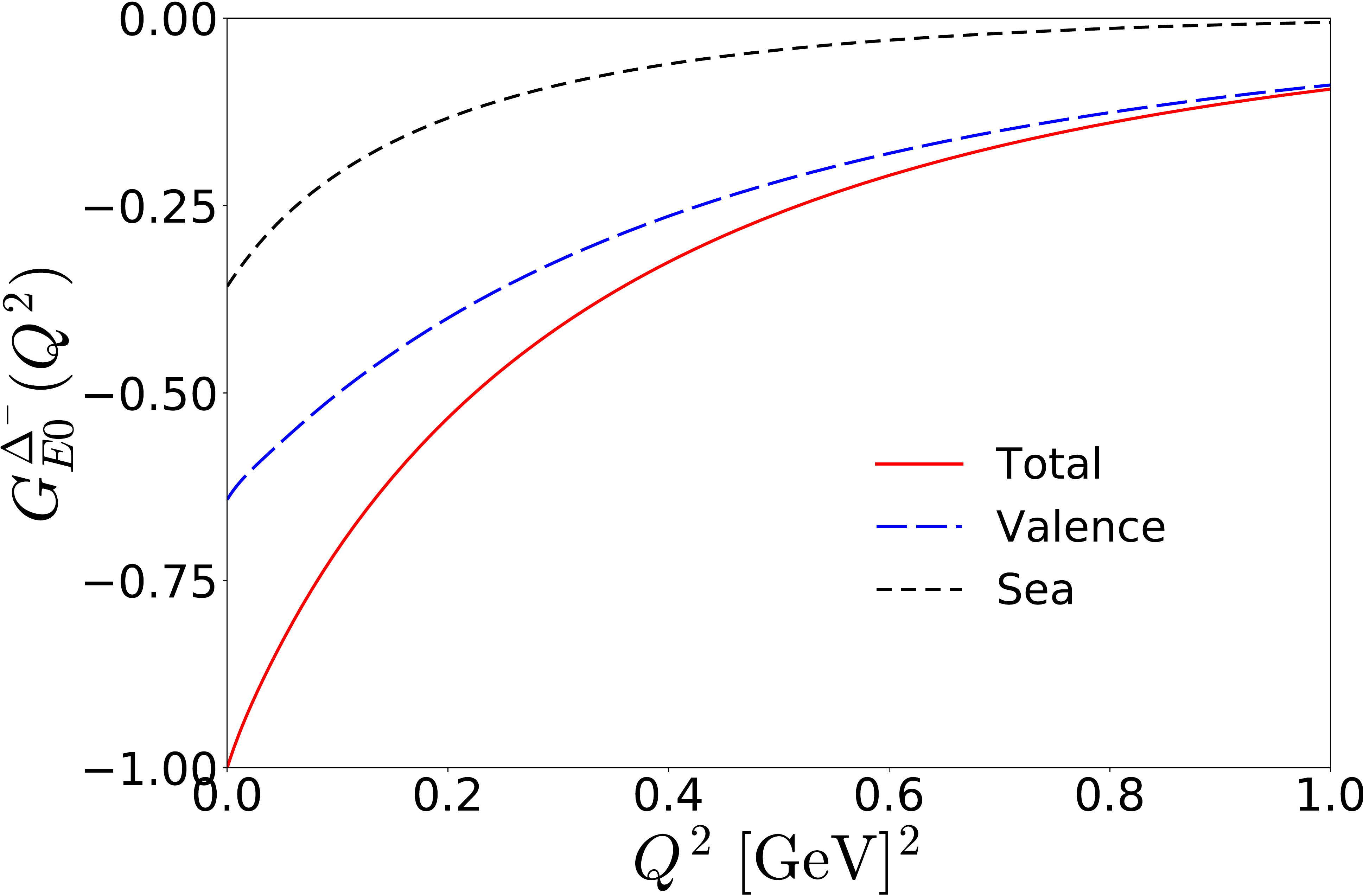}\hspace{0.5cm}
\includegraphics[scale=0.2]{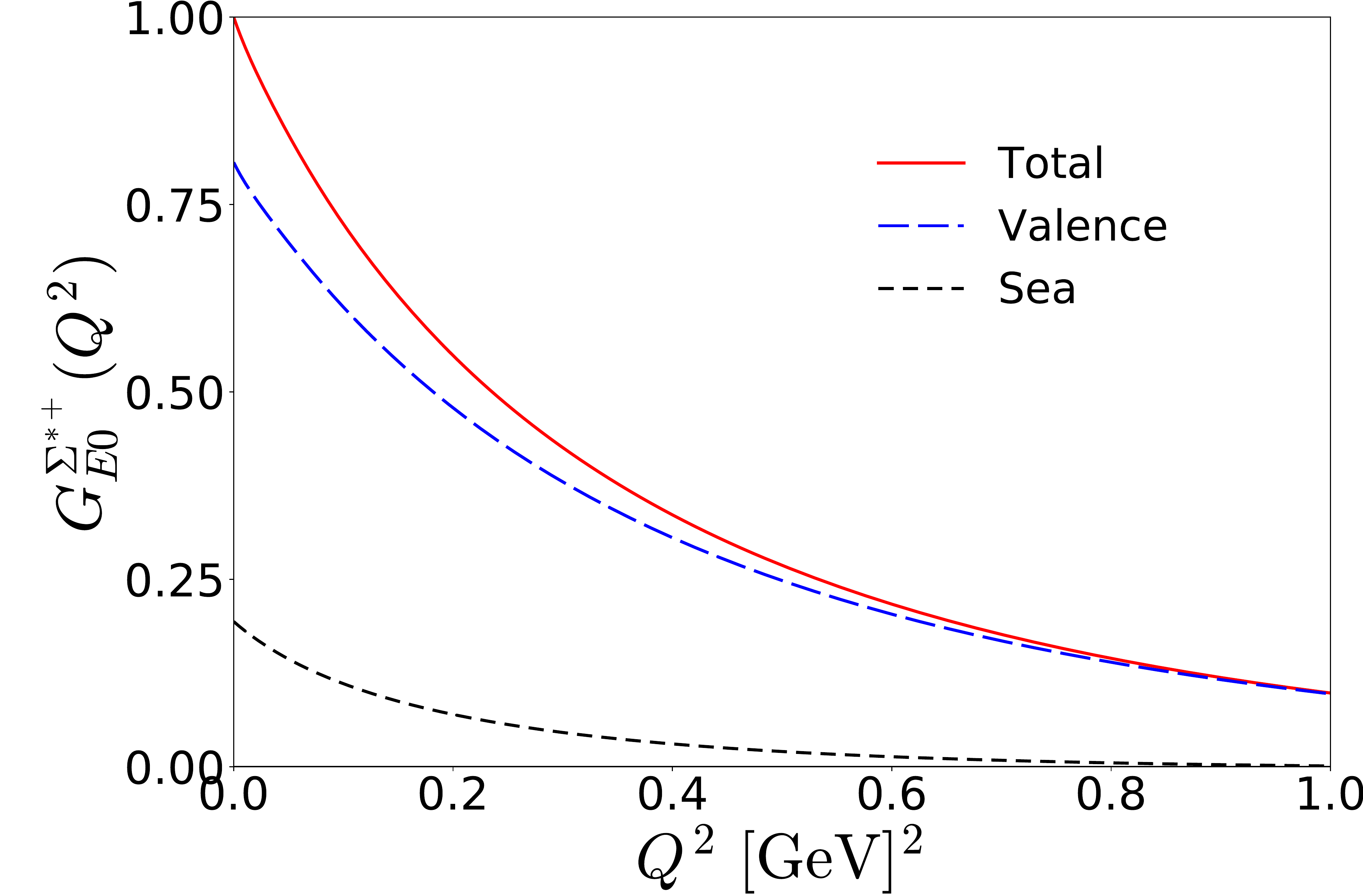}
\includegraphics[scale=0.2]{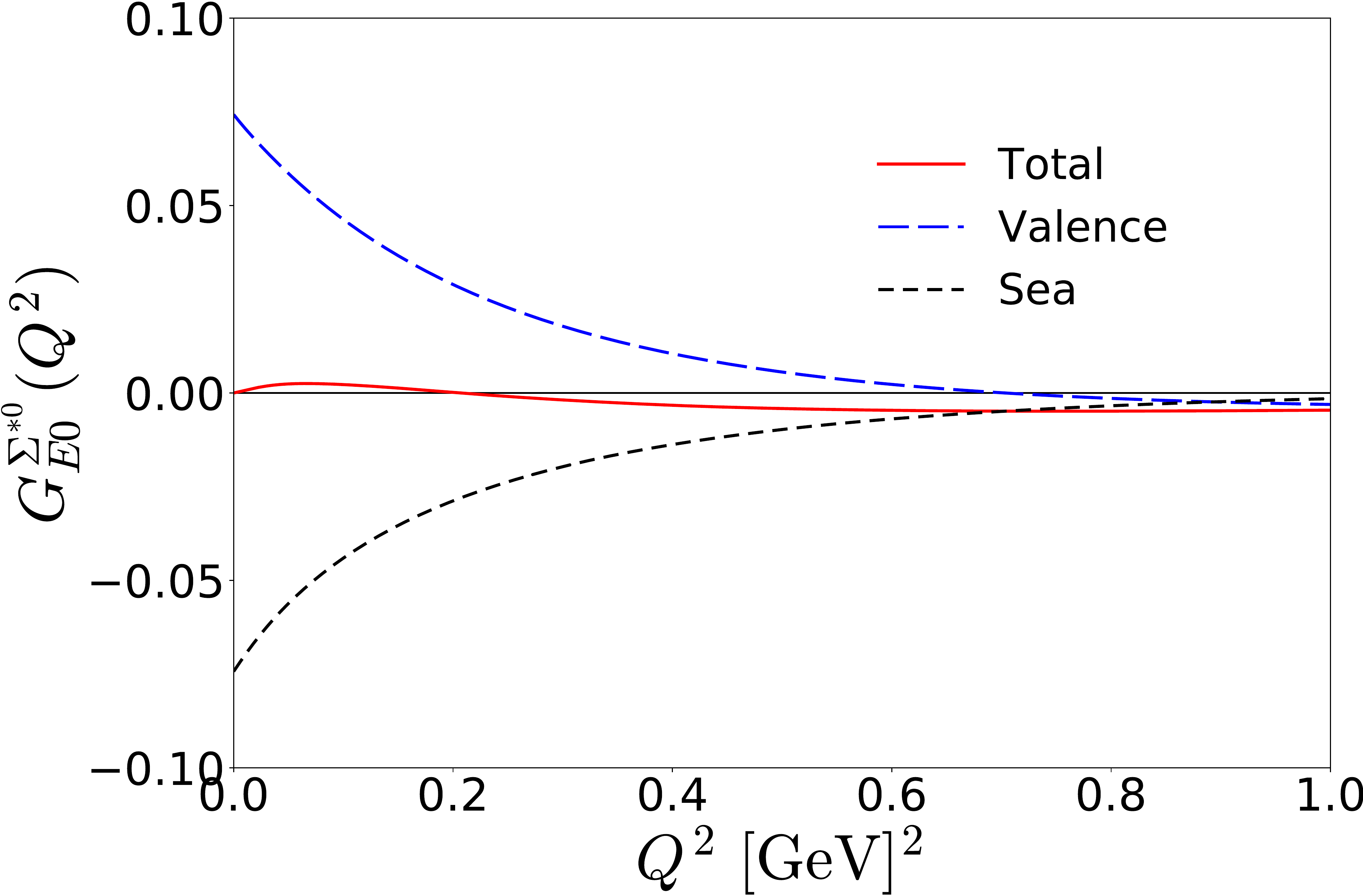}\hspace{0.5cm}
\includegraphics[scale=0.2]{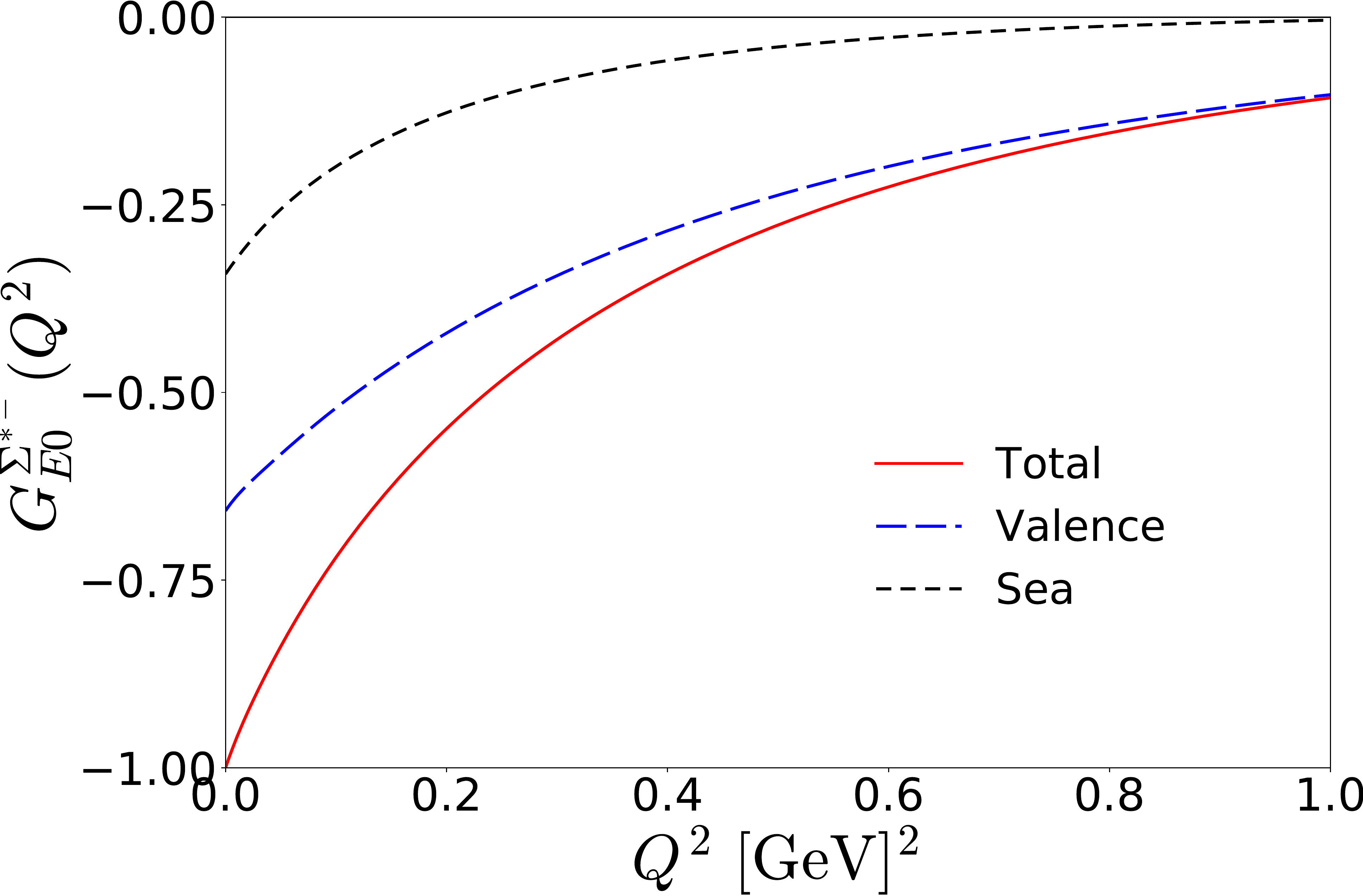}
\includegraphics[scale=0.2]{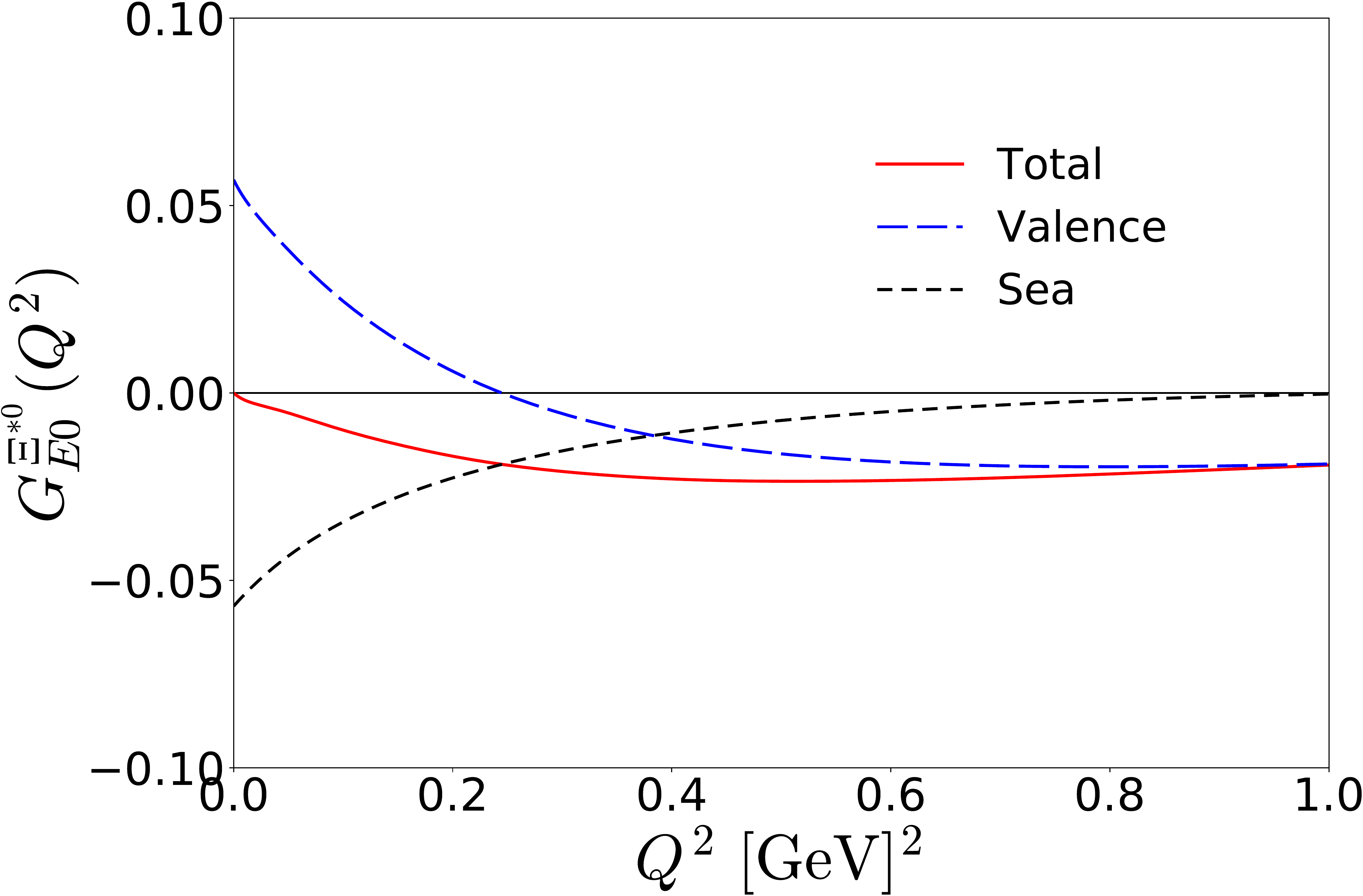}\hspace{0.5cm}
\includegraphics[scale=0.2]{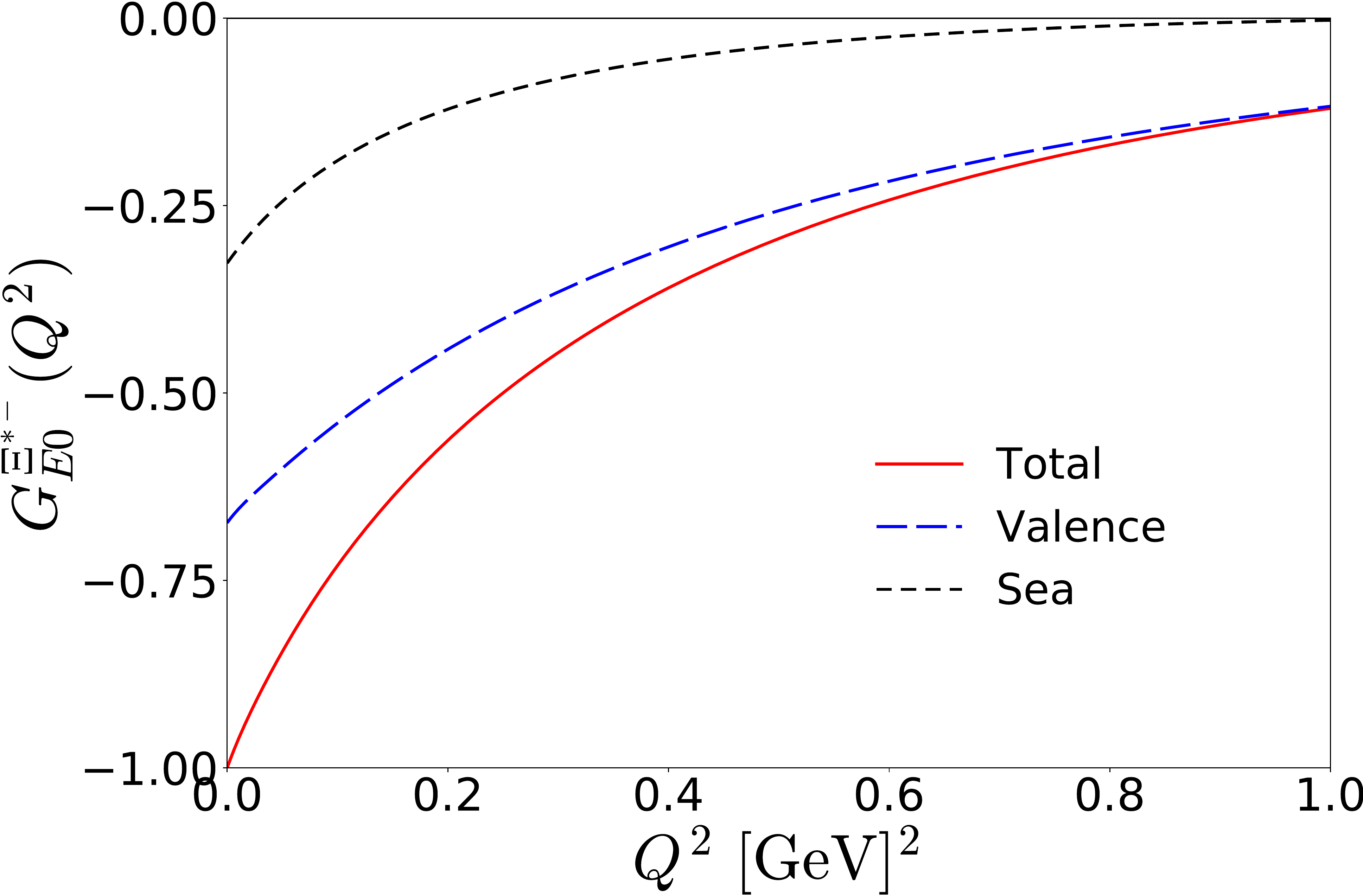}
\caption{Sea and valence contributions of the electric monopole form
  factors of the other members of the baryon decuplet except for the 
  $\Delta^+$ and $\Omega^-$ baryons. Notations are the same as in
  Fig.~\ref{fig:1}.} 
\label{fig:2}
\end{figure}
In Fig.~\ref{fig:2}, we present the results of the electric monopole
form factors for all other baryons except for the $\Delta^+$ and
$\Omega^-$ baryons, decomposing the valence and sea quark 
contributions. Those of the charged baryons show similar tendencies as
the $\Delta^+$ and $\Omega^-$ form factors. However, when it comes to
the electric monopole form factors of the neutral hyperons, the
situation is very different. First of all, since the leading
contributions and rotational $1/N_c$ corrections are proportional to
the charge of the corresponding baryon, which is shown explicitly in 
Eq.~\eqref{eq:E0leading}, they do not contribute to the $E0$ form 
factors of the neutral baryon decuplet. Thus, the $m_s$ corrections
become effectively the main contributions to them.
  
The contribution from the valence quarks is almost canceled by the
sea-quark contribution in the case of the $\Delta^0$ electric monopole
form factor. This cancellation makes it possible to 
satisfy the charge conservation at $Q^2=0$. For the $\Sigma^{*0}$, the 
cancellation is even stronger and complicated. While the valence-quark
part wins the sea-quark contribution, it becomes weaker as $Q^2$
increases. When $Q^2$ grows more than $0.2\,\mathrm{GeV}^2$, the sea
quarks takes control of the $\Sigma^{*0}$ $E0$ form factor. Thus,
its $Q^2$ dependence seems rather different from that of the
$\Delta^0$ form factor that looks similar to the neutron electric form
factor (see Fig.~\ref{fig:13} in Appendix~\ref{app:c}). The electric
monopole form factor of the $\Xi^{*0}$ baryon is even more
interesting. The magnitude of the sea-quark contribution is larger
than the valence part in lower $Q^2$ regions and the valence-quark
contribution turns negative as $Q^2$ increases. As a result, the
$\Xi^{*0}$ $E0$ form factor is negative over all the $Q^2$ regions.   

 \begin{figure}[htp]
 \includegraphics[scale=0.2]{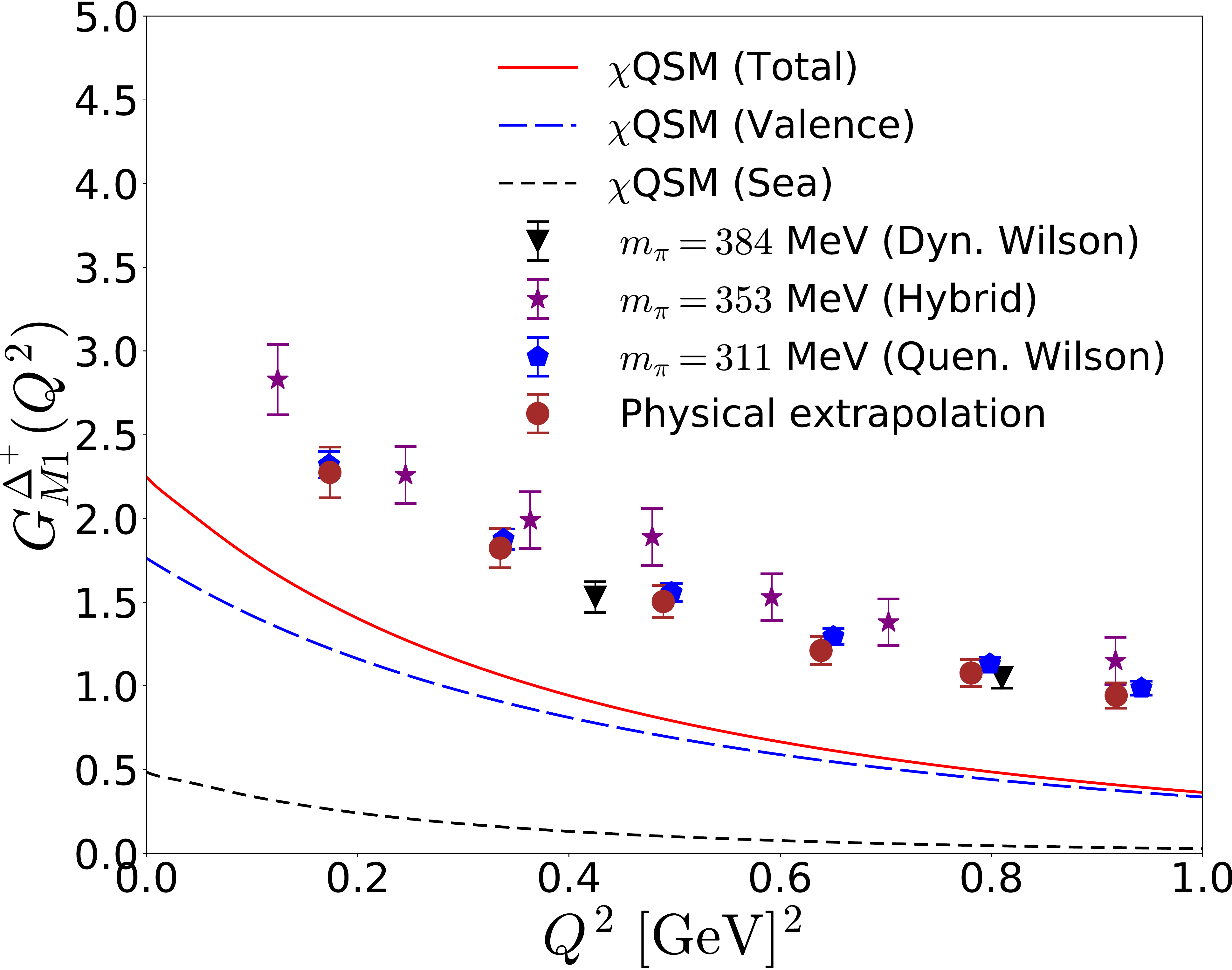}\hspace{0.5cm}
  \includegraphics[scale=0.2]{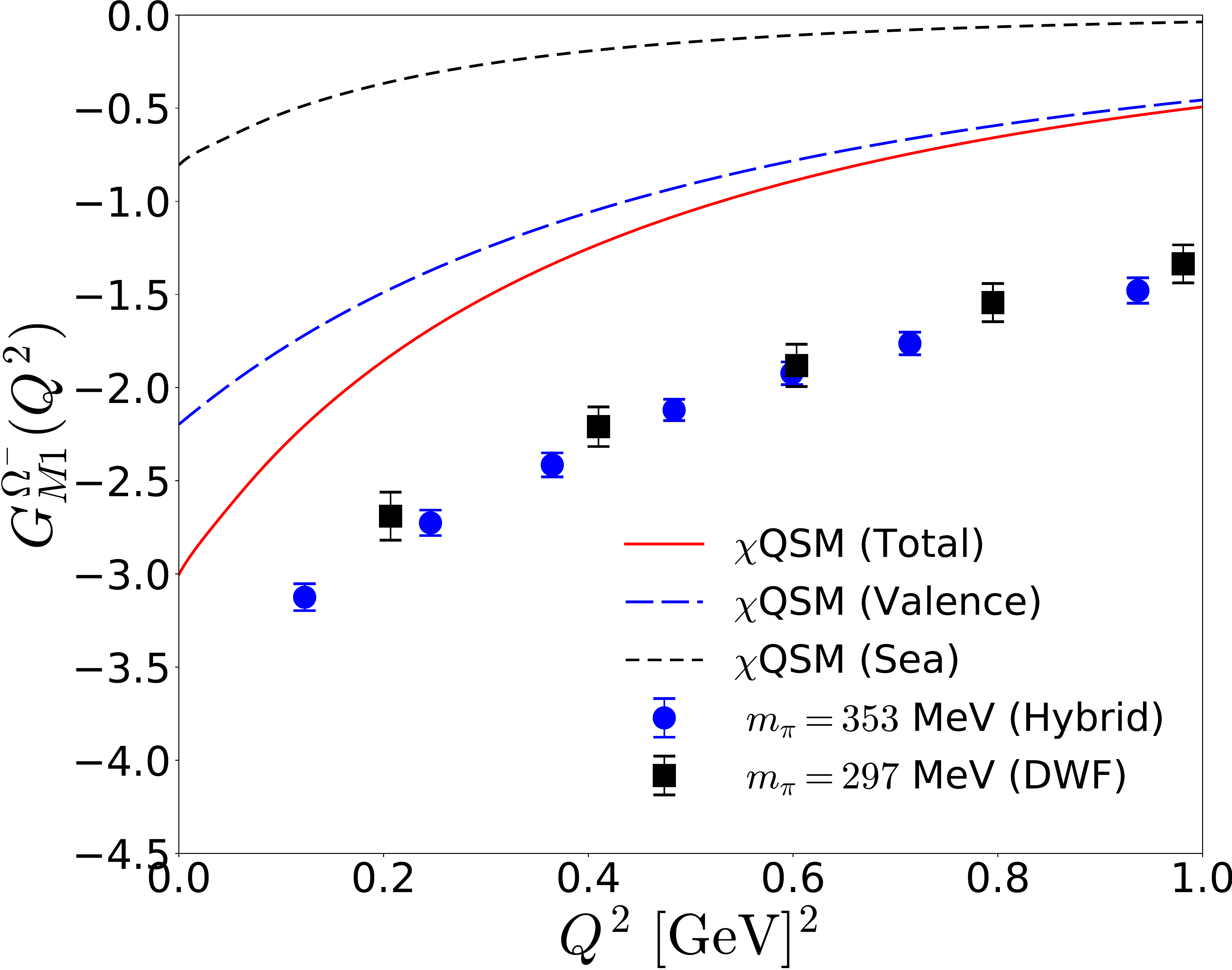}
\caption{Contributions of the valence and sea quarks to the magnetic
  dipole form factors of the $\Delta^+$ isobar and $\Omega^-$ in the
  left and right panels, respectively. Notations are the same as in
  Fig.~\ref{fig:1}. The lattice data are taken from 
  Refs.~\cite{Alexandrou:2007we, Alexandrou:2009hs, Alexandrou:2010jv}.}    
\label{fig:3}
\end{figure}
Before we compare the present results of the magnetic dipole form
factors of the baryon decuplet with those of the lattice calculation,
we want to mention that we follow the definition employed by the lattice
calculation. In Refs.~\cite{Alexandrou:2007we, Alexandrou:2009hs,
    Alexandrou:2010jv}, the following definition for the magnetic dipole
  moment of the $\Delta^+$ and $\Omega^-$ was used
  \begin{align}
\mu_{\Delta^+} = G_{M1}^{\Delta^+} (0) \left(\frac{e}{2M_\Delta}\right) =
    G_{M1}^{\Delta^+}(0) \left(\frac{M_N}{M_\Delta}\right)\mu_N,\;\;\;   
\mu_{\Omega^-} = G_{M1}^{\Omega^-} (0) \left(\frac{e}{2M_\Omega}\right) =
    G_{M1}^{\Omega^-} (0) \left(\frac{M_N}{M_\Omega}\right)\mu_N,    
  \end{align}
where $M_N$, $M_\Delta$, and $M_{\Omega}$ denote the masses of the
nucleon, the $\Delta$ isobar, and the $\Omega^-$. $\mu_N$ stands for
the nuclear magneton. Thus, the $M1$ form 
factors of the $\Delta^+$ and $\Omega^-$ in the lattice calculation
are scaled by the mass of the corresponding baryon. In order to
compare the present results with the lattice ones, we must use the
same definition of the $M1$ form factors of the baryon decuplet.
We want to note that the present values of the $M1$ form factors of
the baryon decuplet do not give the corresponding magnetic dipole
moments that are presented conventionally in unit of the nuclear
magneton.  

Figure~\ref{fig:3} depicts the results of the magnetic dipole form
factors of the $\Delta^+$ and $\Omega^-$ in the left and right panels,
respectively. The sea-quark polarization contributes to the $\Delta^+$
$M1$ form factor only by about 20~\%, while it enhances the $\Omega^-$
form factor by approximately 30~\%. Thus, we find the similar tendency
as in the case of the electric monopole form factors shown in
Fig.~\ref{fig:1}. In comparison with the lattice data, the present
results again fall off faster than the lattice ones. 

  \begin{figure}[htp]
  \includegraphics[scale=0.2]{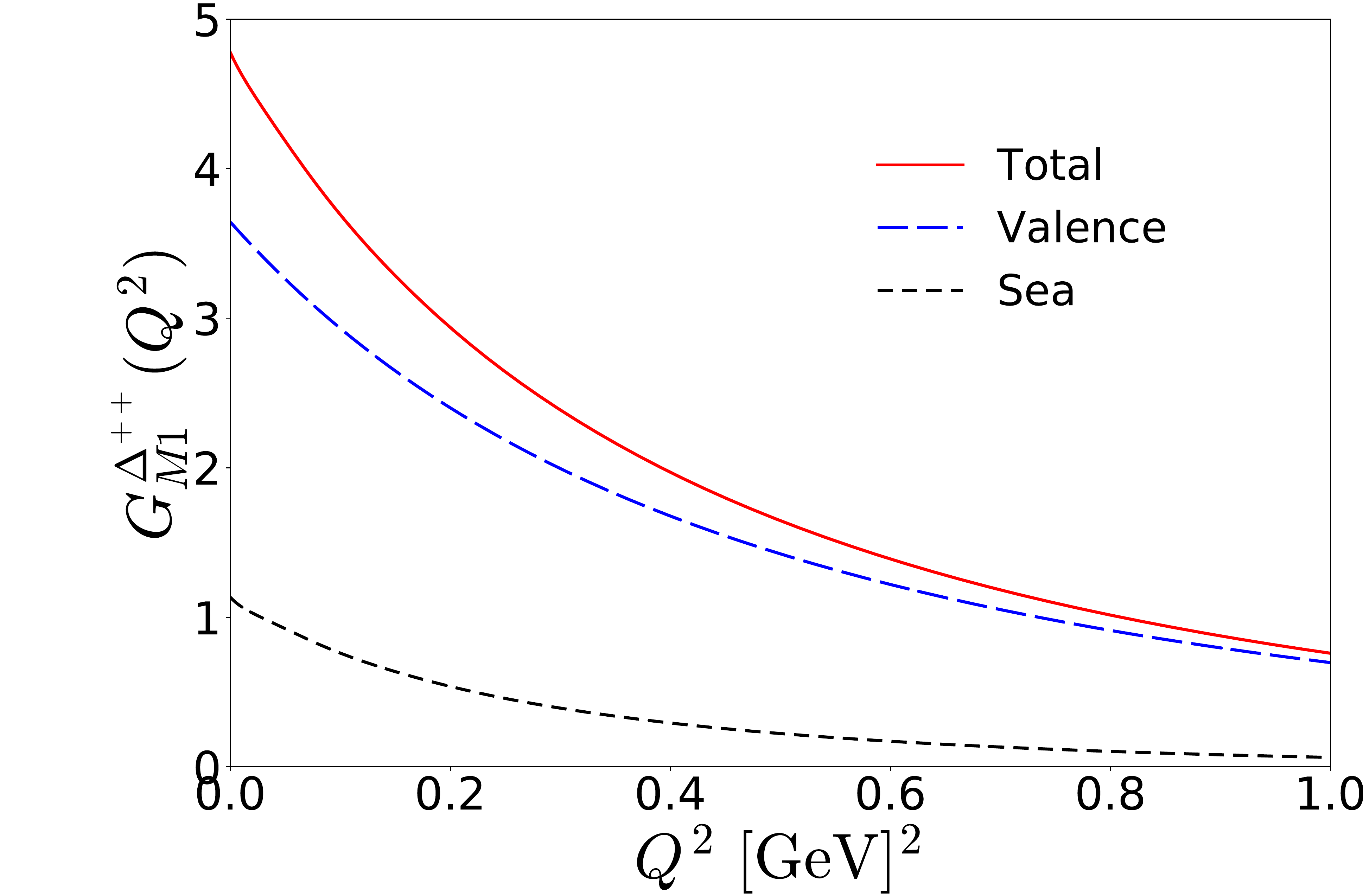} \hspace{0.5cm}
  \includegraphics[scale=0.2]{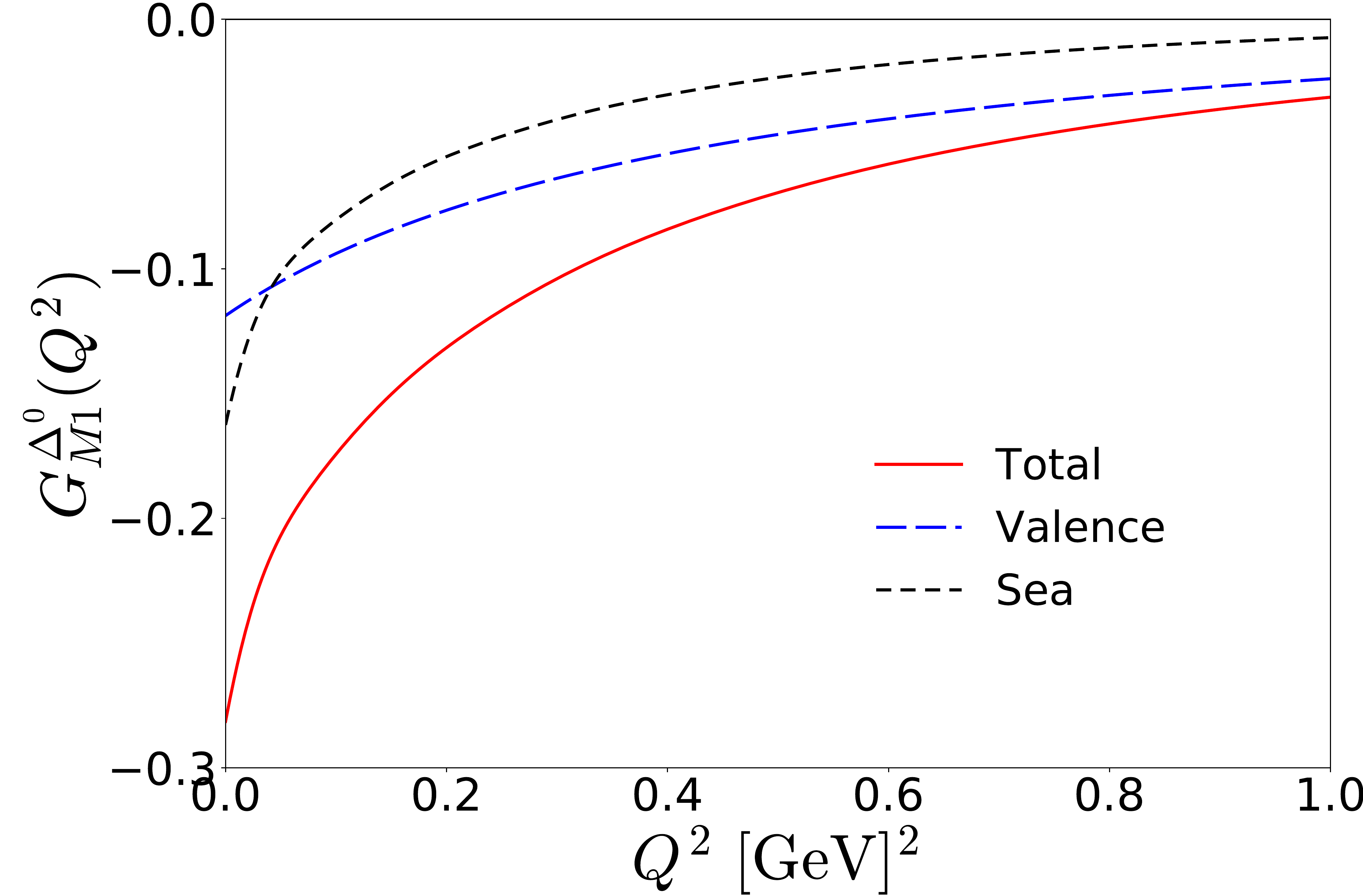}
  \includegraphics[scale=0.2]{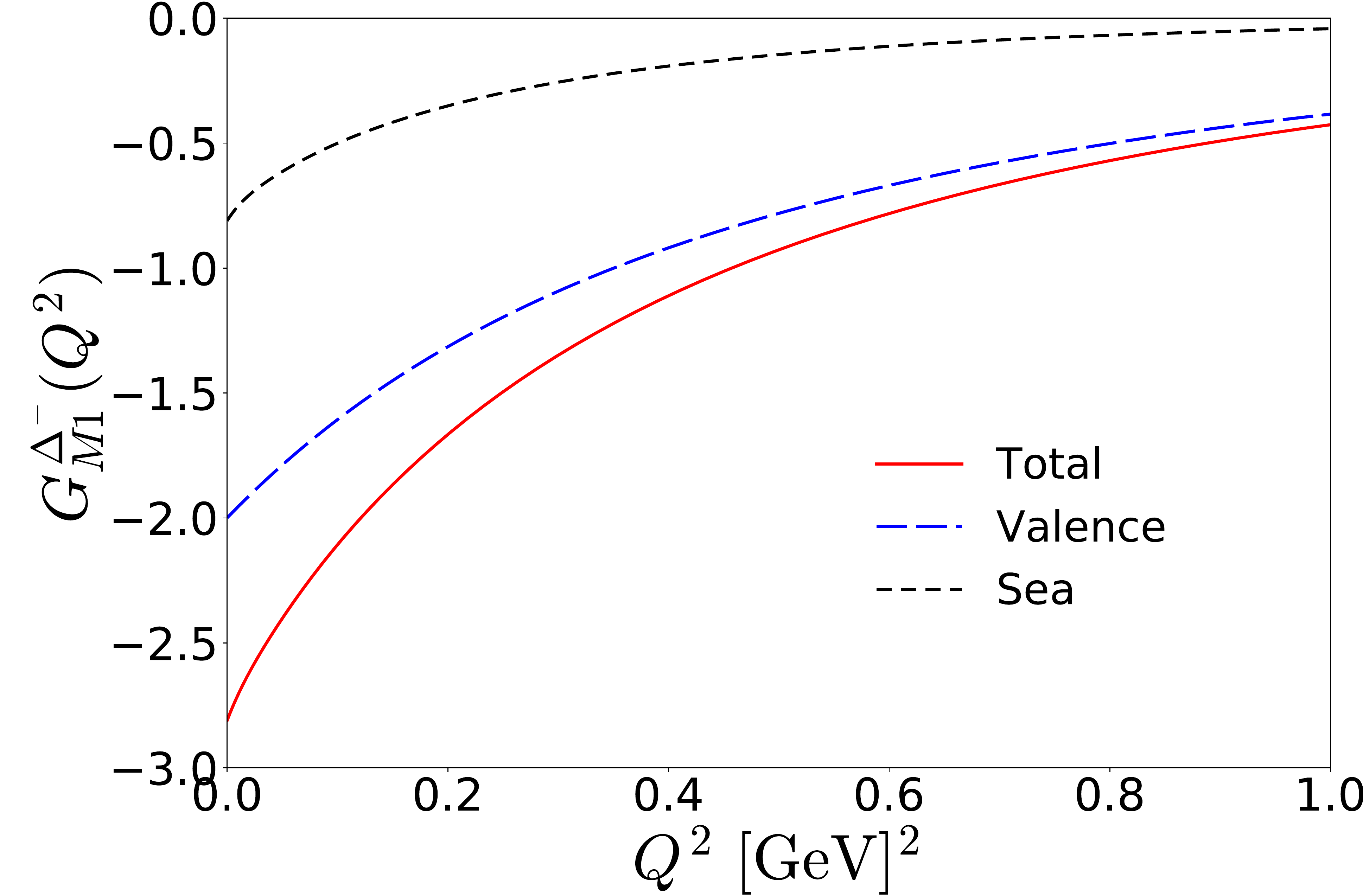}\hspace{0.5cm}
  \includegraphics[scale=0.2]{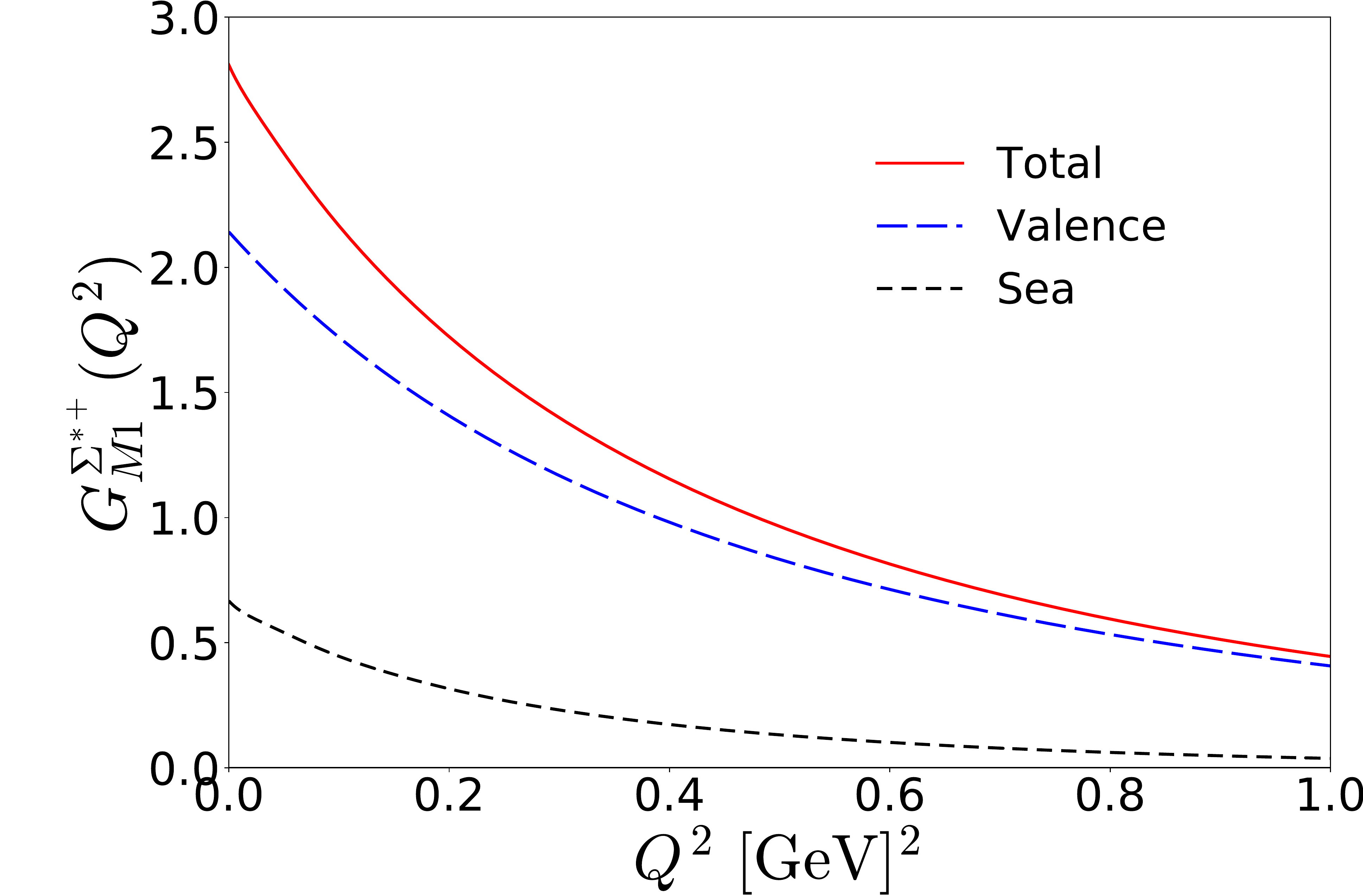}
  \includegraphics[scale=0.2]{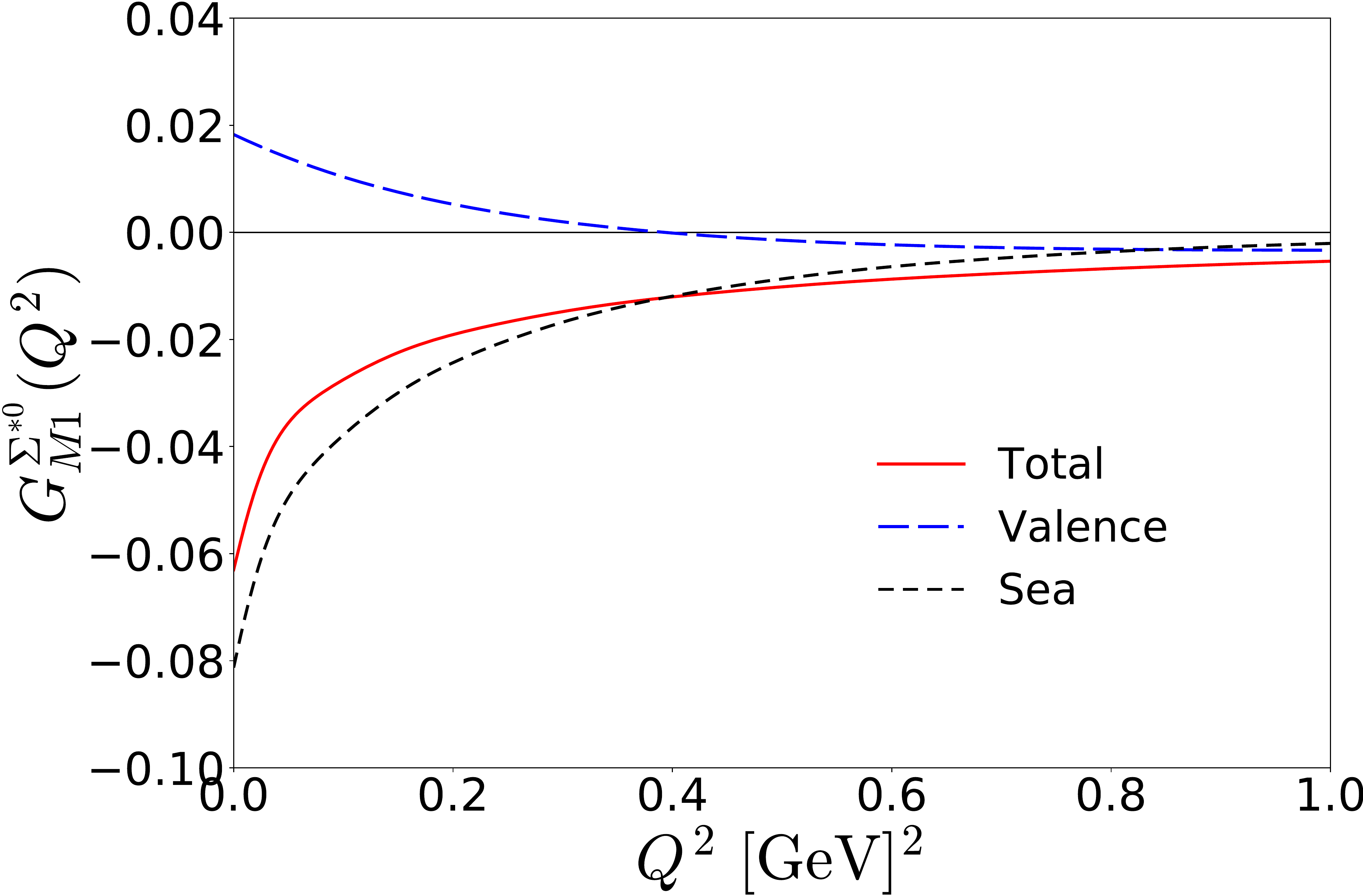}\hspace{0.5cm}
  \includegraphics[scale=0.2]{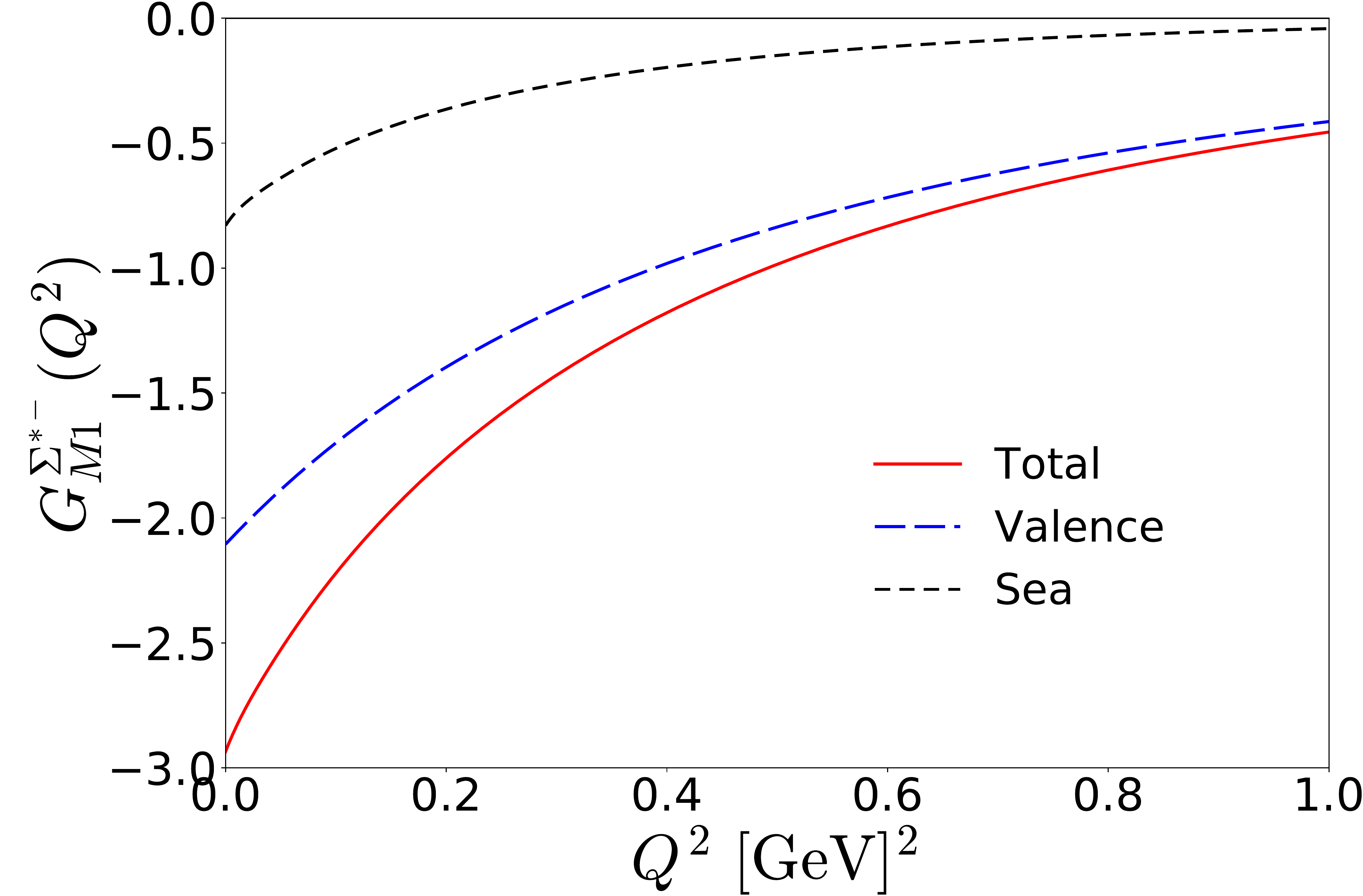}
  \includegraphics[scale=0.2]{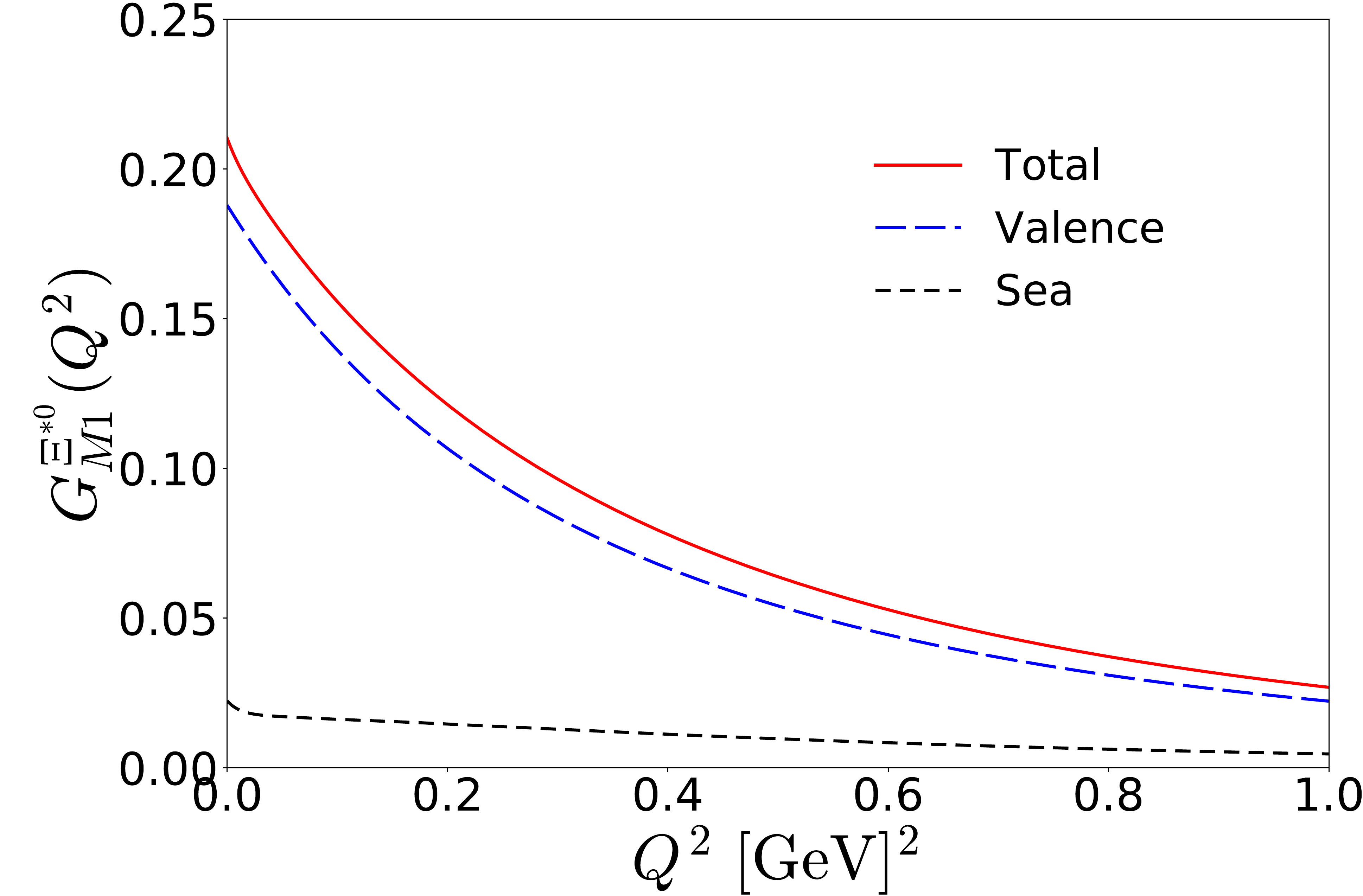}\hspace{0.5cm}
  \includegraphics[scale=0.2]{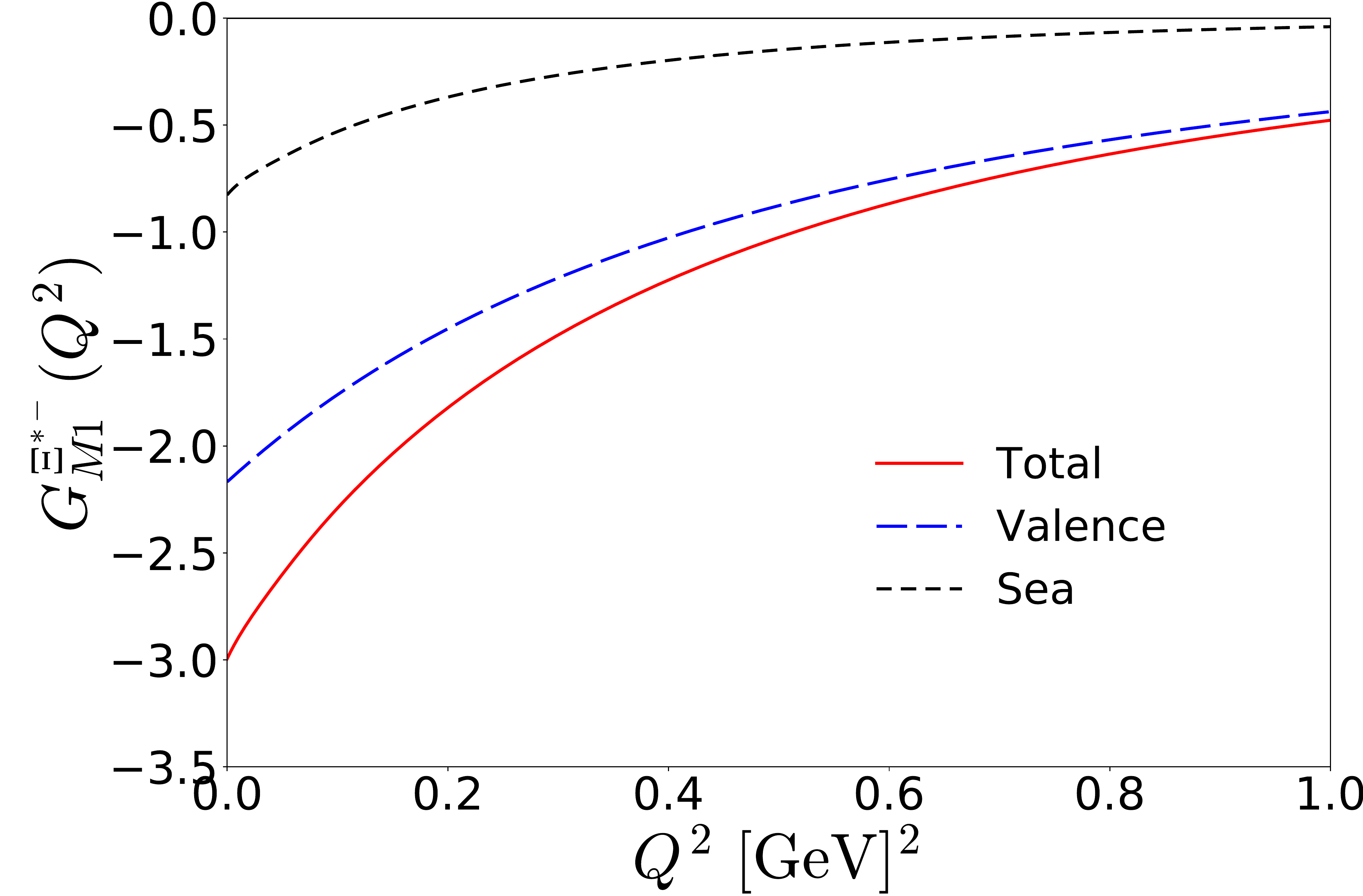}
\caption{Sea and valence contributions to the magnetic dipole form
  factors of the other members of the baryon decuplet except for the 
  $\Delta^+$ and $\Omega^-$ baryons. Notations are the same as in
  Fig.~\ref{fig:1}.}
\label{fig:4}
\end{figure}
In Fig.~\ref{fig:4}, we draw the contributions of the valence and sea
quarks on the magnetic dipole form factors of the baryon decuplet
except for the $\Delta^+$ and $\Omega^-$ form factors. Being similar
to the results of the $\Delta^+$ and $\Omega^-$ $M1$ form factors
drawn in Fig.~\ref{fig:3}, those of the charged baryon decuplet
exhibit similar $Q^2$ dependence. The sizes of the valence and sea
quark contributions show the similar tendency. However, the $M1$ form
factors of the charged baryon decuplet show rather different
behaviors. While the sea-quark polarizations contribute to the
$\Xi^{*0}$ $M1$ form factor, they dominate over the valence-quark
contributions for the $\Sigma^{*0}$ one. On the other hand, the
sea-quark effects are larger than the valence-quark ones only at the
low $Q^2$ region and fall off faster than the valence-quark
contributions.  

 \begin{figure}[htp]
 \includegraphics[scale=0.2]{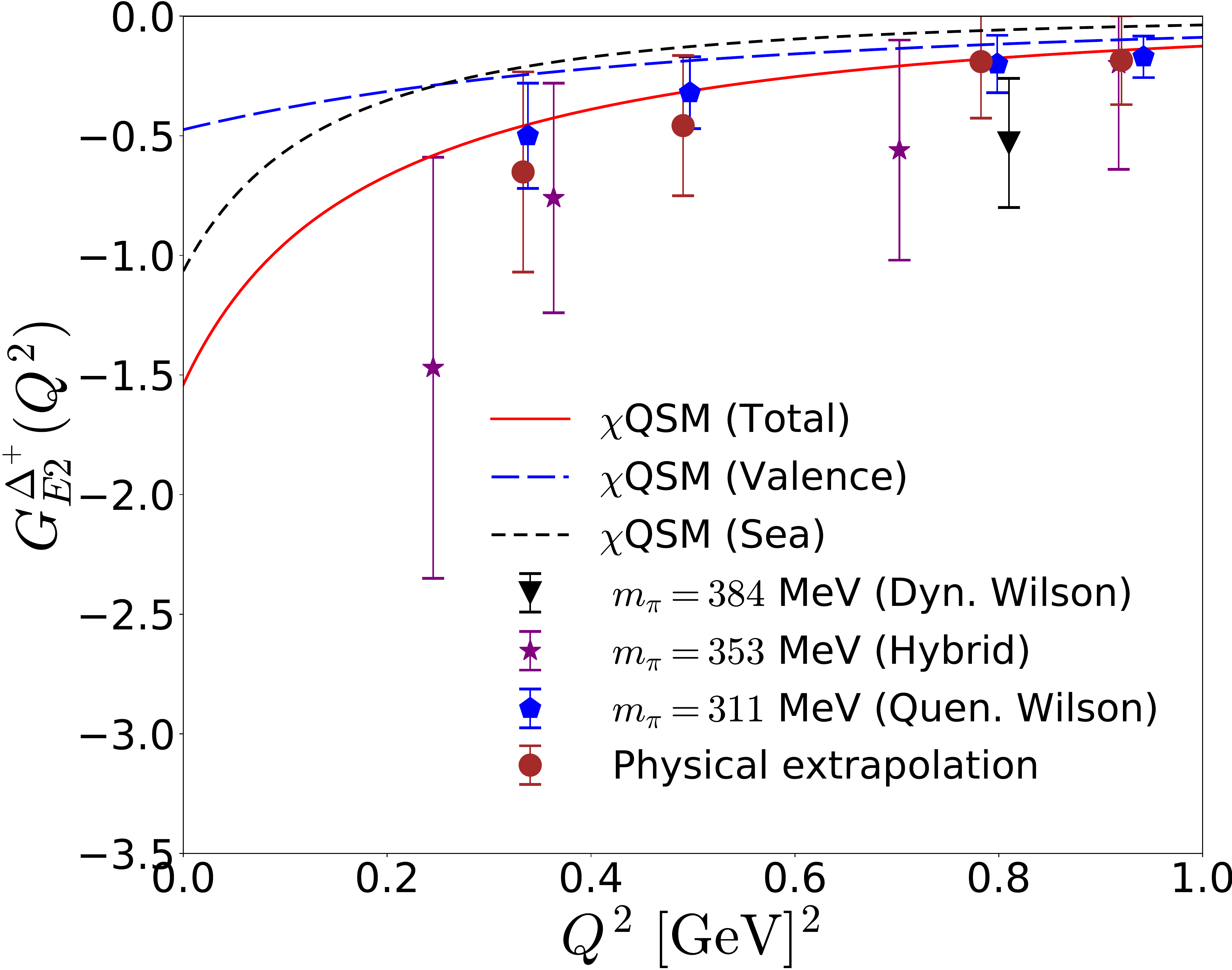}\hspace{0.5cm}
 \includegraphics[scale=0.2]{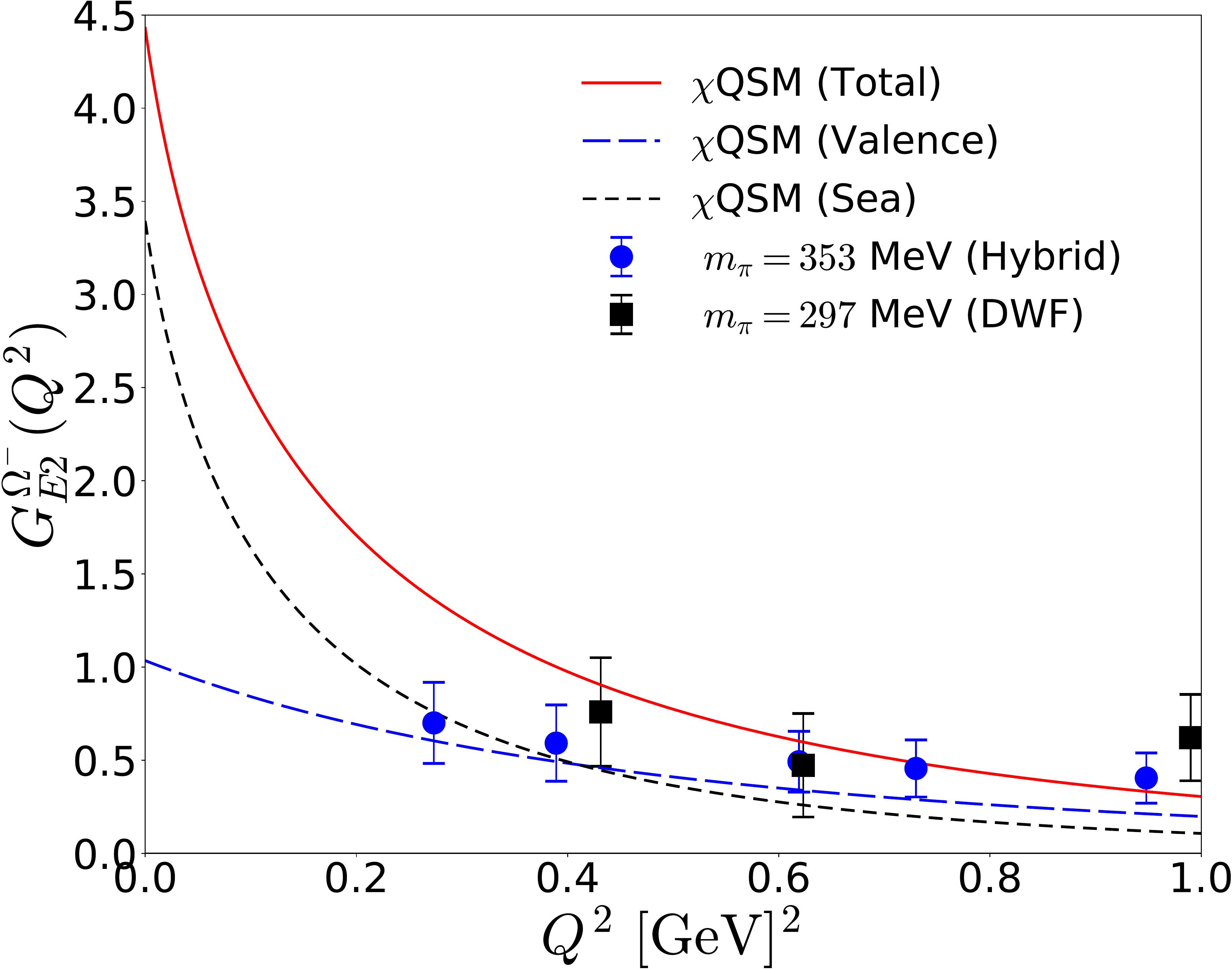}
\caption{Contributions of the valence and sea quarks to the electric
  quadrupole form factors of the $\Delta^+$ isobar and $\Omega^-$ in the
  left and right panels, respectively. Notations are the same as in
  Fig.~\ref{fig:1}. The lattice data are taken from 
 Refs.~\cite{Alexandrou:2007we,Alexandrou:2009hs,Alexandrou:2010jv}.}   
\label{fig:5}
\end{figure}
We use the following definition of the electric quadrupole form
factors of the baryon decuplet used in the lattice calculation
\begin{align}
Q_{B_{10}} = G_{E2}^{B_{10}} (0) \left(\frac{|e|}{M_{B_{10}}^2}\right),   
\end{align}
so that we are able to compare the present results with the lattice
data. In Fig.~\ref{fig:5}, we draw the results of the $E2$ form
factors of the $\Delta^+$ and $\Omega^-$ in the left and right panels,
respectively. The contribution of the sea-quark polarization to the
$\Delta^+$ $E2$ form factor is dominant over that of the valence
quarks in lower $Q^2$ regions ($0\le Q^2 \le 0.3\,\mathrm{GeV}^2$),
which is opposite to the case of both the $E0$ and $M1$ form factors
of the $\Delta^+$, as already discussed in
Ref.~\cite{Ledwig:2008es}. In particular, the sea-quark polarization
gives more than 90~\% contribution to the $\Delta^+$ $E2$ form factor
in the vicinity of $Q^2=0$. When $Q^2$ further increases ($Q^2 > 0.3
\,\mathrm{GeV}^2$), the valence-quark effects outdo the sea-quark
contribution. Interestingly, though the sea-quark polarization
contributes also dominantly to the $\Omega^-$ $E2$ form factor but it
provides approximately 70~\%. The $Q^2$ dependence of the $\Omega^-$
$E2$ form factor is similar to the case of the $\Delta^+$ one. 
However, the sign of the $\Omega^-$ $E2$ form factor is positive while
that of the $\Delta^+$ is negative. It indicates that the charge
distribution of the $\Delta^+$ is in an oblate shape ($Q_{\Delta^+}<0$)
whereas that of the $\Omega^-$ looks prolate
($Q_{\Omega^-}>0$). Though the absolute value of
$G_{E2}^{\Omega^-}(0)$ at $Q^2=0$ seems larger than that of
$G_{E2}^{\Delta^+}(0)$, the magnitude of the $\Omega^-$ electric 
quadrupole moment is in fact slightly smaller than that of the
$\Delta^+$, as will be shown later in Table~\ref{tab:4}. 
The discussion given above has important physical implications. The
sea-quark polarizations or the pion clouds in a conventional
terminology are the main reason for the deformation of a decuplet
baryon. It also indicates that the valence quarks sit on the inner part
of a decuplet baryon whereas the pion clouds govern the charge
distribution in the outer region of the baryon.

In contrast to the cases of the $E0$ and $M1$ form factors, the
lattice data~\cite{Alexandrou:2007we,Alexandrou:2009hs,
Alexandrou:2010jv} contain large numerical uncertainties for the $E2$
form factors of the $\Delta^+$ and $\Omega^-$. Moreover, there
exist the lattice date only in relatively larger $Q^2$
regions. Nevertheless, the present results are in good agreement with
them, as illustrated in Fig~\ref{fig:5}.     

  \begin{figure}[htp]
  \includegraphics[scale=0.2]{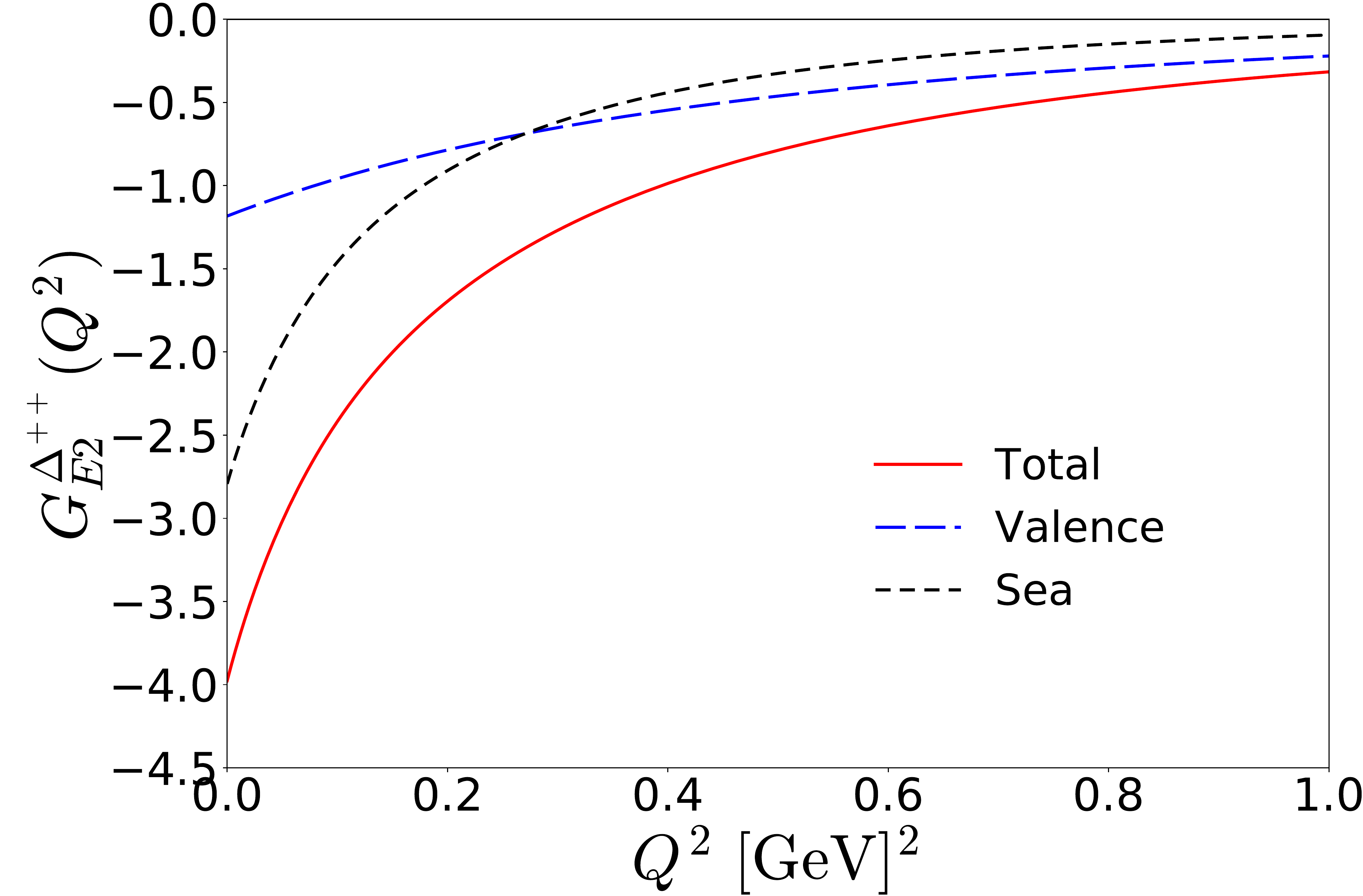} \hspace{0.5cm}
  \includegraphics[scale=0.2]{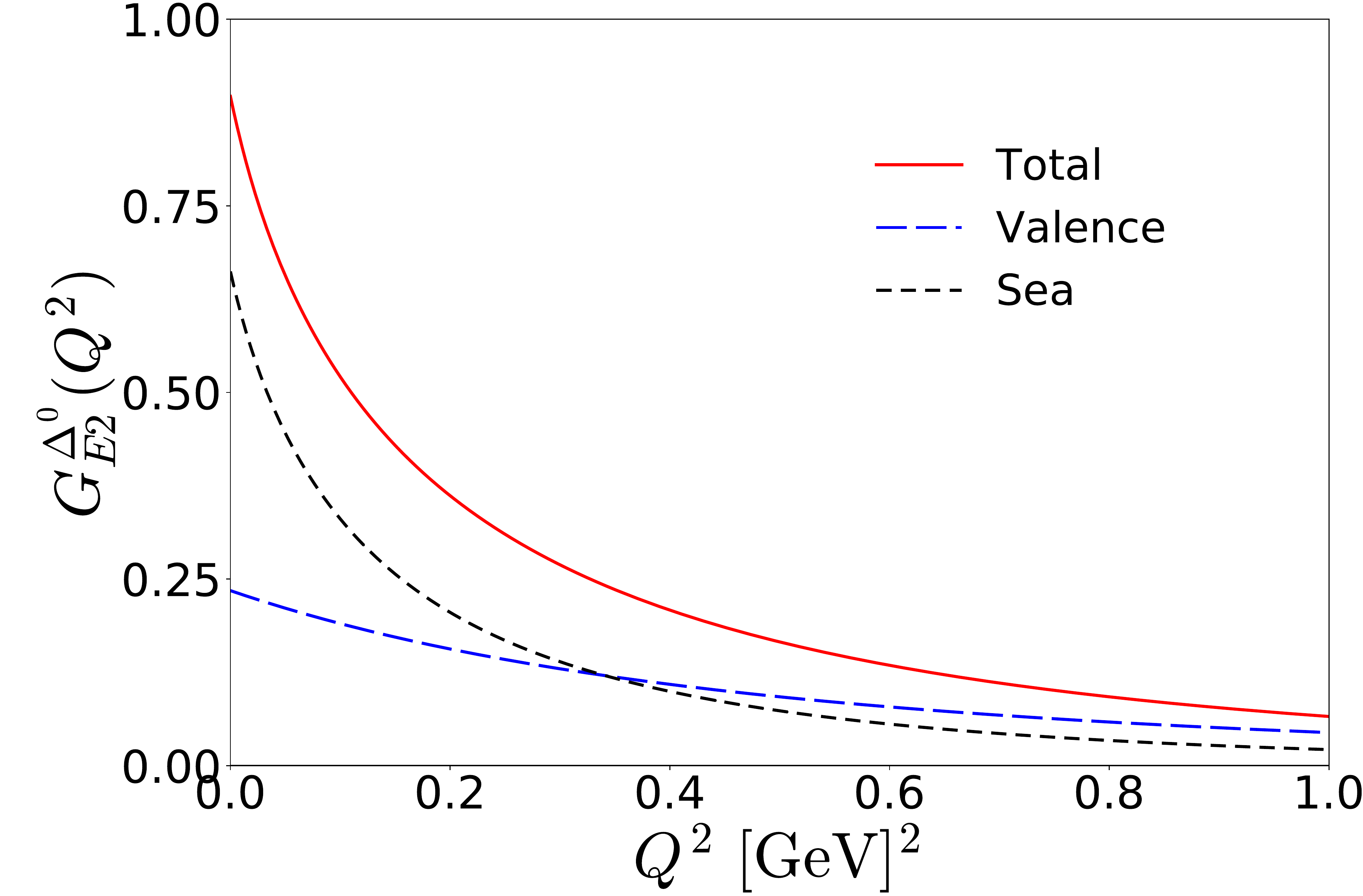}
  \includegraphics[scale=0.2]{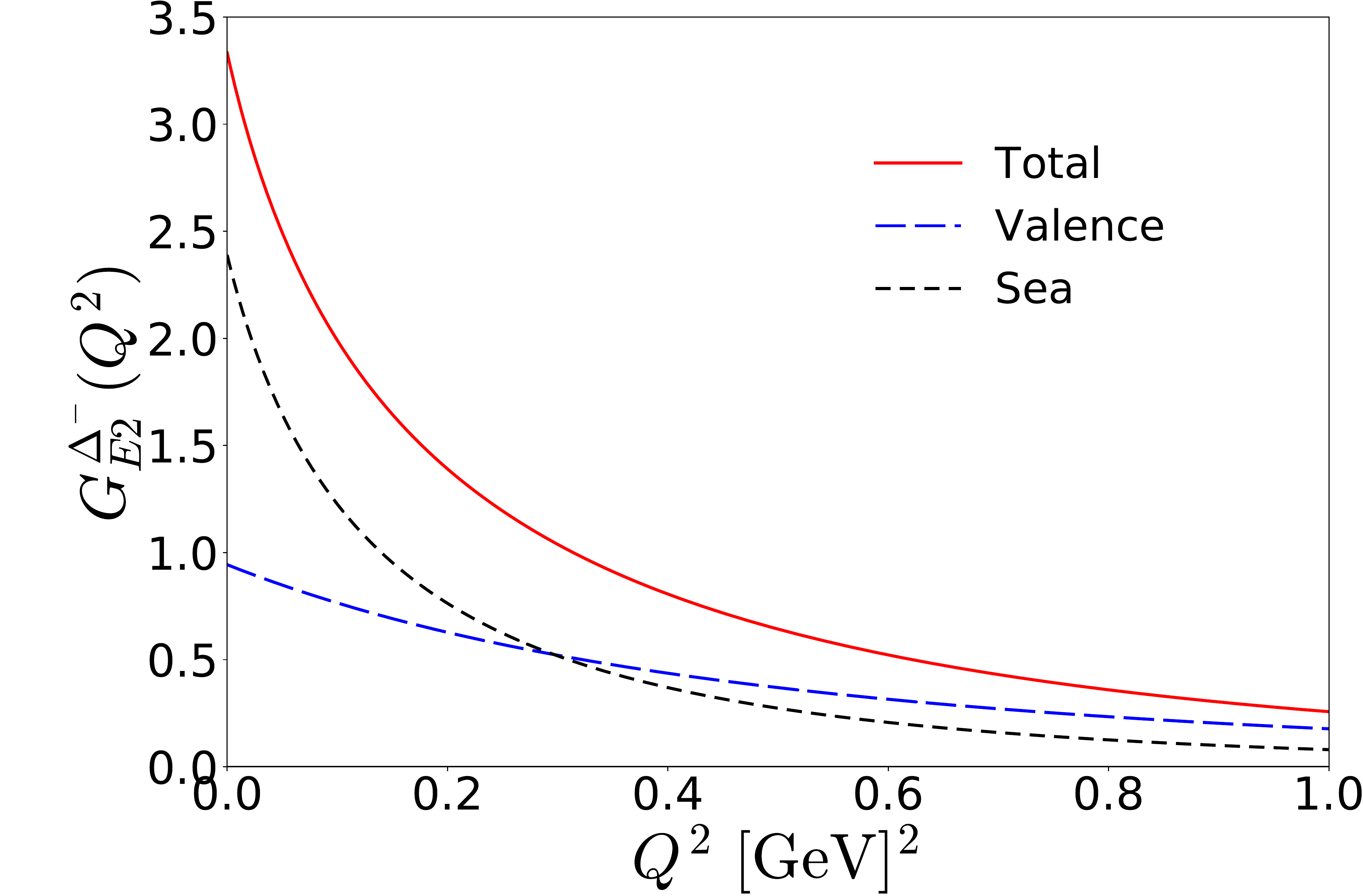}\hspace{0.5cm}
  \includegraphics[scale=0.2]{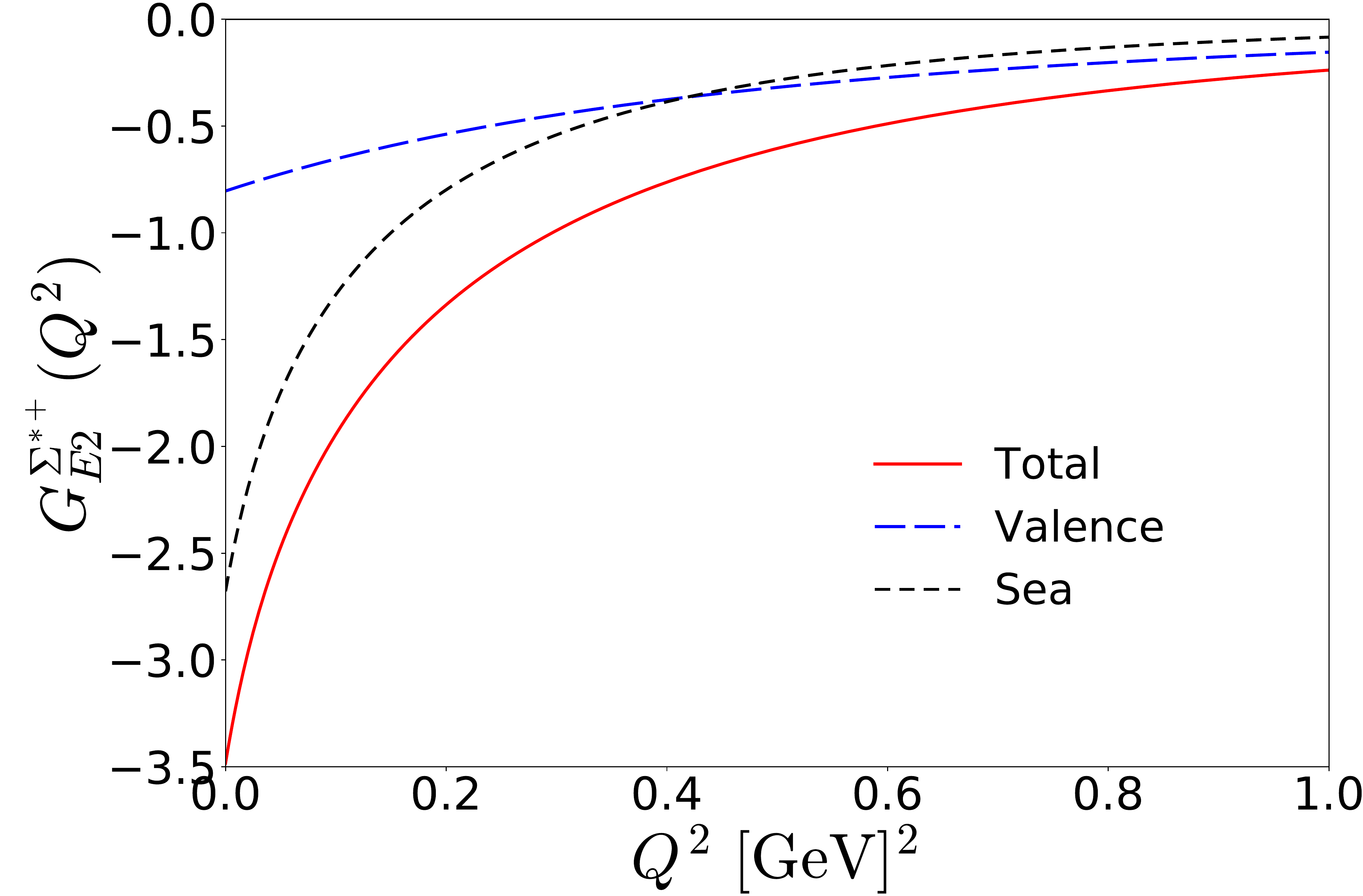}
  \includegraphics[scale=0.2]{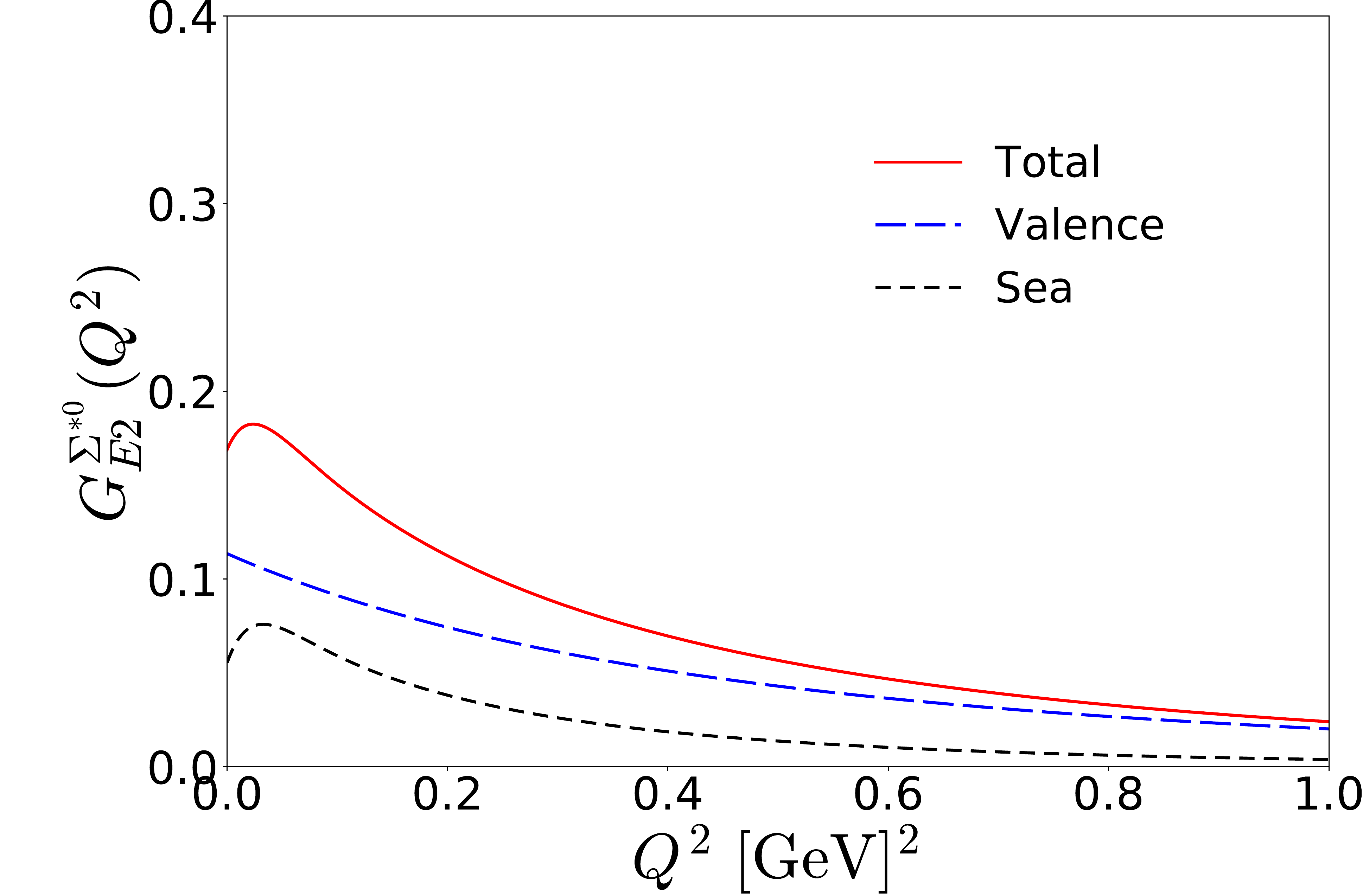}\hspace{0.5cm}
  \includegraphics[scale=0.2]{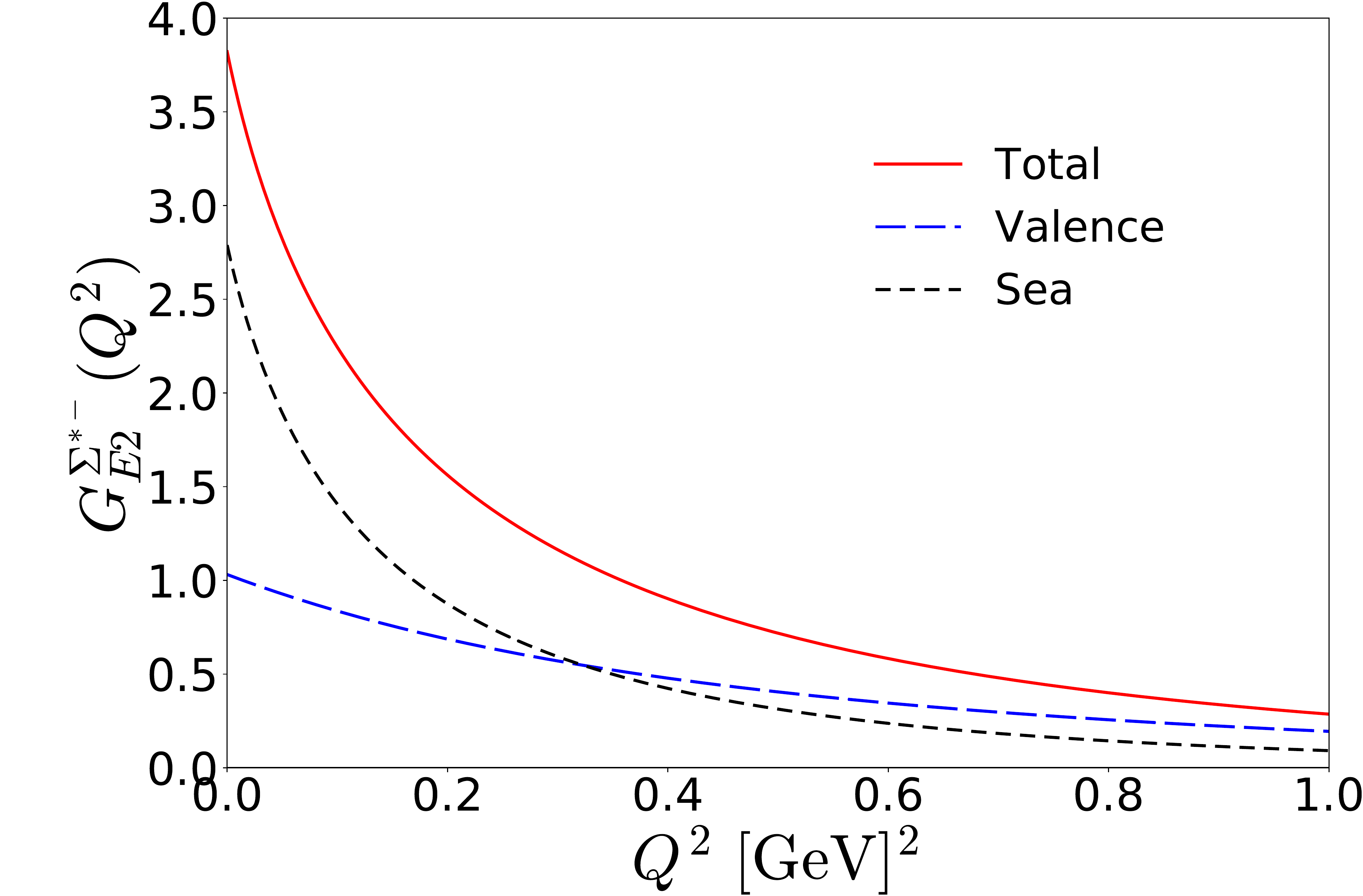}
  \includegraphics[scale=0.2]{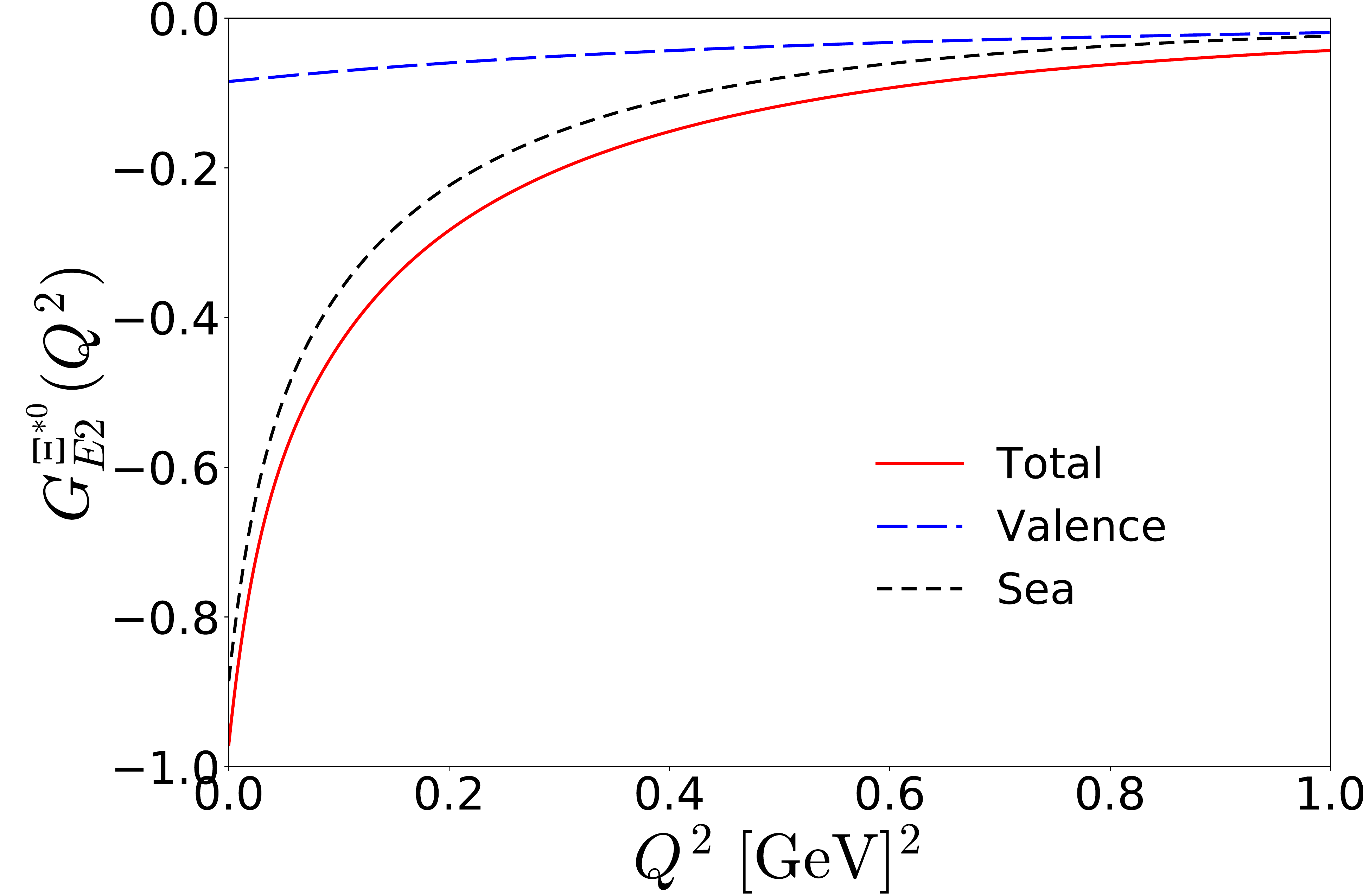}\hspace{0.5cm}
  \includegraphics[scale=0.2]{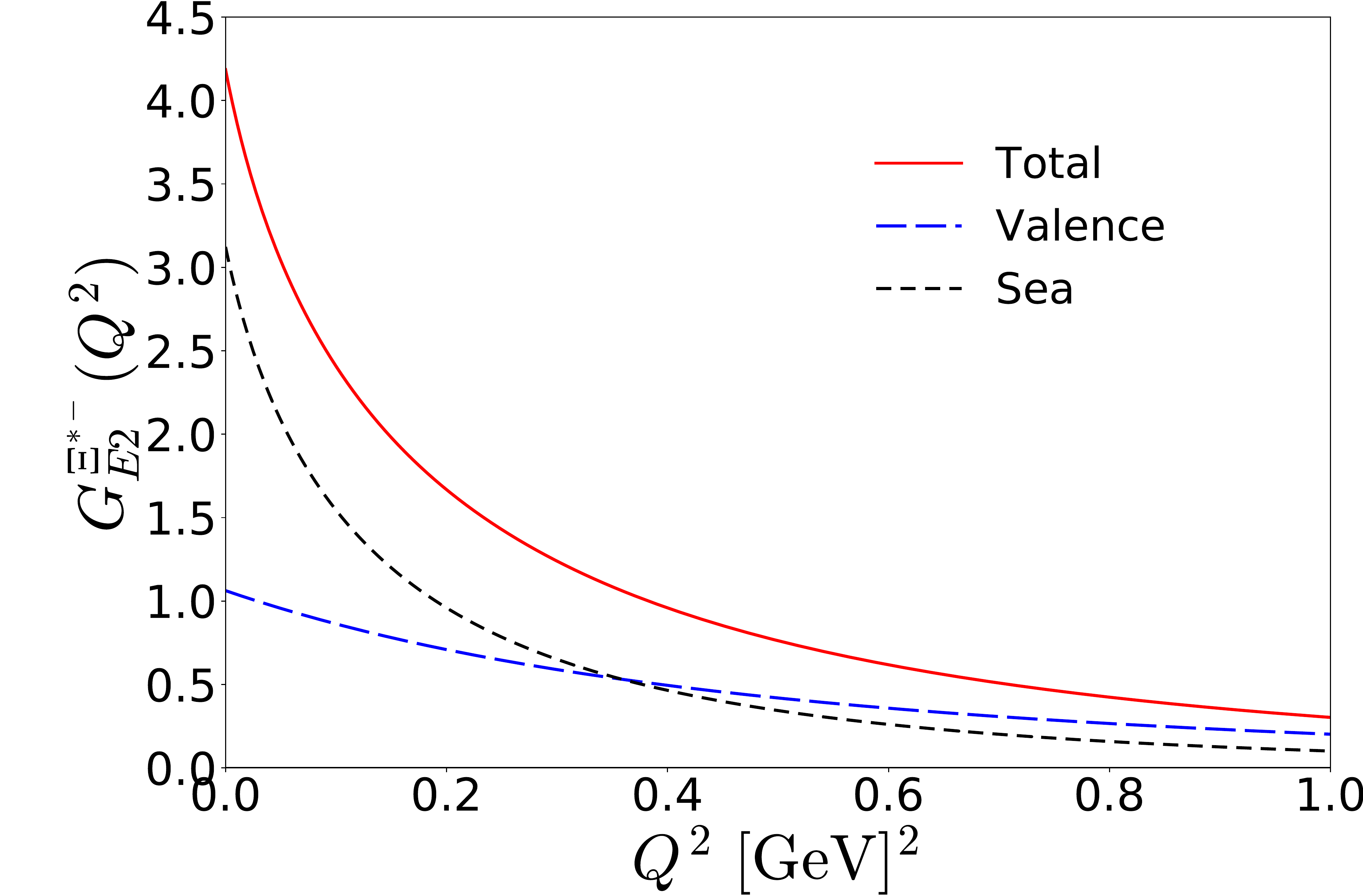}
\caption{Contributions of the valence and sea quarks to the electric
  quadrupole form factors of the other members of the baryon
  decuplet except for the $\Delta^+$ and $\Omega^-$ baryons. Notations
  are the same as in Fig.~\ref{fig:1}.}
\label{fig:6}
\end{figure}
In Fig.~\ref{fig:6}, we draw the results of the electric quadrupole
form factors of the baryon decuplet except for the $\Delta^+$ and
$\Omega^-$. As in the cases of the $\Delta^+$ and $\Omega^-$, the
sea-quark contributions generally dominate over the valence-quark ones
in lower $Q^2$ regions and fall off faster than the valence-quark
ones, so that the sea-quark contributions becomes smaller than the
valence-quark contributions in higher $Q^2$ regions. The positively
charged baryon decuplet have in general negative values of the $E2$
form factors, while it is consistently other way around in the case of
the negatively charged decuplet baryons. As discussed previously, the
distributions of the positively charged baryon decuplet take an
cushion shape ($Q_B<0$) whereas the negative ones look like a
rugby-ball shape ($Q_B>0$). 

The $E2$ form factors of the neutral baryon decuplet arise only from
the contribution of the $m_s$ corrections. As expressed in 
Eq.~\eqref{eq:E2final}, the leading-order contributions to the neutral
$E2$ form factors vanish, because they are proportional to the
corresponding charges. Thus, the $m_{s}$ corrections play the leading
role.

\subsection{Effects of flavor SU(3) symmetry breaking}
  \begin{figure}[htp]
  \includegraphics[scale=0.2]{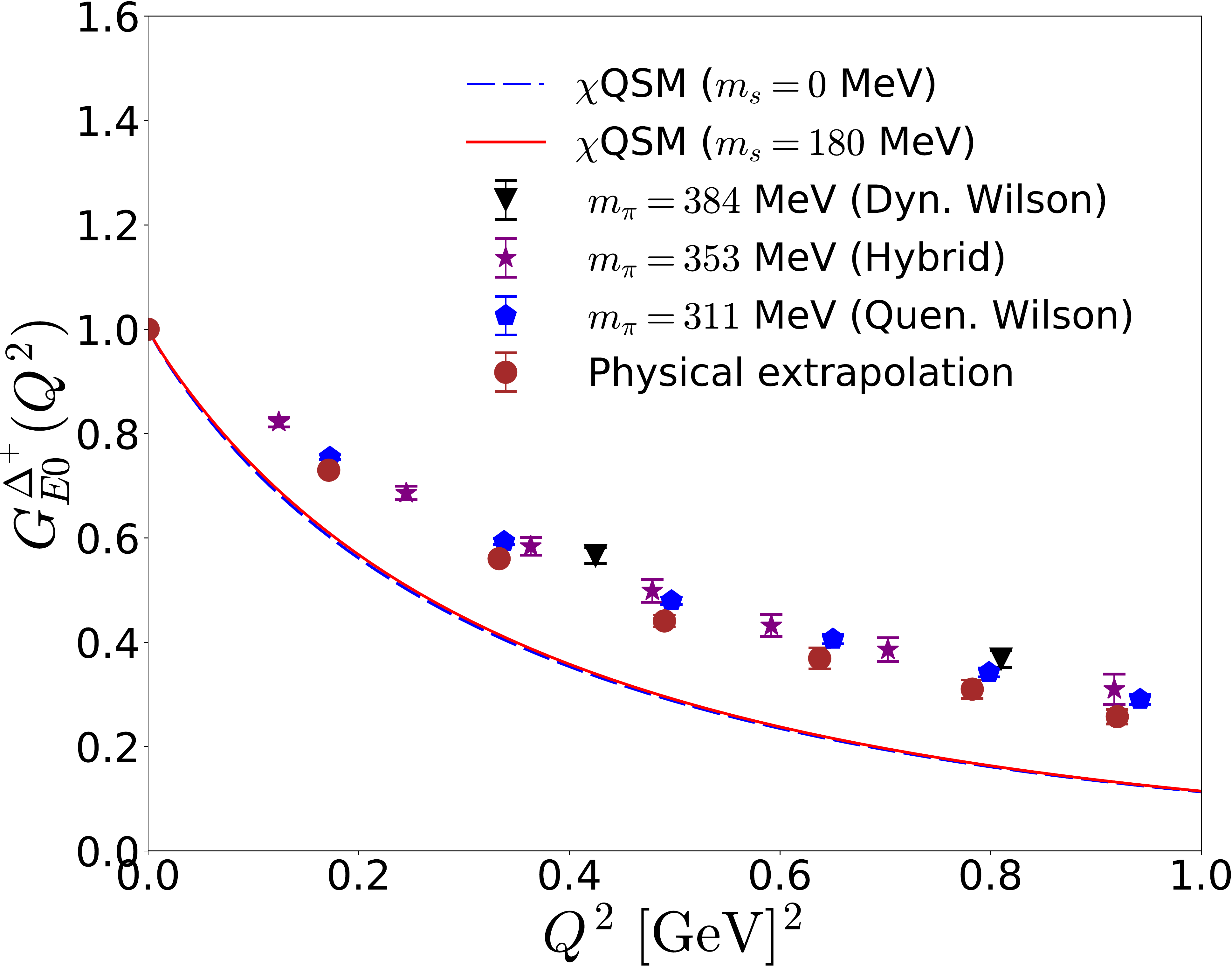}\hspace{0.5cm}
  \includegraphics[scale=0.2]{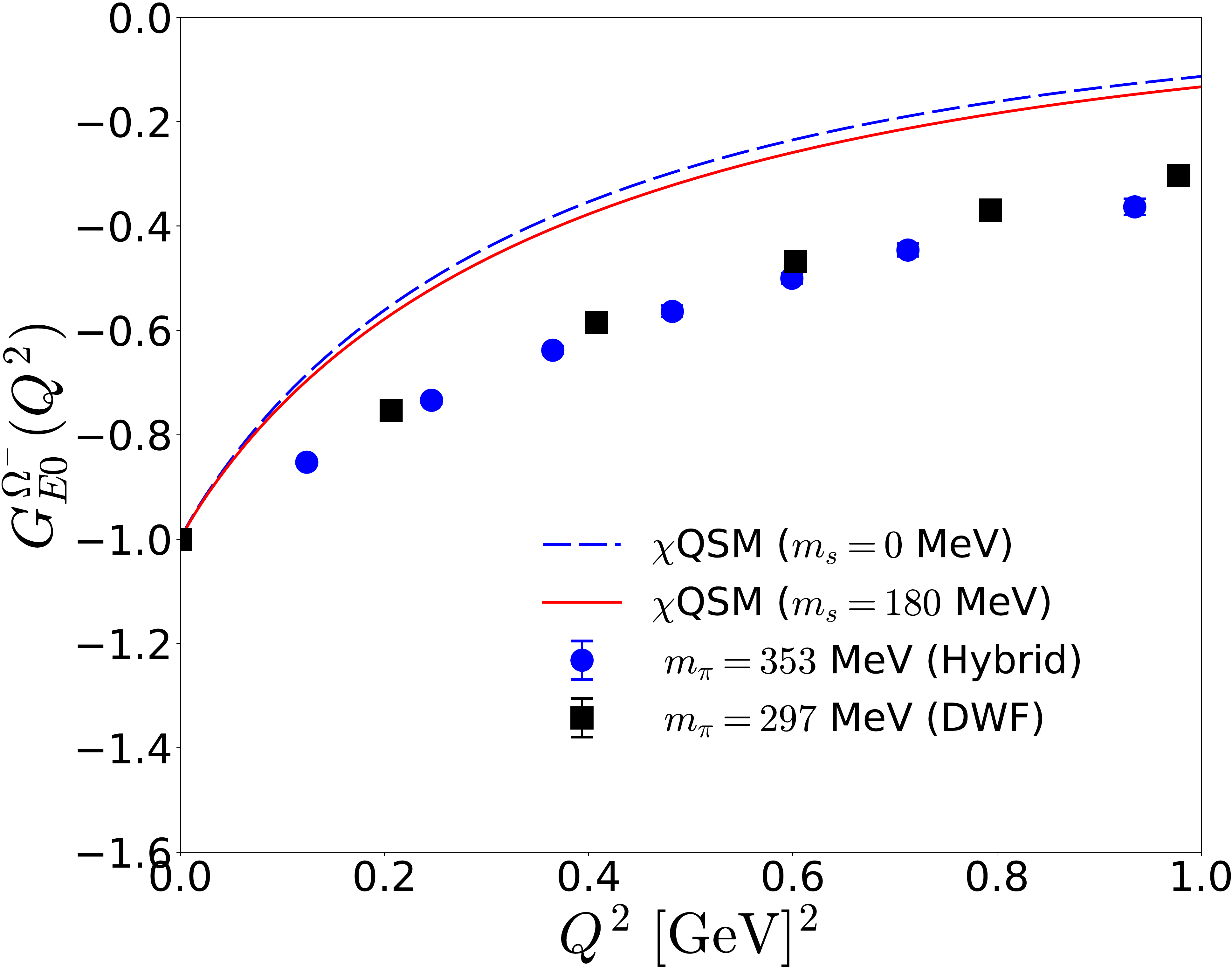}
  \caption{Effects of the explicit flavor SU(3) symmetry breaking on
    the electric monopole form factors of the $\Delta^+$ isobar and
    $\Omega^-$ in the left and right panels, respectively. The dashed
    curves depict the $E0$ form factors without the effects of the
    explicit flavor SU(3) symmetry breaking, whereas the solid ones 
    present those with the effects. The lattice data
    are taken from Refs.~\cite{Alexandrou:2007we, Alexandrou:2009hs,
      Alexandrou:2010jv}.}  
\label{fig:7}
\end{figure}
 
  \begin{figure}[htp]
  \includegraphics[scale=0.2]{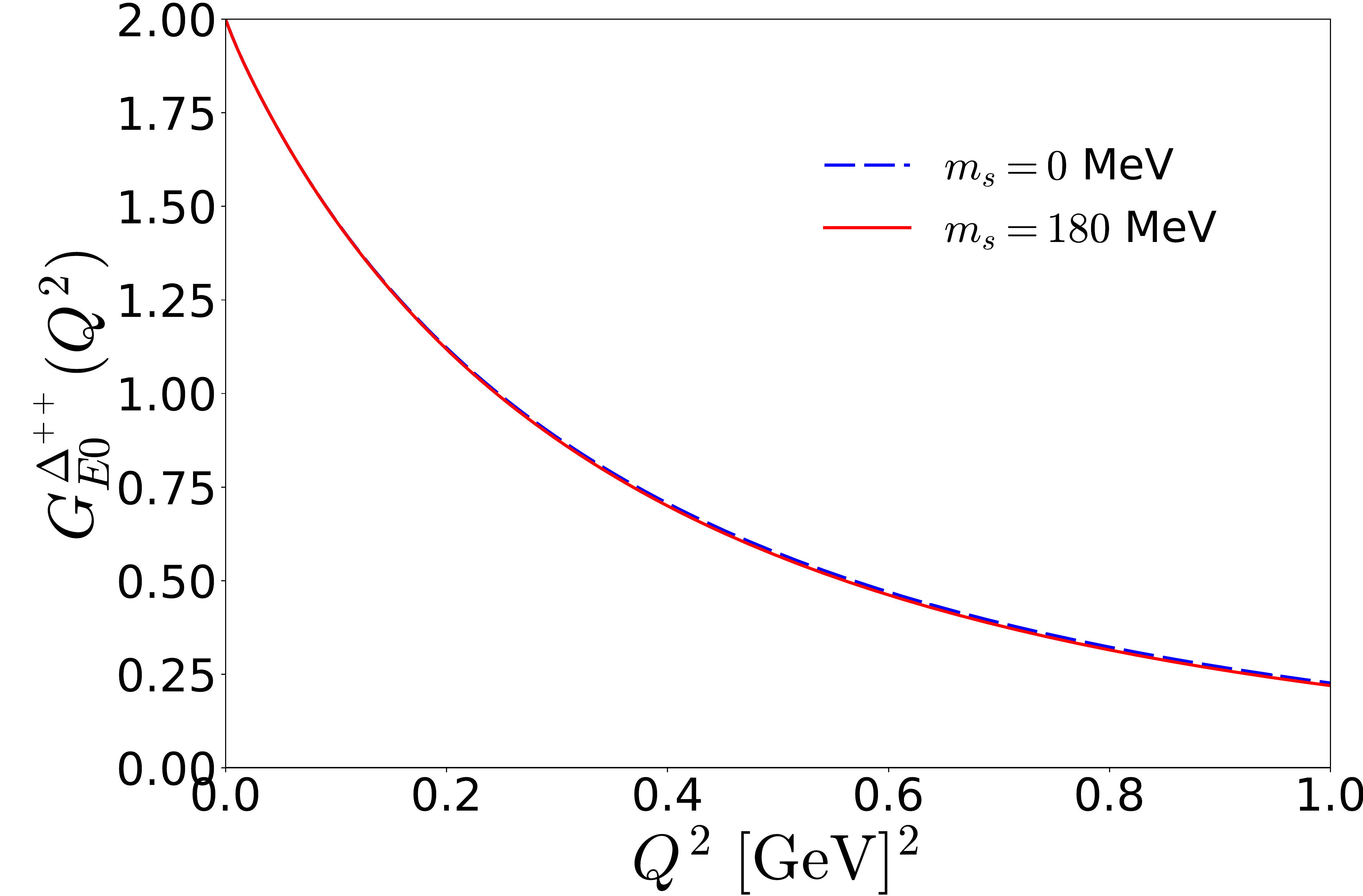} \hspace{0.5cm}
  \includegraphics[scale=0.2]{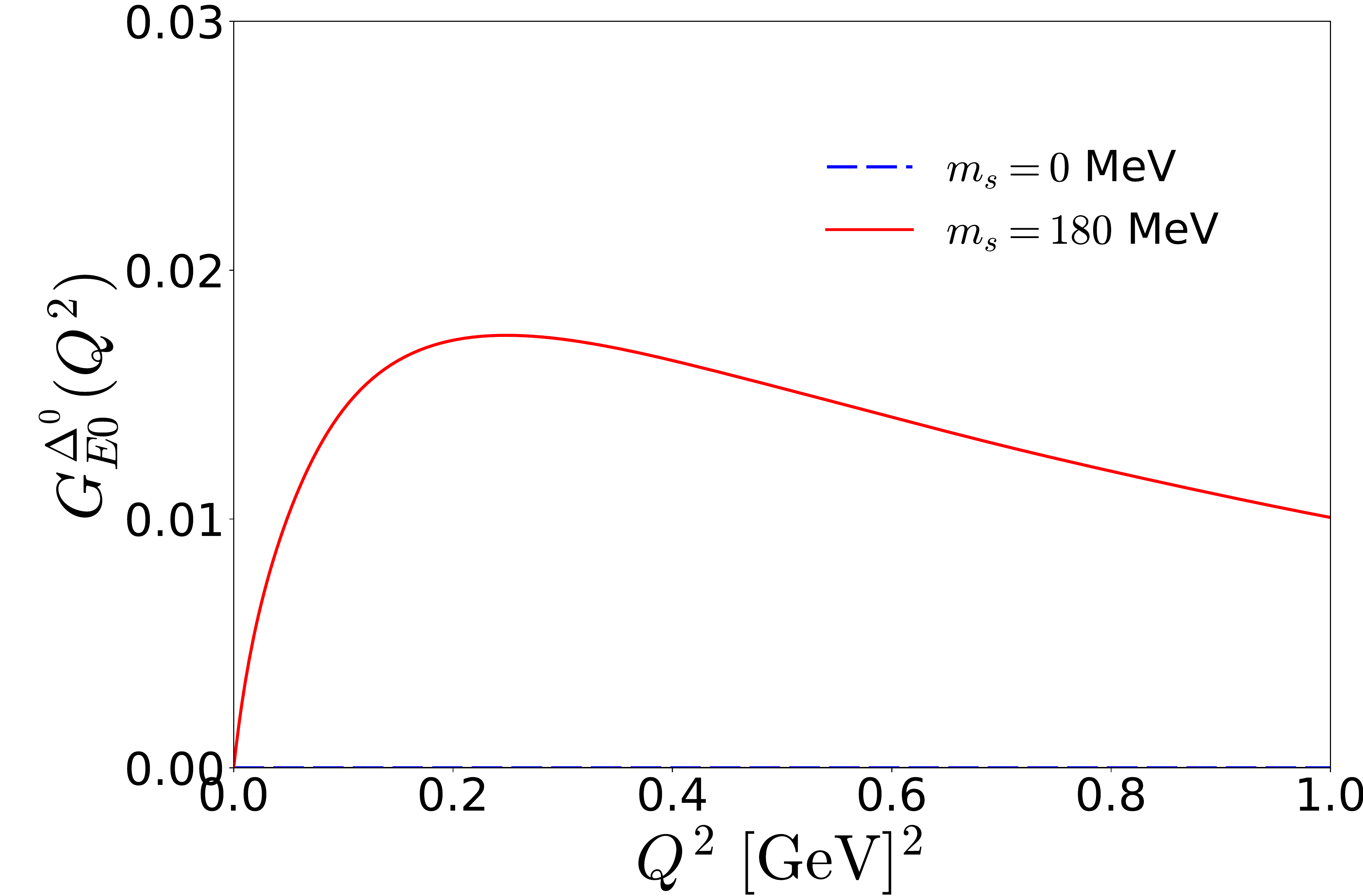}
  \includegraphics[scale=0.2]{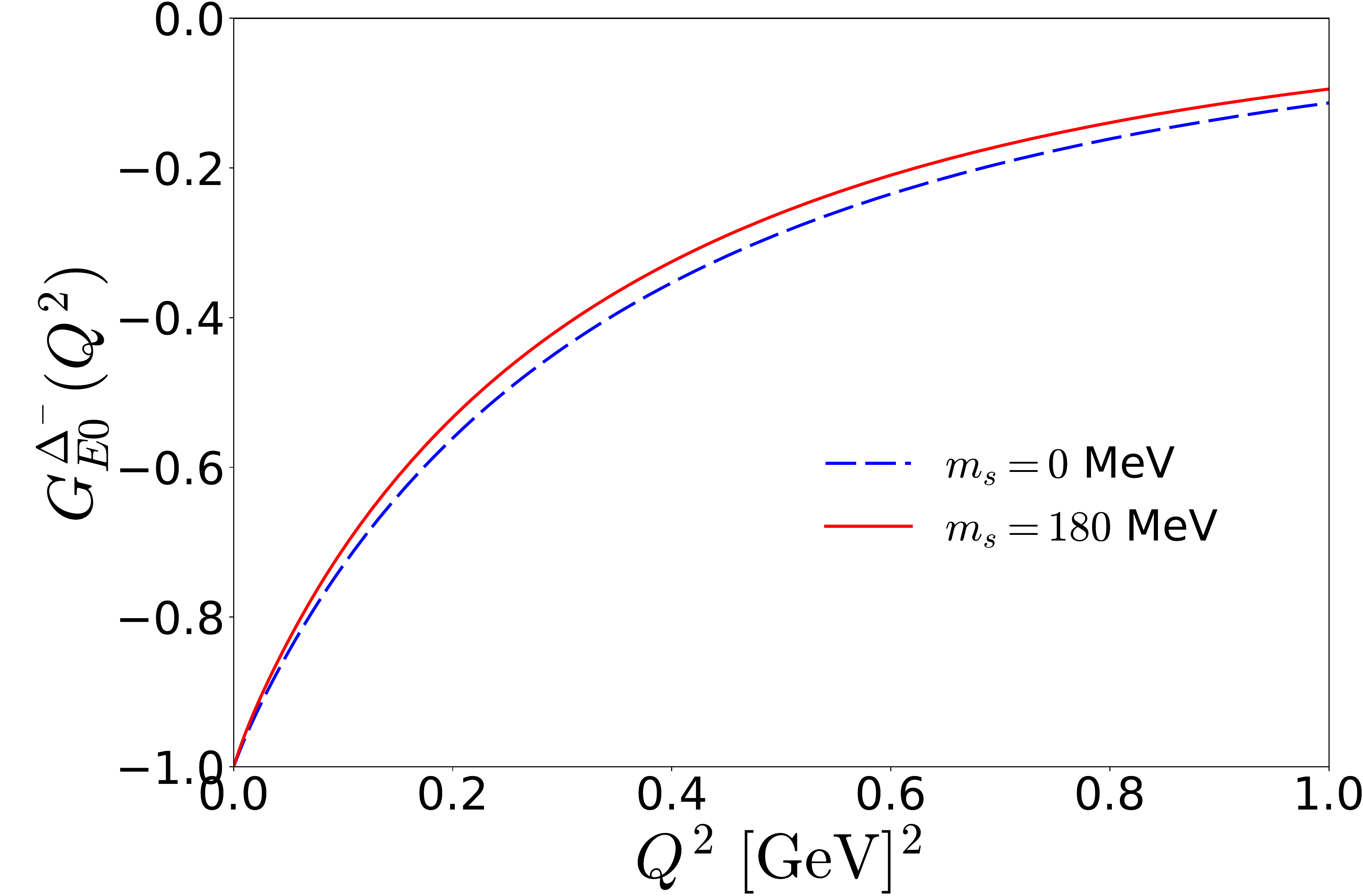}\hspace{0.5cm}
  \includegraphics[scale=0.2]{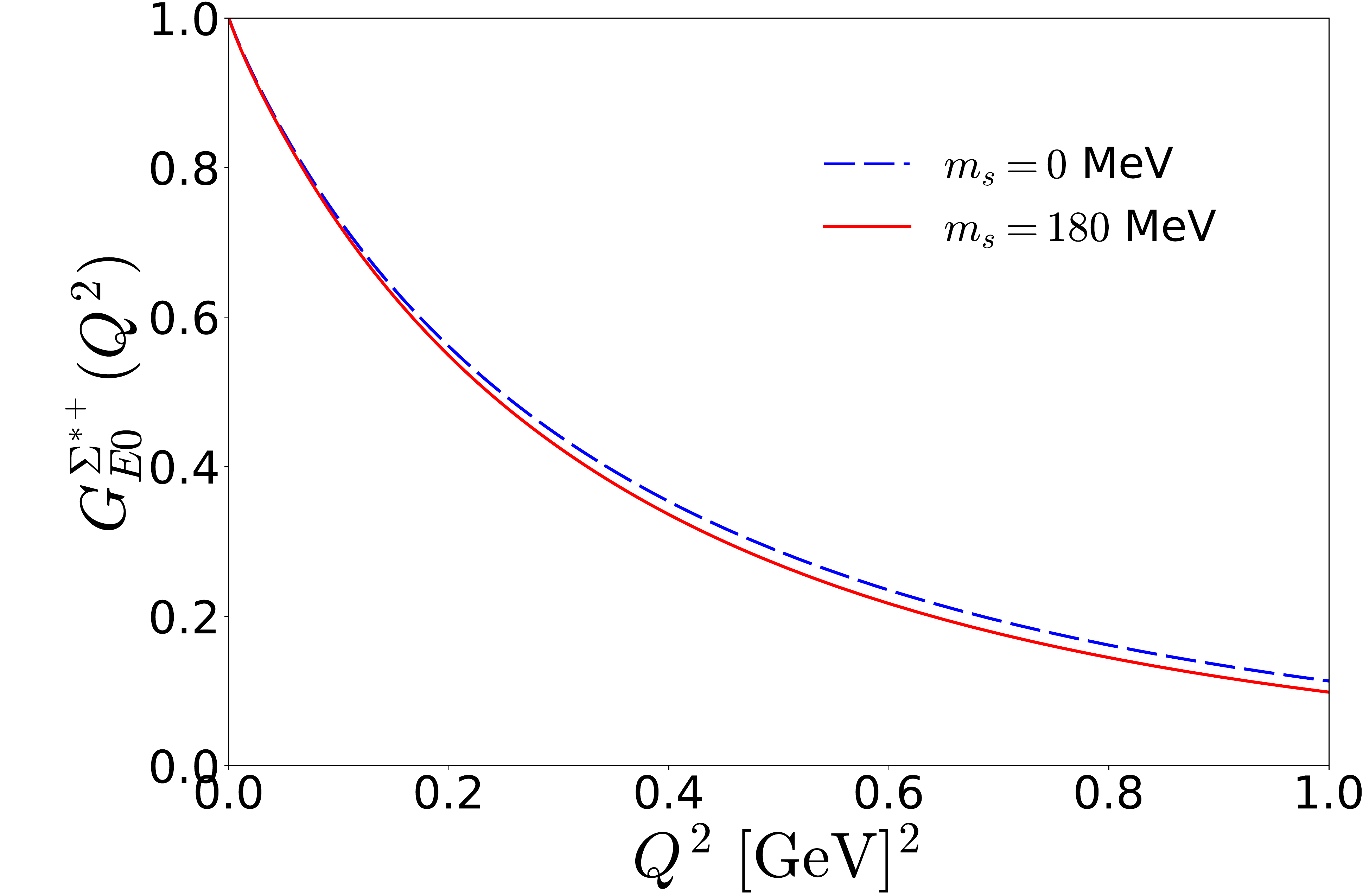}
  \includegraphics[scale=0.2]{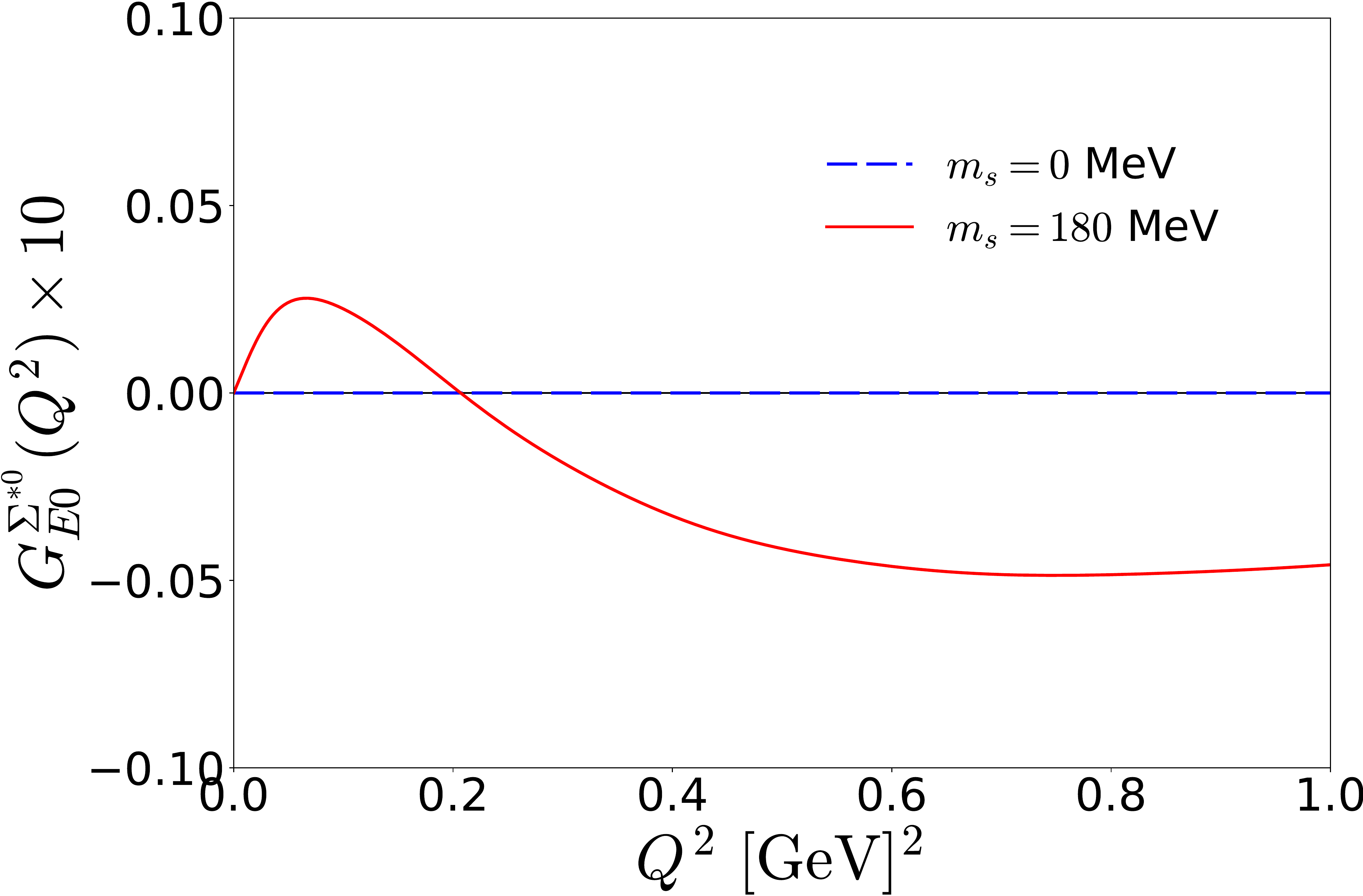}\hspace{0.5cm}
  \includegraphics[scale=0.2]{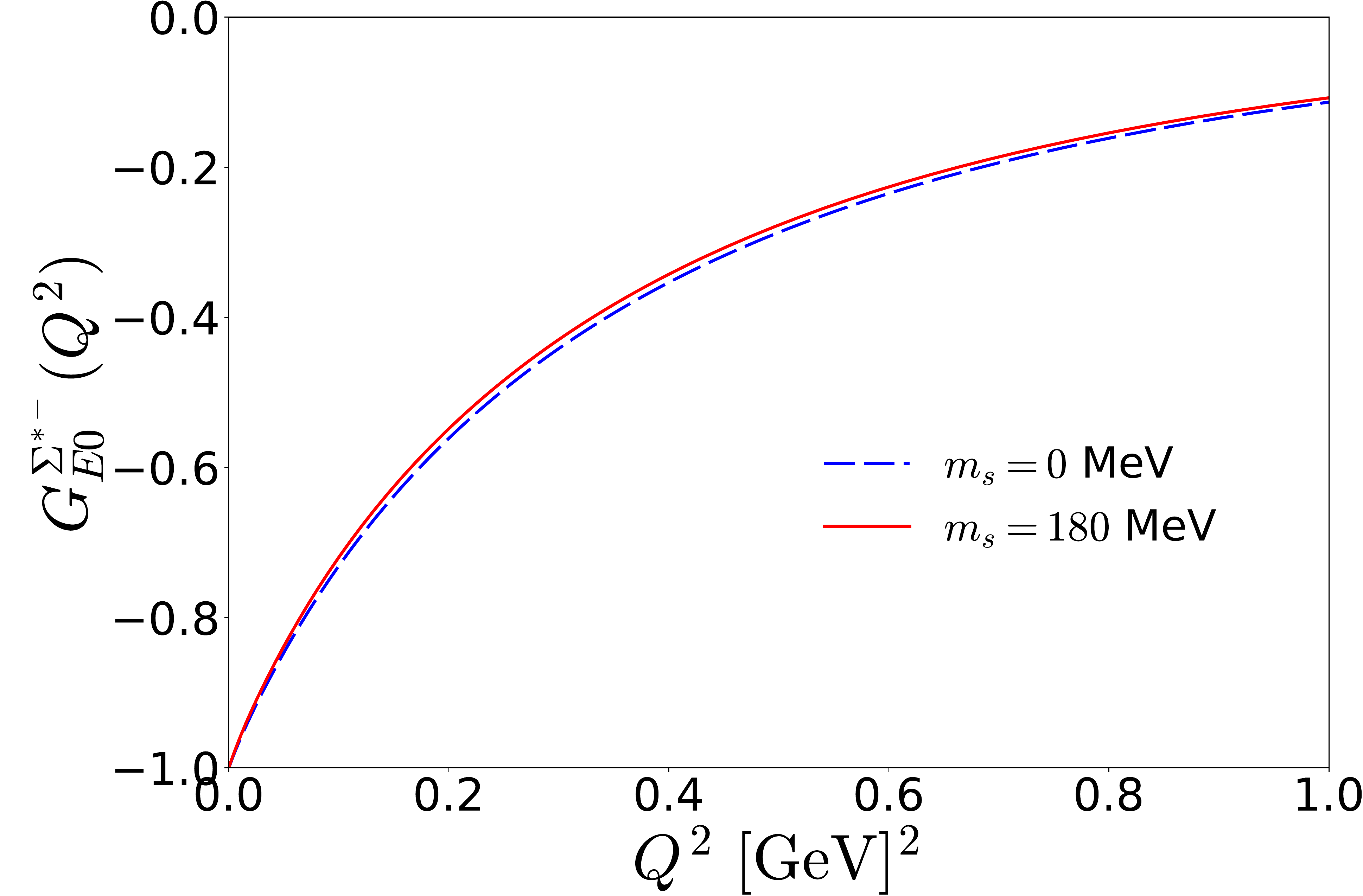}
  \includegraphics[scale=0.2]{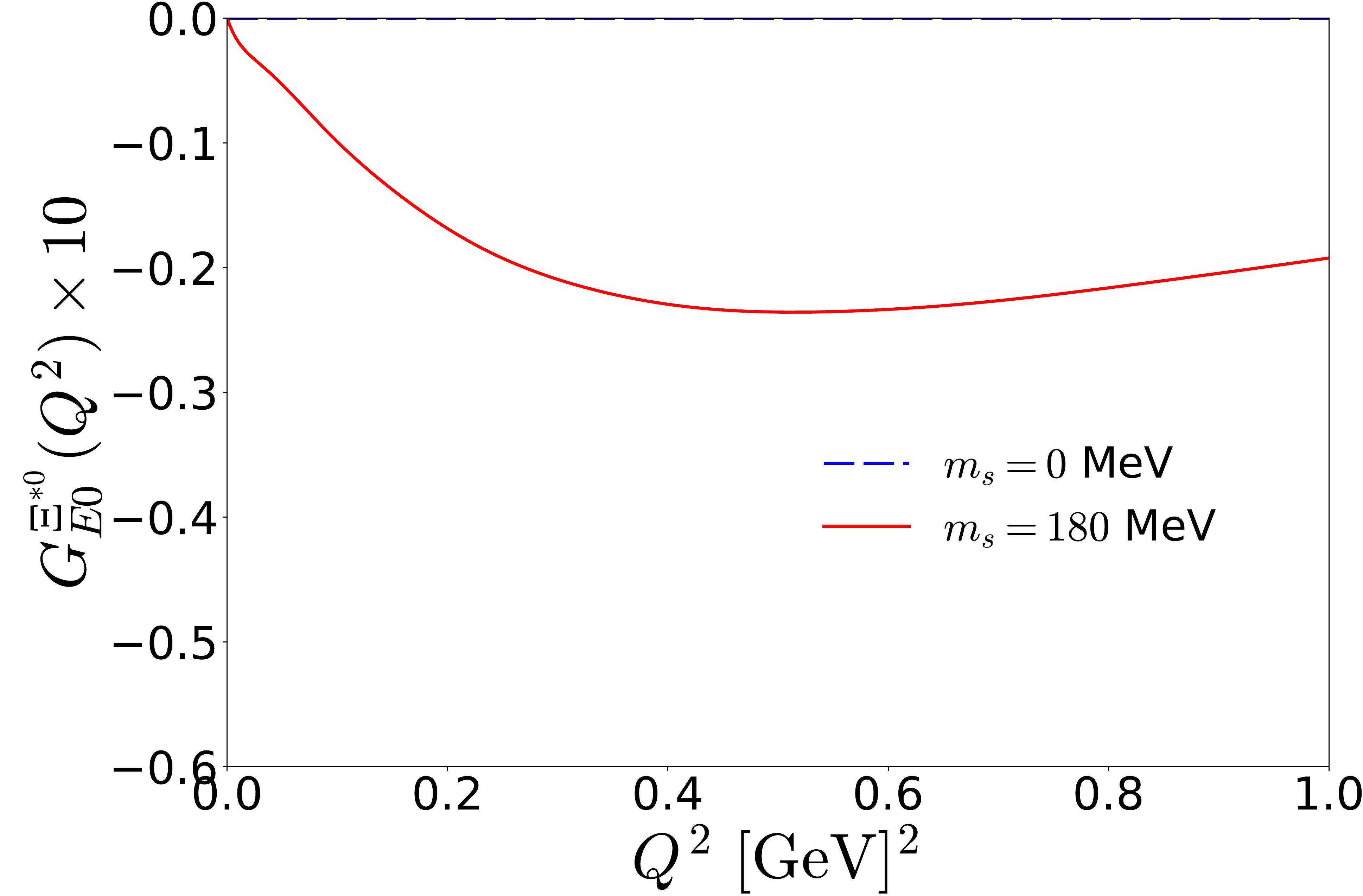}\hspace{0.5cm}
  \includegraphics[scale=0.2]{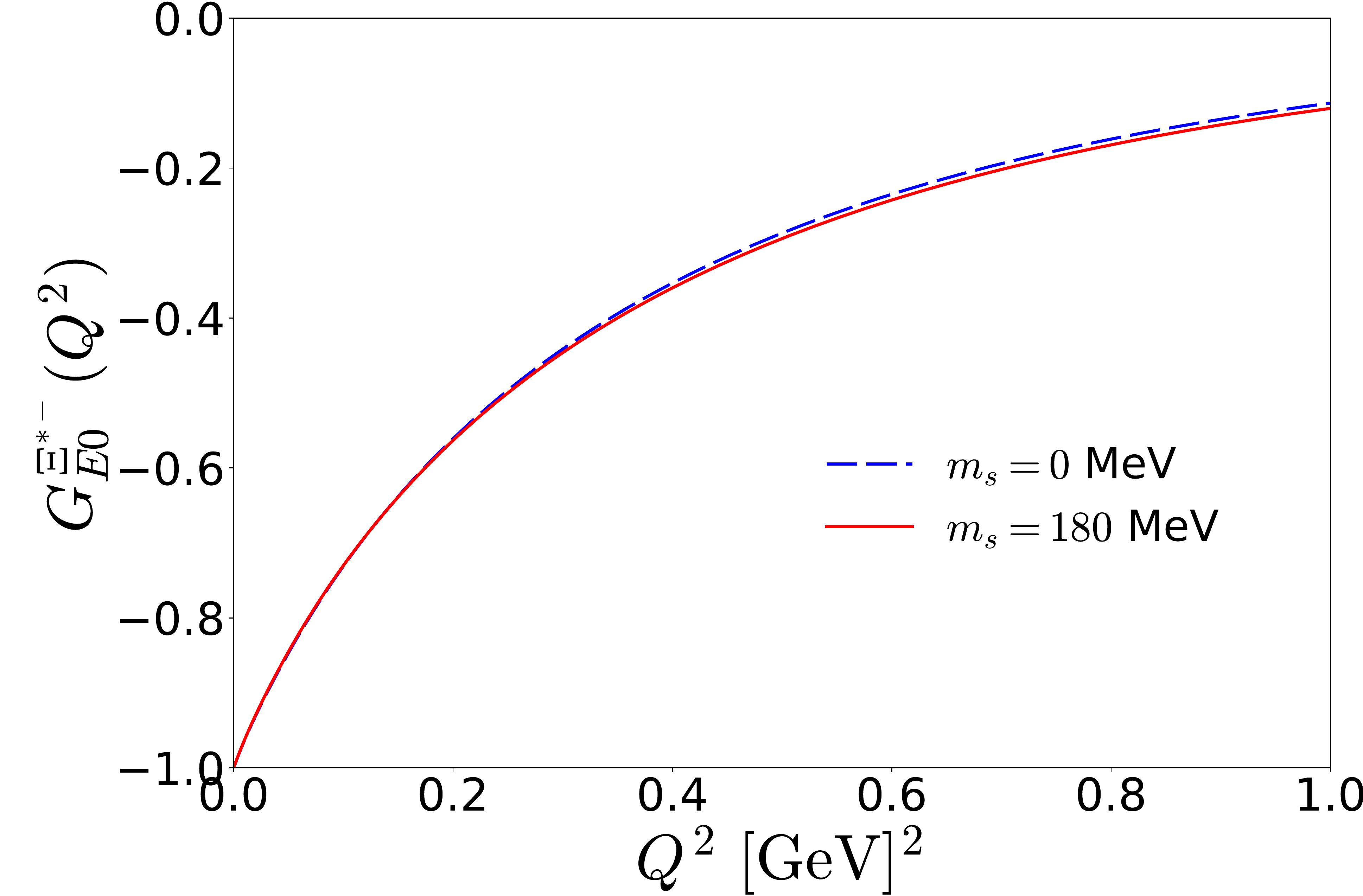}
  \caption{Effects of the explicit flavor SU(3) symmetry breaking on
    the electric monopole form factors of the baryon decuplet except
    for the $\Delta^+$ and $\Omega^-$ baryons. Notations are the same
    as in Fig.~\ref{fig:7}.}
\label{fig:8}
\end{figure}

We now examine the effects of the linear $m_s$ corrections on the $E0$
form factors of the $\Delta^+$ and $\Omega^-$ baryons. As shown
explicitly in Fig.~\ref{fig:7}, the linear $m_s$ corrections are
almost negligible in the case of the $\Delta^+$ baryon. They also do
not contribute much to the $G_{E0}^{\Omega^-}$ form factor. In the
case of the neutral $E0$ form factor, however, the leading contributions  
together with the rotational $1/N_c$ effects vanish, because they are
proportional to the corresponding charge as shown in
Eq.~\eqref{eq:E0leading}. In Fig.~\ref{fig:8}, we draw the $E0$ form
factors of all the members in the baryon decuplet except for the
$\Delta^+$ and $\Omega^-$. In general, the effects of the linear $m_s$
corrections are rather small on the $E0$ form factors of the charged
baryons. However, it is of great interest to see the results of the
neutral $E0$ form factors. In this case, the linear $m_s$ corrections 
become the leading-order contribution. 

  \begin{figure}[htp]
  \includegraphics[scale=0.2]{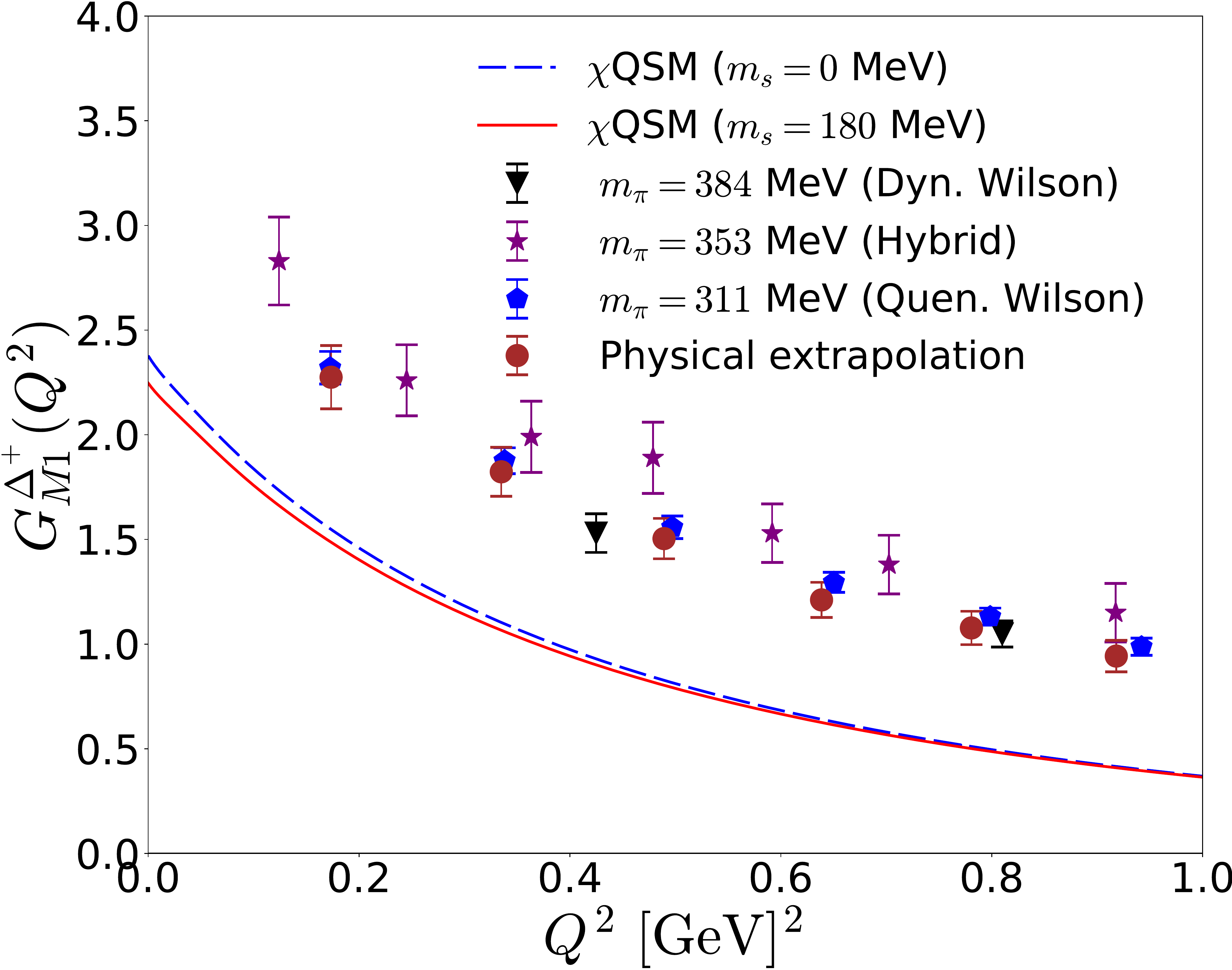}\hspace{0.5cm}
  \includegraphics[scale=0.2]{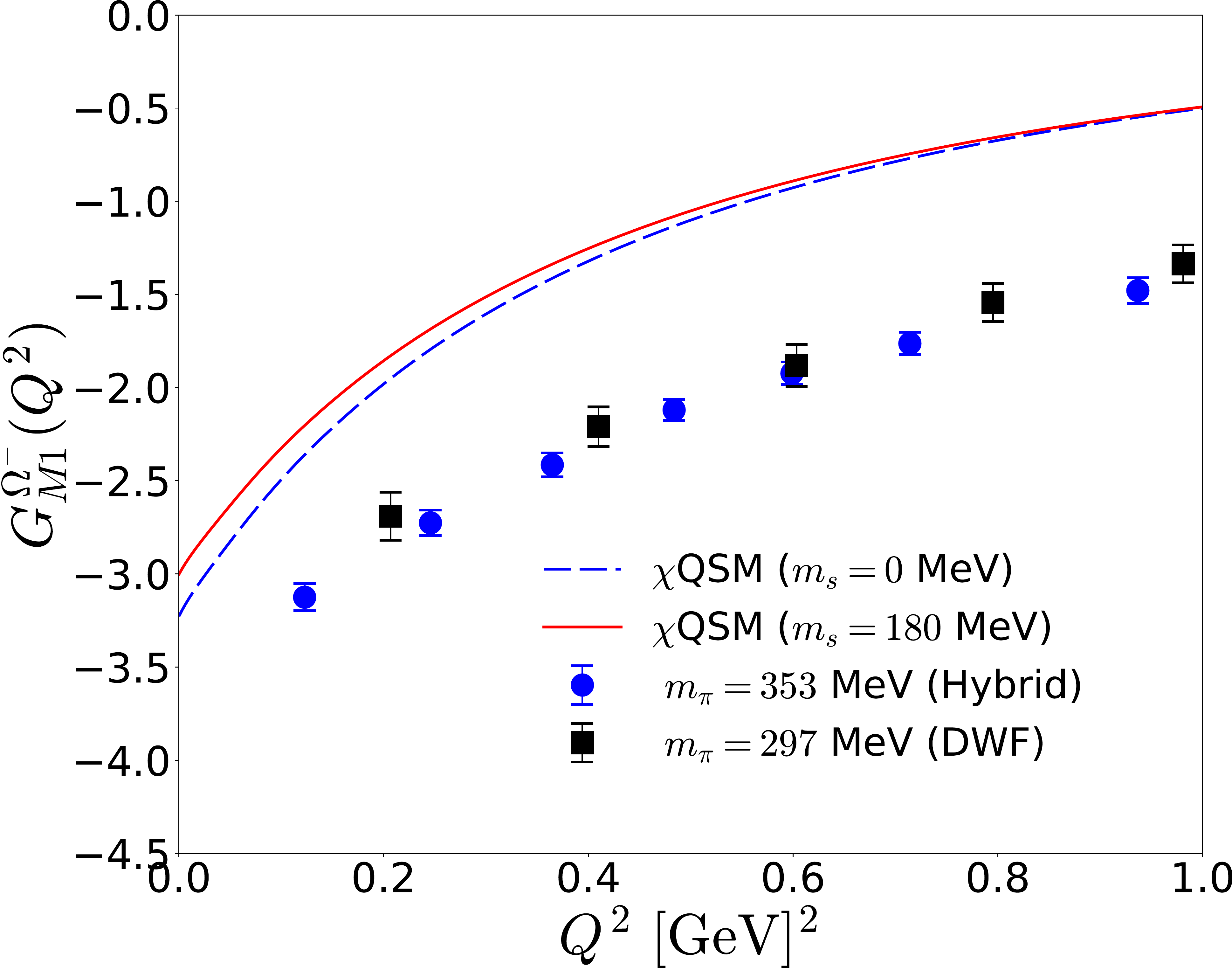}
  \caption{Effects of the explicit flavor SU(3) symmetry breaking
  on the magnetic dipole form factors of the $\Delta^+$ isobar and
   $\Omega^-$ in the left and right panels, respectively. The dashed
   curves depict the E0 form factors without the effects of the
   explicit flavor SU(3) symmetry breaking, whereas the solid ones 
   present the $E0$ form factors with the 
   effects. The lattice data 
   are taken from Refs.~\cite{Alexandrou:2007we,Alexandrou:2009hs,
     Alexandrou:2010jv}. Notations are the same
    as in Fig.~\ref{fig:7}.} 
\label{fig:9}
\end{figure}

  \begin{figure}[htp]
  \includegraphics[scale=0.2]{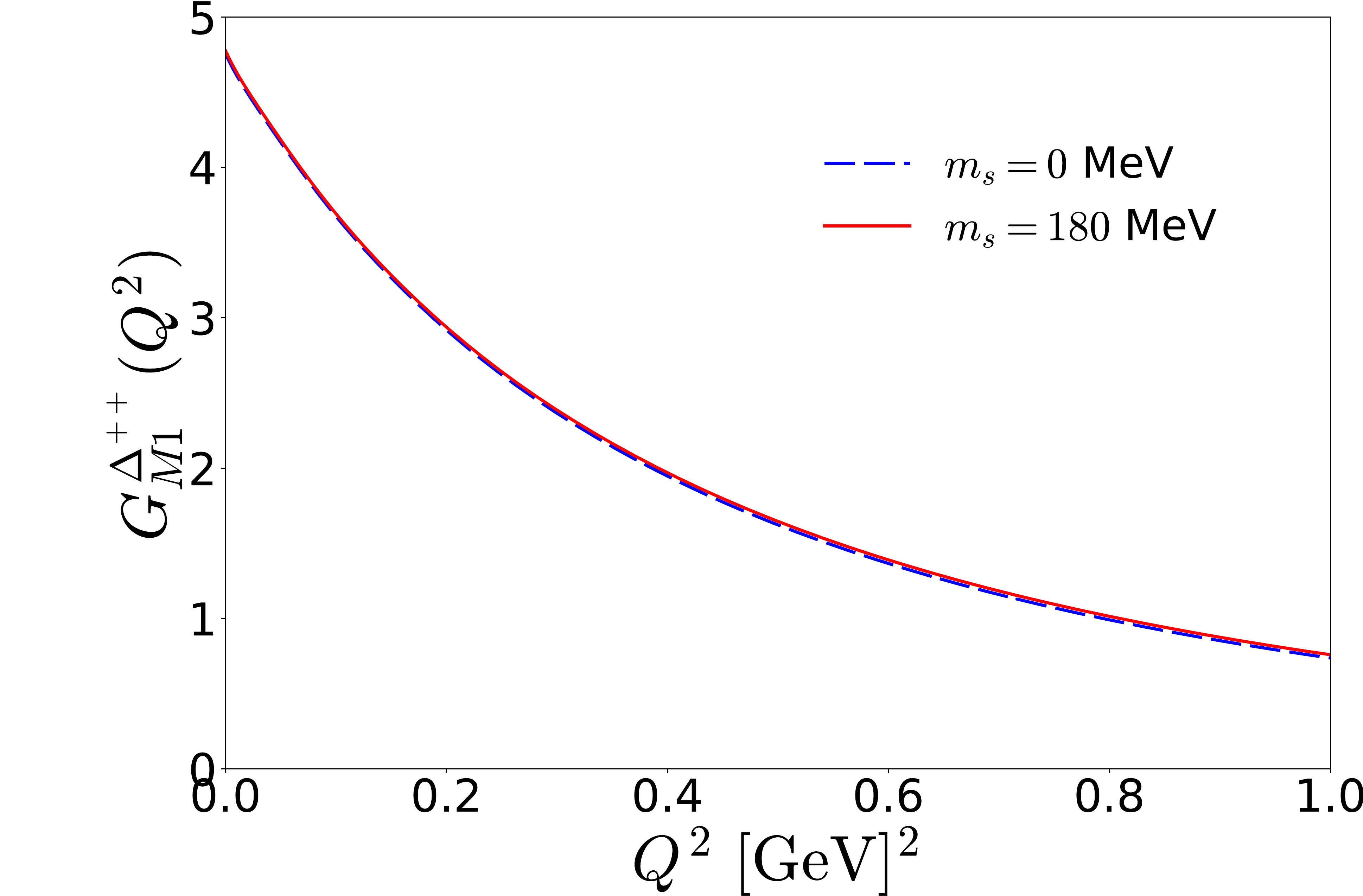} \hspace{0.5cm}
  \includegraphics[scale=0.2]{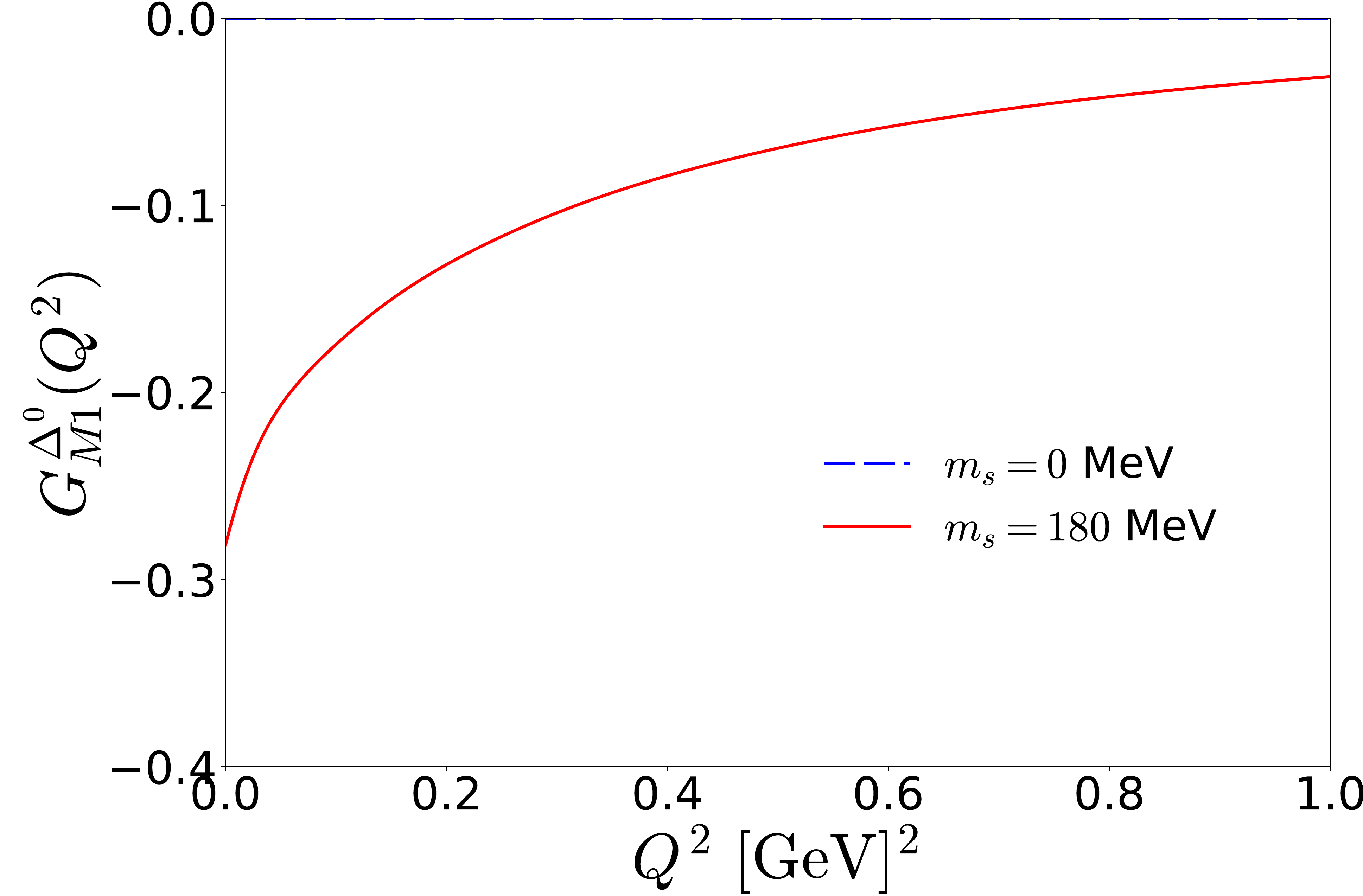}
  \includegraphics[scale=0.2]{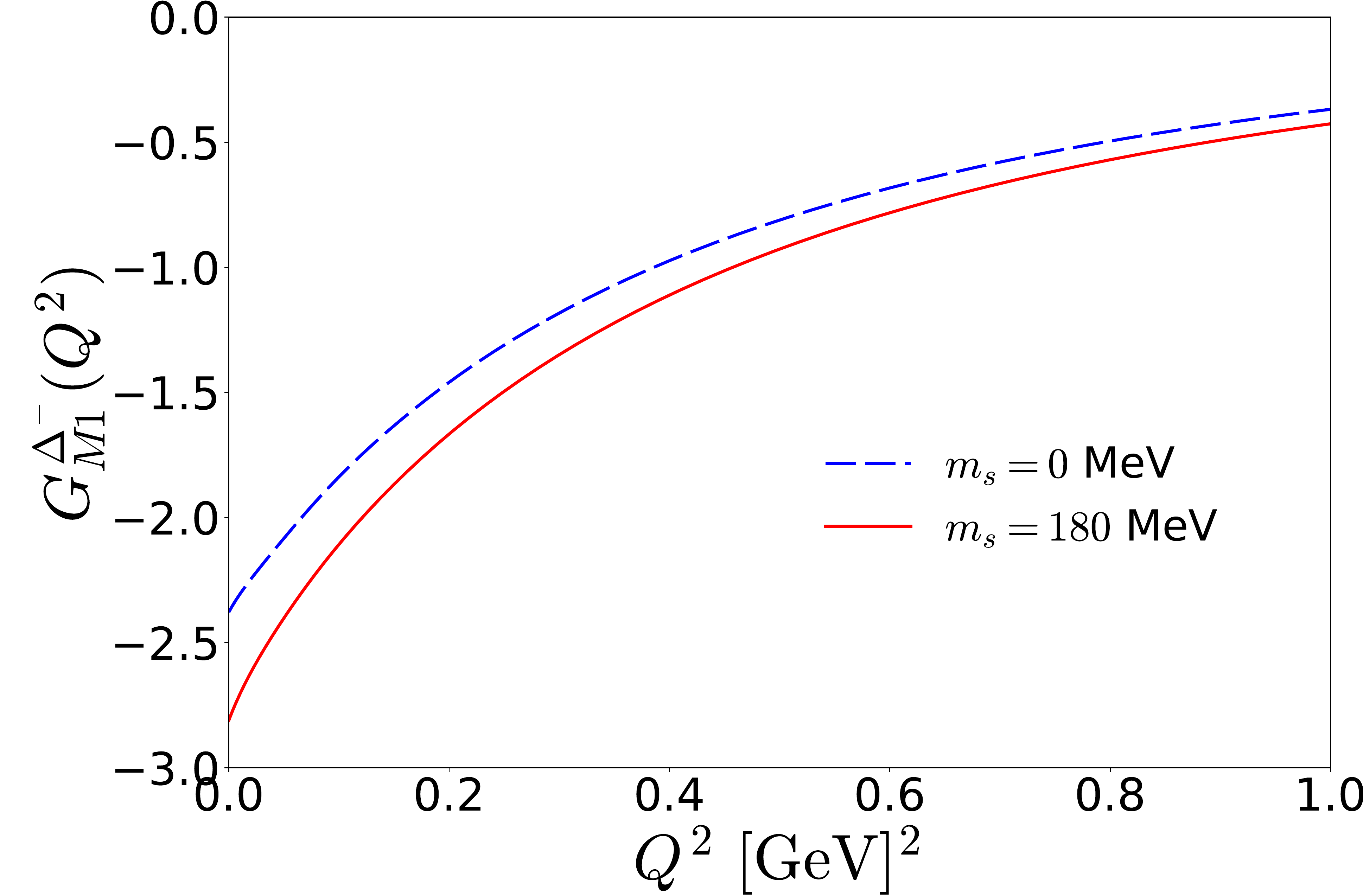}\hspace{0.5cm}
  \includegraphics[scale=0.2]{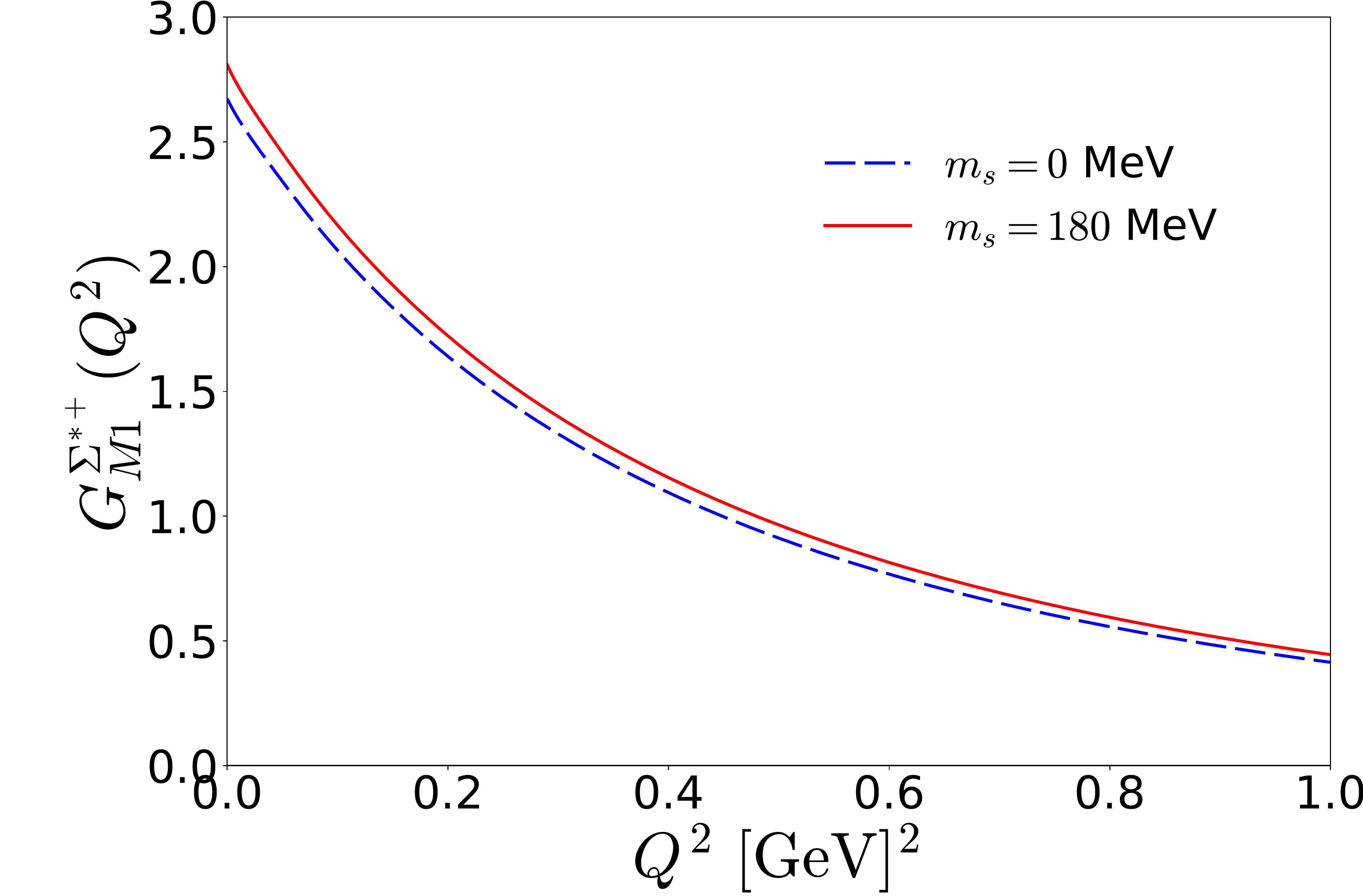}
  \includegraphics[scale=0.2]{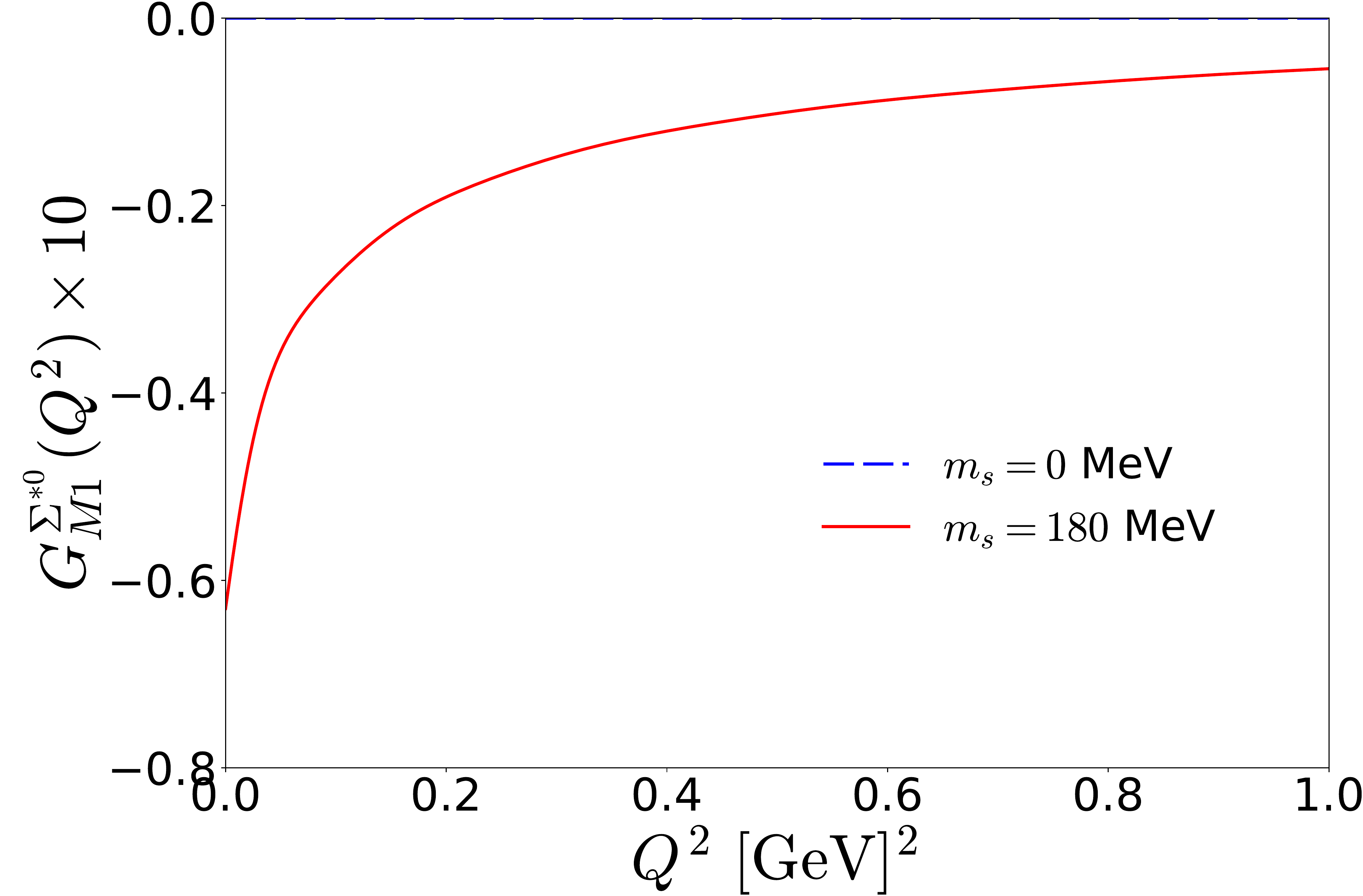}\hspace{0.5cm}
  \includegraphics[scale=0.2]{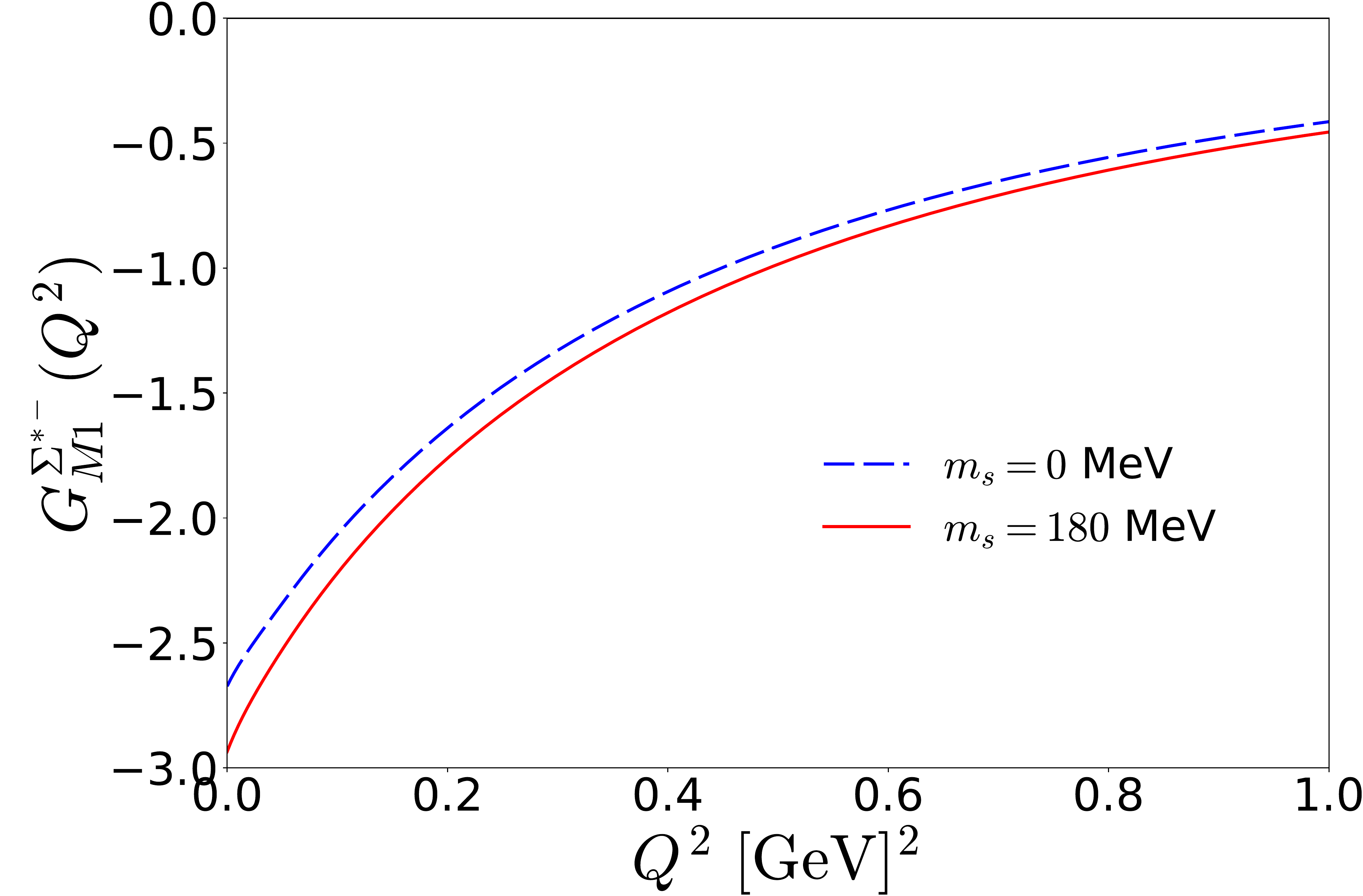}
  \includegraphics[scale=0.2]{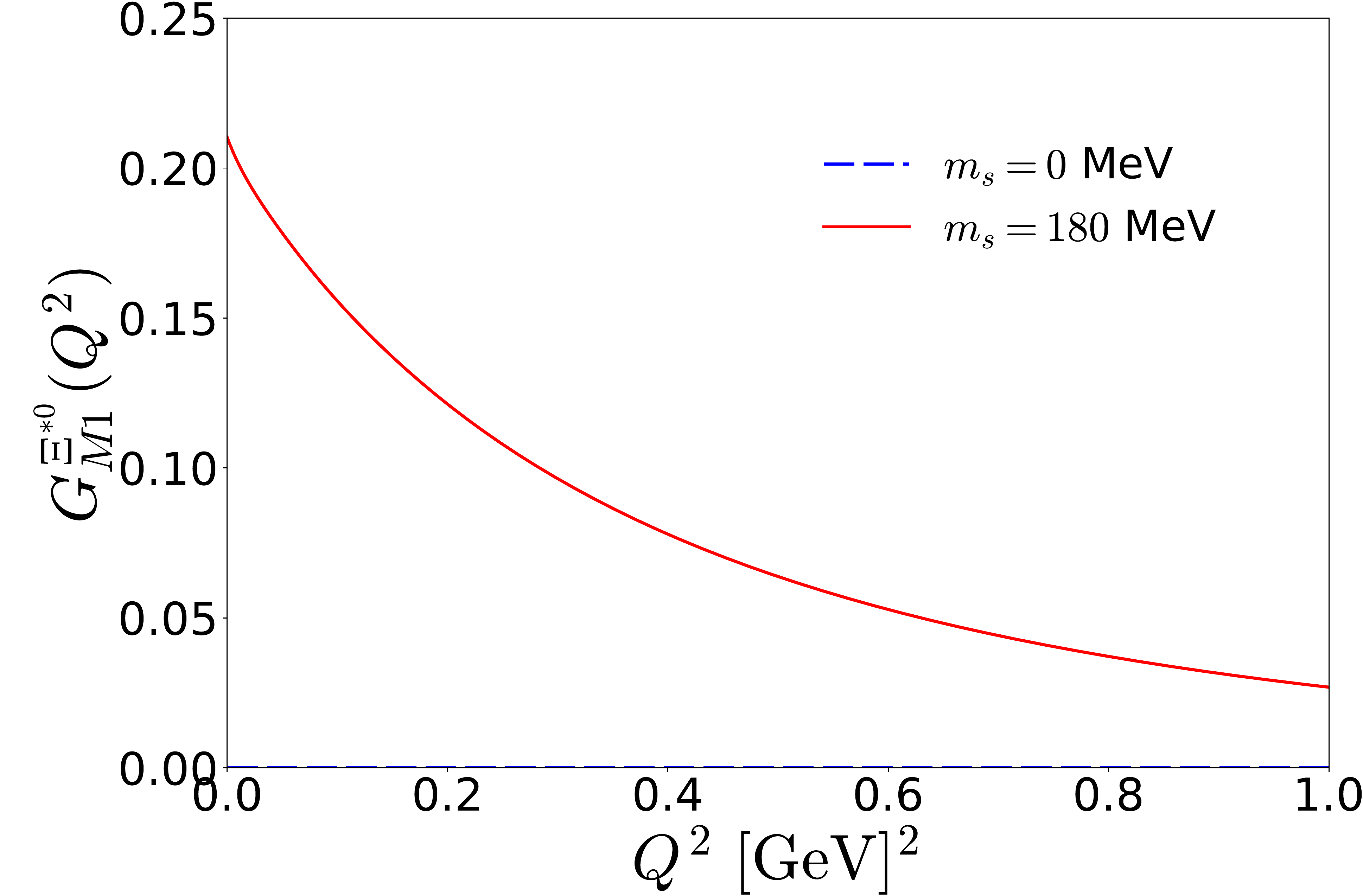}\hspace{0.5cm}
  \includegraphics[scale=0.2]{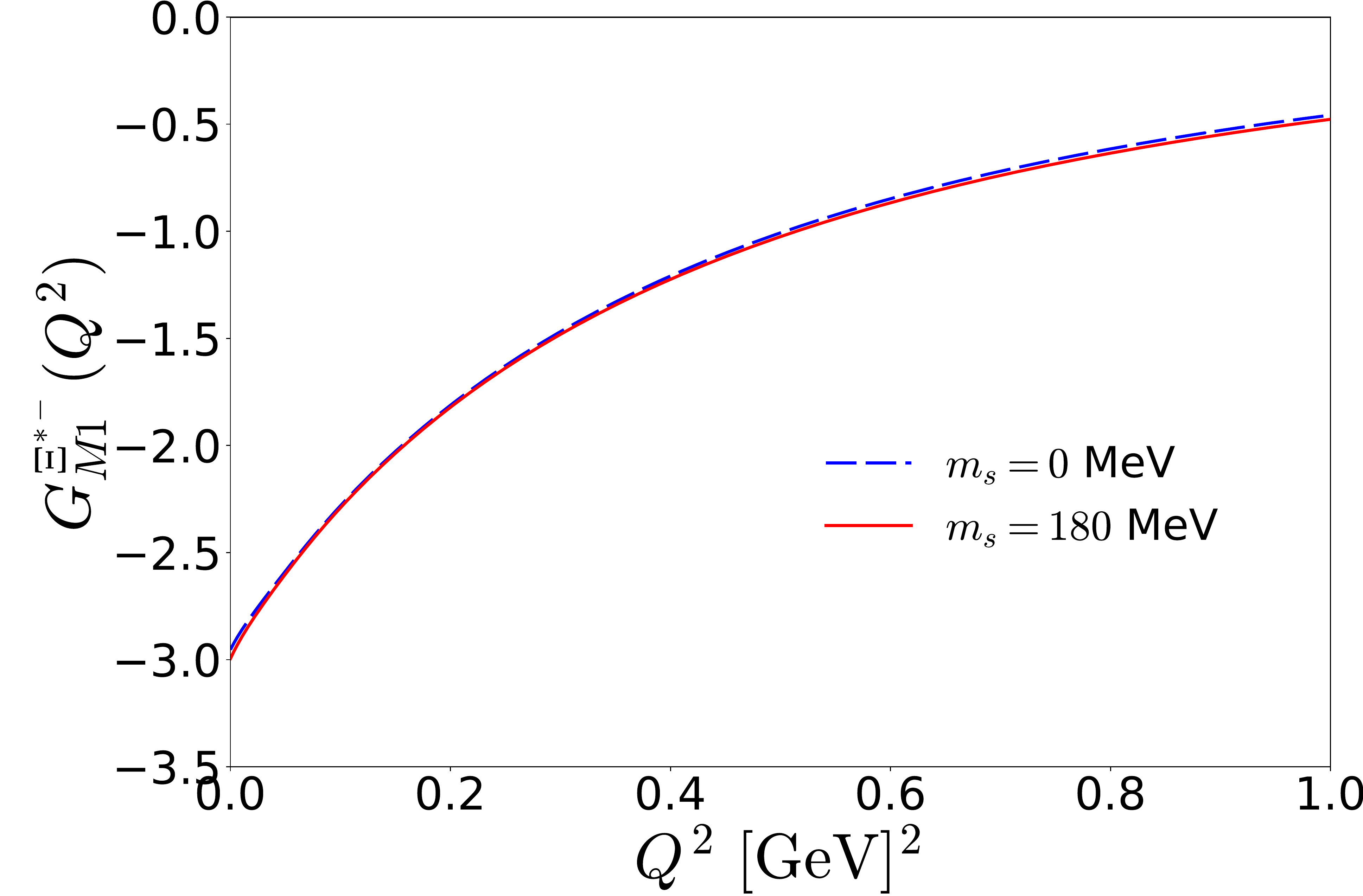}
  \caption{Effects of the explicit flavor SU(3) symmetry breaking on
    the magnetic dipole form factors of the baryon decuplet except
    for the $\Delta^+$ and $\Omega^-$ baryons. Notations are the same
    as in Fig.~\ref{fig:7}.}
\label{fig:10} 
\end{figure}
As depicted in Fig.~\ref{fig:9}, the effects of the flavor SU(3)
symmetry breaking are also small on the $\Delta^+$ and $\Omega^-$ $M1$
form factors. However, the result of the magnetic dipole form factor
of the $\Delta^-$ is enhanced by almost $15~\%$ with the linear $m_s$
corrections included. The reason can be found easily by examining the
expressions of the $m_s$ corrections written in
Eqs.~\eqref{eq:M1mslinear1} and \eqref{eq:M1final}. In the case of
$\Delta^-$, the charge-dependent terms and other ones are
constructively added, so that the $m_s$ corrections are strengthened. 
As for the $M1$ form factors of the neutral baryon decuplet, the
leading-order and rotational $1/N_c$ contributions vanish, since they
are proportional to the corresponding charge. Thus, the linear $m_s$
corrections again play the leading role in the case of the 
neutral form factors of the baryon decuplet.

  \begin{figure}[htp]
  \includegraphics[scale=0.2]{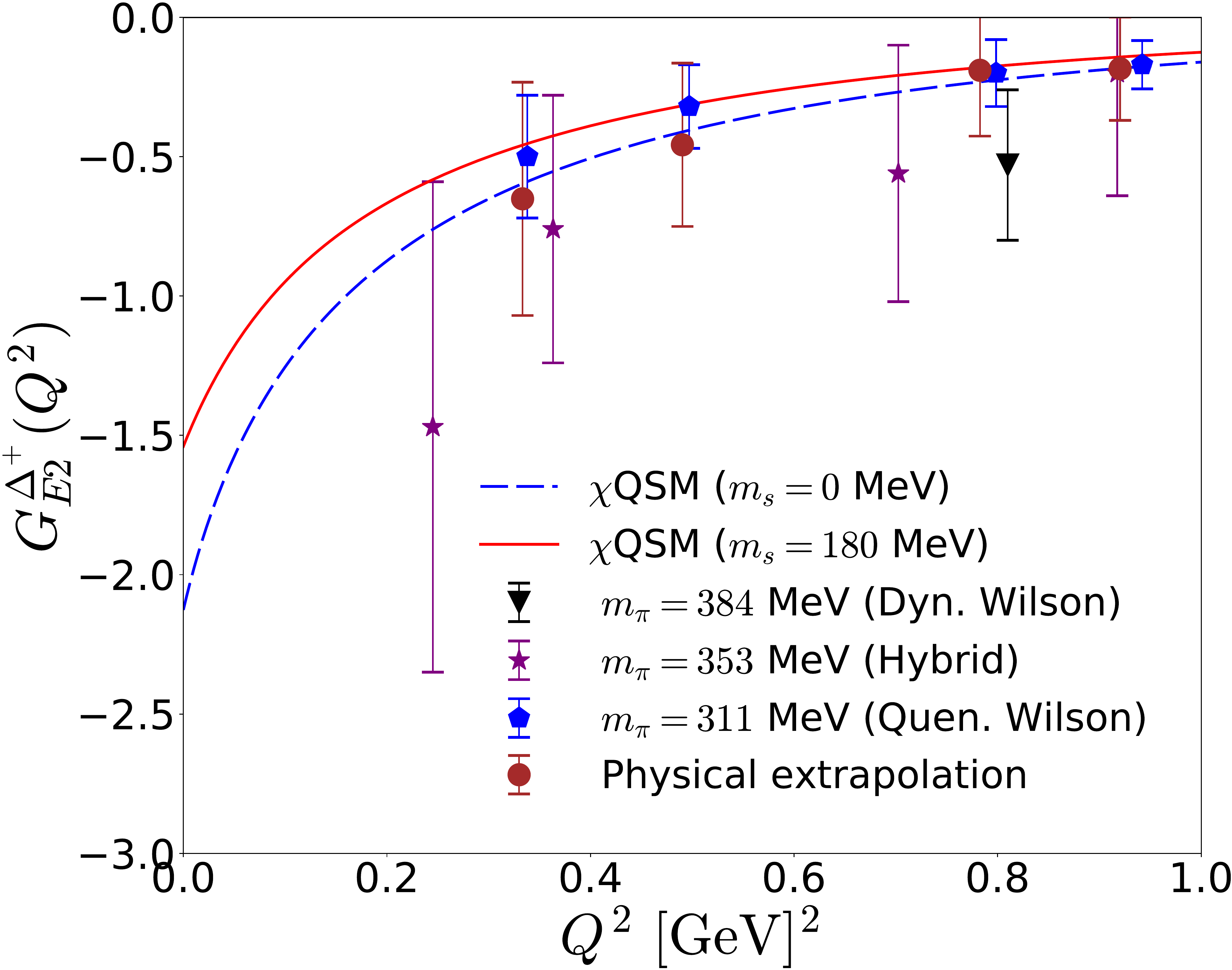}\hspace{0.5cm}
  \includegraphics[scale=0.2]{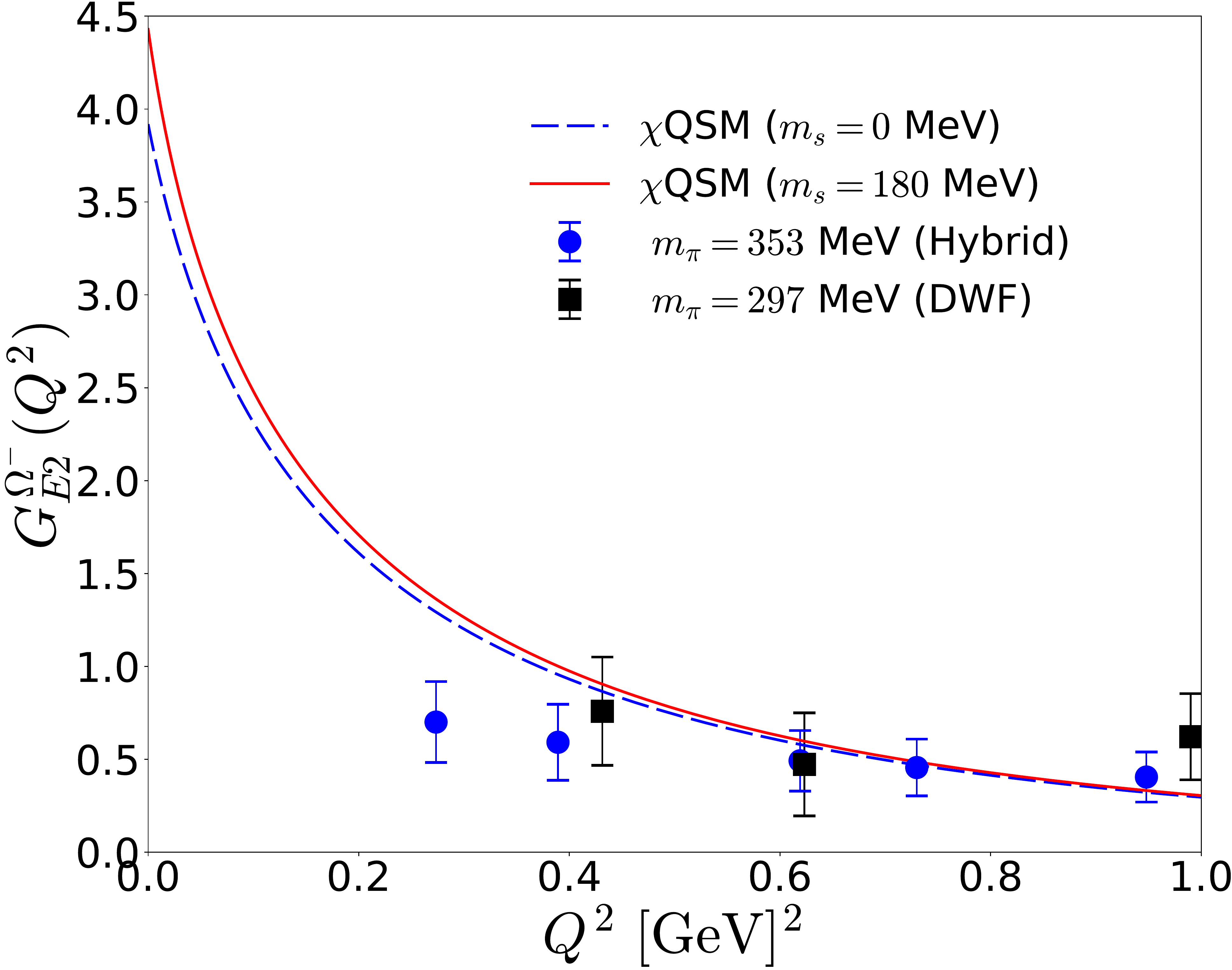}
  \caption{Effects of the explicit flavor SU(3) symmetry breaking on
    the electric quadrupole form factors of the $\Delta^+$ isobar and
    $\Omega^-$ in the left and right panels, respectively. The dashed
    curves depict the E2 form factors without the effects of the
    explicit flavor SU(3) symmetry breaking, whereas the solid ones 
    present the E2 form factors with the effects. The lattice data
    are taken from Refs.~\cite{Alexandrou:2007we,Alexandrou:2009hs,
Alexandrou:2010jv}. Notations are the same as in Fig.~\ref{fig:7}.} 
\label{fig:11}
\end{figure}

  \begin{figure}[htp]
  \includegraphics[scale=0.2]{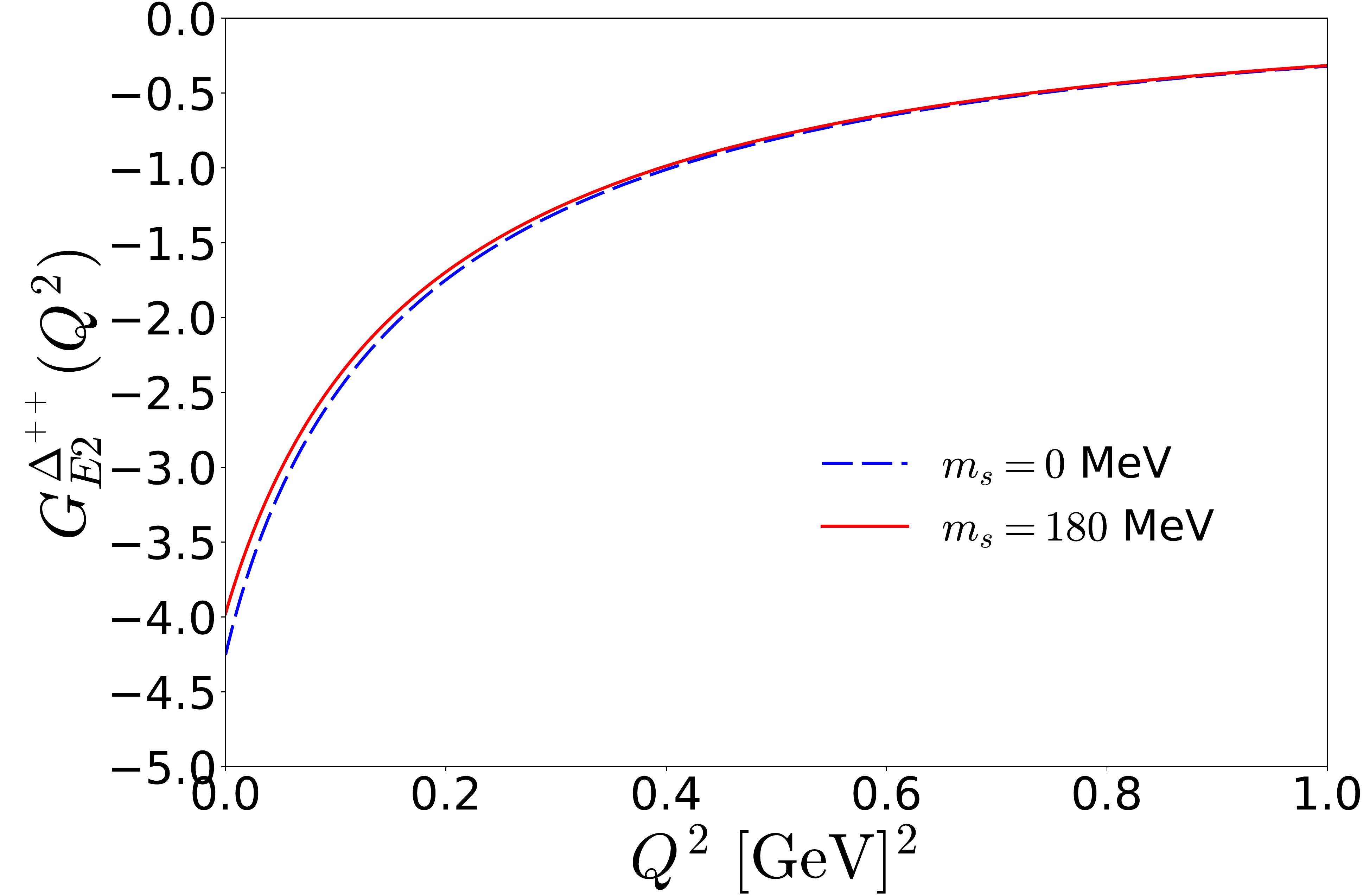}\hspace{0.5cm}
  \includegraphics[scale=0.2]{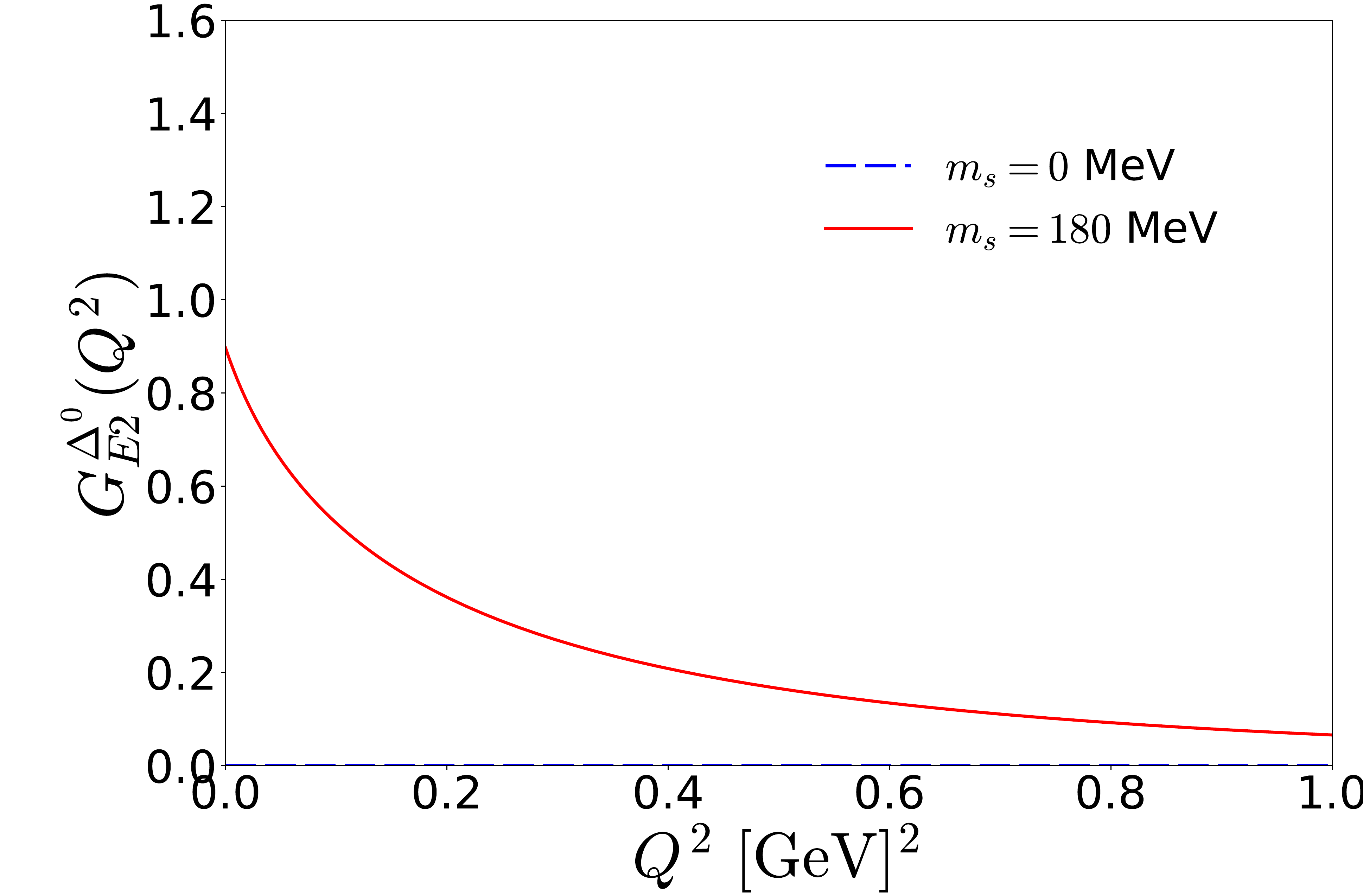}
  \includegraphics[scale=0.2]{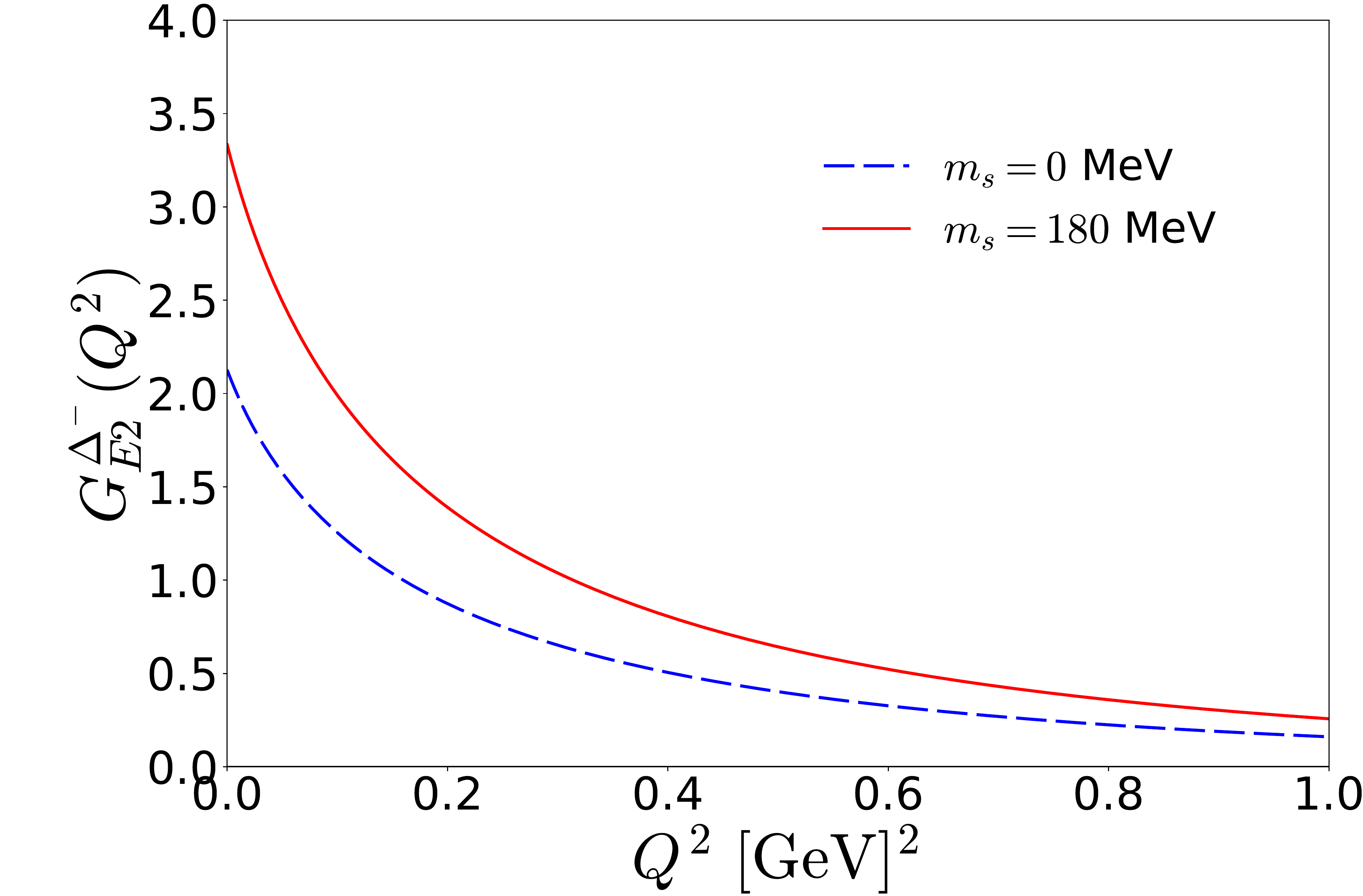}\hspace{0.5cm}
  \includegraphics[scale=0.2]{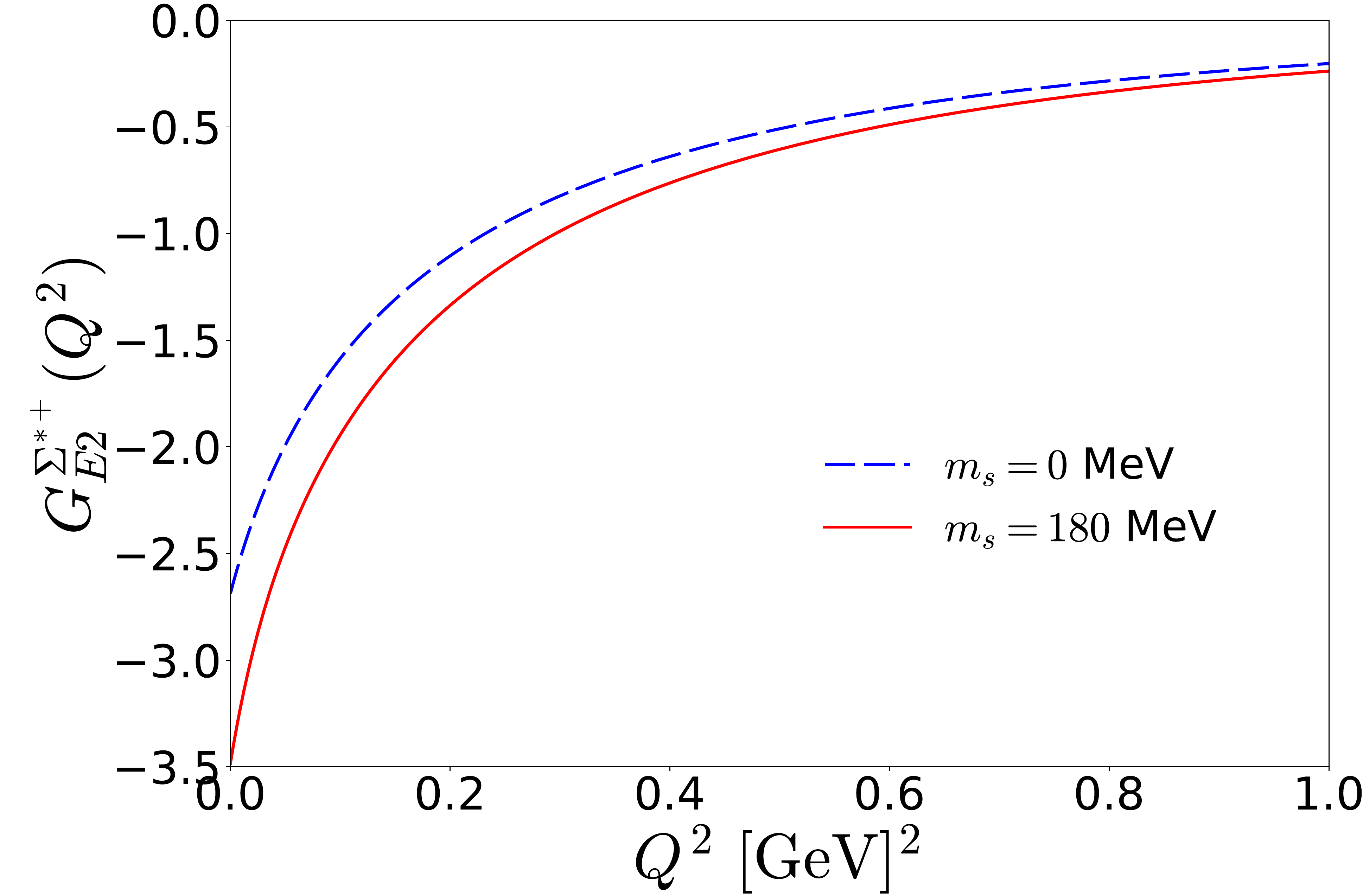}
  \includegraphics[scale=0.2]{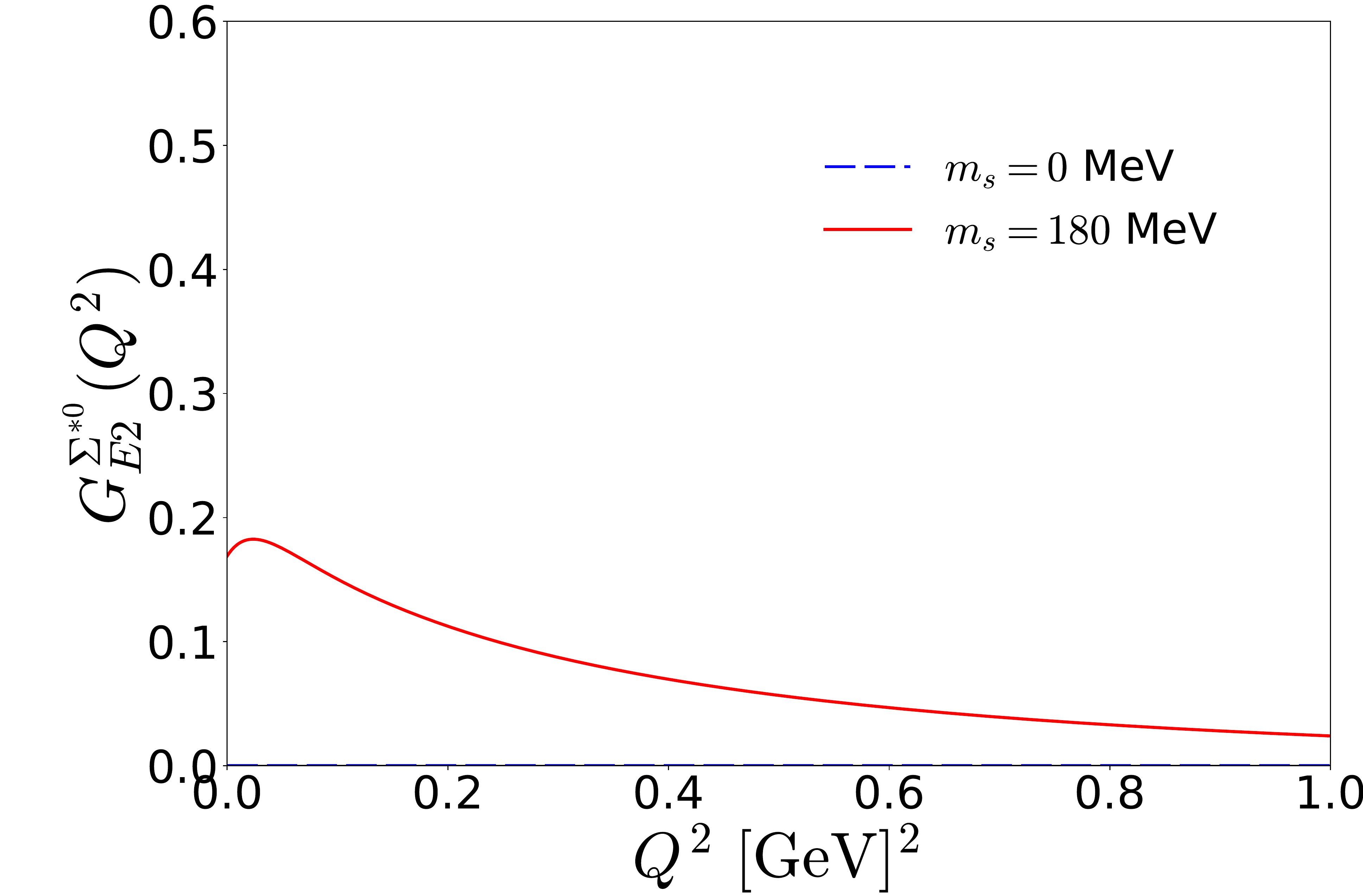}\hspace{0.5cm}
  \includegraphics[scale=0.2]{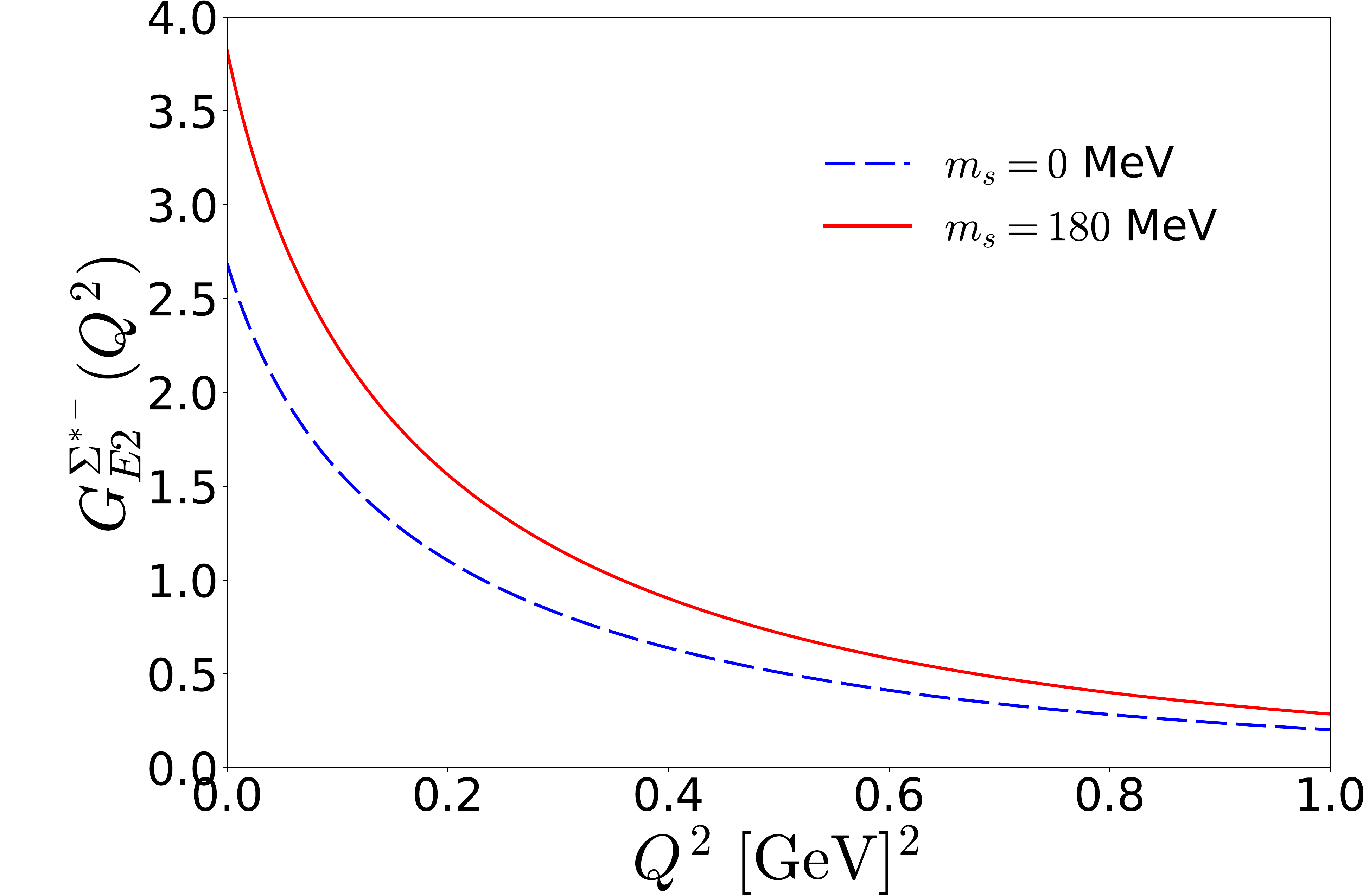}
  \includegraphics[scale=0.2]{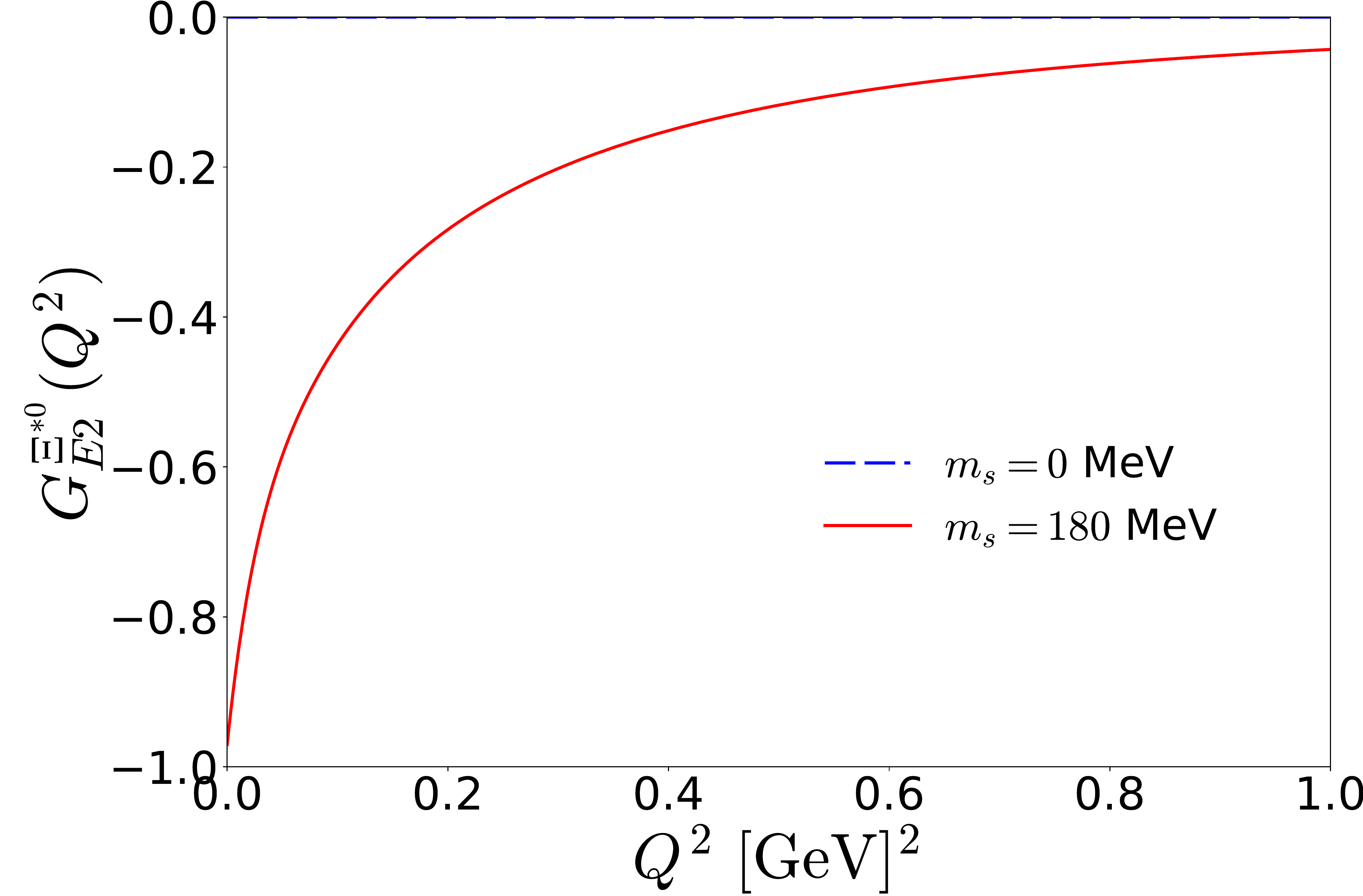}\hspace{0.5cm}
  \includegraphics[scale=0.2]{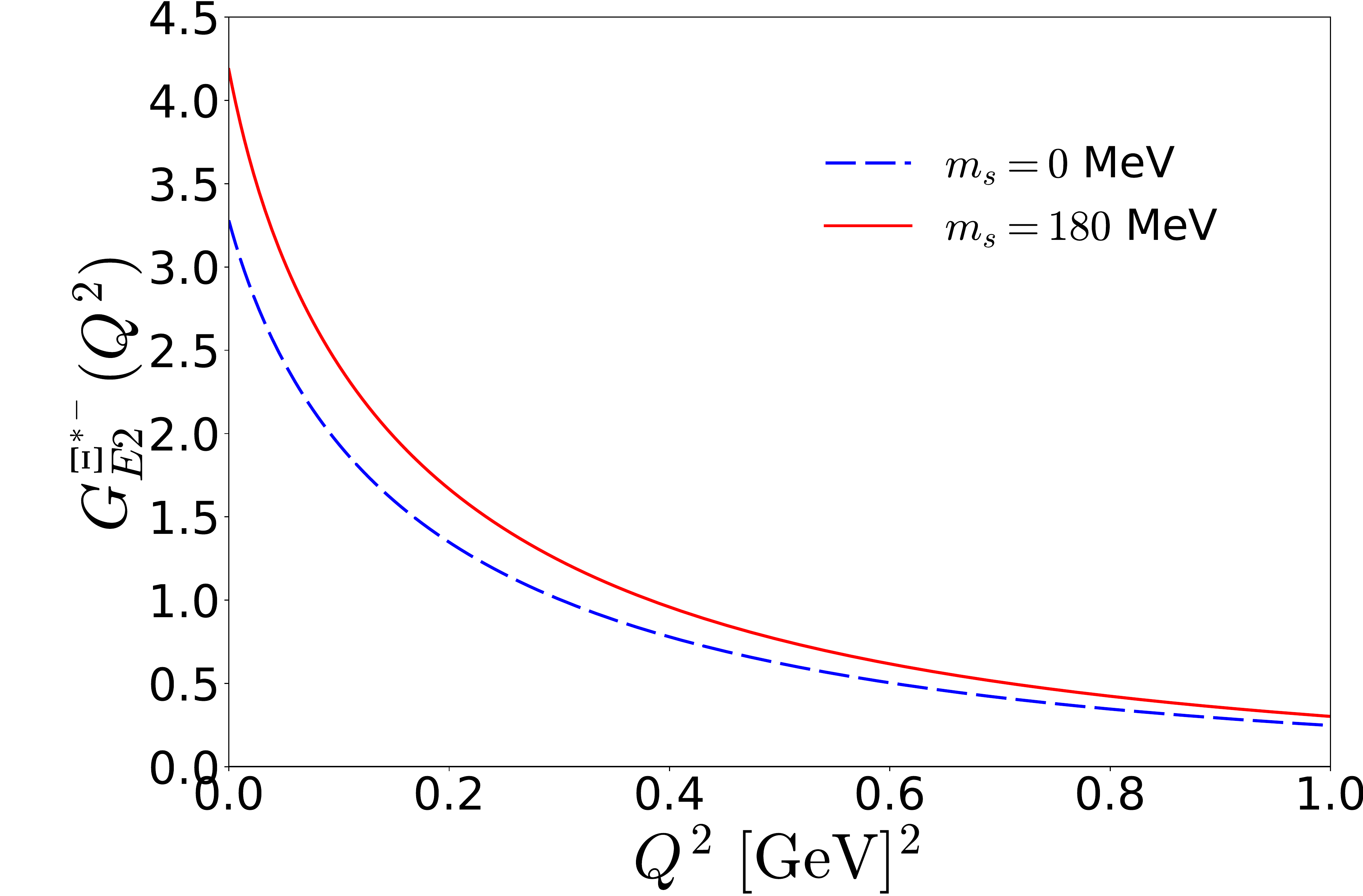}
  \caption{Effects of the explicit flavor SU(3) symmetry breaking on
    the electric quadrupole form factors of the baryon decuplet except
    for the $\Delta^+$ and $\Omega^-$ baryons. Notations are the same
    as in Fig.~\ref{fig:7}.}
\label{fig:12}
\end{figure}
Figure~\ref{fig:11} shows the results of the electric
quadrupole form factors of the $\Delta^+$ and $\Omega^-$
baryons. Interestingly, the $m_s$ corrections are quite sizable to the
$\Delta^+$ $E2$ form factor whereas they are marginal to the
$\Omega^-$ one. We can understand this difference from
Eq.~\eqref{eq:E2final}. 

\subsection{Charge radii, magnetic dipole moments, magnetic radii, and
electric quadrupole moments}

\begin{table}
\caption{Charge radii of the baryon decuplet in comparison with those
  from lattice
  QCD~\cite{Alexandrou:2007we,Boinepalli:2009sq,Alexandrou:2010jv}, 
the  chiral constutuent-quark model~\cite{Berger:2004yi}, the chiral quark 
  model~\cite{Wagner:2000ii}, large $N_{c}$~\cite{Buchmann:2002et},
  combined chiral and $1/N_{c}$
  expansion~\cite{Flores-Mendieta:2015wir}, covariant
  $\chi$PT~\cite{Geng:2009ys} and $\chi$PT~\cite{Li:2016ezv}.} 
\label{tab:1}
 \begin{tabular}{ c | c c c c c c c c c c } 
  \hline 
    \hline 
 $\langle r^{2}\rangle_{E}  $& $\Delta^{++}$  &  $\Delta^{+}$ &
 $\Delta^{0}$ & $\Delta^{-}$  &  $\Sigma^{*+}$ &  $\Sigma^{*0}$ &  
$\Sigma^{*-}$ &  $\Xi^{*0}$ &  $\Xi^{*-}$ &  $\Omega^{-}$ \\
  \hline 
$m_{s} = 180$ &$0.826 $ & $ 0.792 $  &  $-0.069$ & $0.930$
& $0.843$  &  $-0.024$  & $0.891$ & $0.021$ & $0.852$ & $0.813$\\  
$m_{s} = 0$ &$0.832$ & $0.832$  &  $0$ & $0.832$
& $0.832$  &  $0$ & $0.832$ &  $0$ & $0.832$ & $0.832$\\  
 \hline 
LQCD~\cite{Alexandrou:2007we,Alexandrou:2010jv} &$-$ & $0.477(8)$  &  $-$ & $-$
& $-$  &  $-$ & $-$ &  $-$ & $-$ & $0.348(52)$\\ 
LQCD~\cite{Boinepalli:2009sq} &$-$ & $0.410(57)$  &  $0$ & $-$
& $0.399(45)$  &  $0.020(7)$ & $0.360(32)$ &  $0.043(10)$ & $0.330(20)$ & $0.307(15)$\\ 
\hline
$\chi$PT~\cite{Geng:2009ys} &$0.325(22)$ & $0.328(21)$  &  $0.006(1)$ & $0.316(23)$
& $0.315(21)$  &  $0$ & $0.315(21)$ &  $-0.006(1)$ & $0.312(18)$ & $0.307(15)$\\ 
$\chi$PT~\cite{Li:2016ezv} &$0.30(11)$ & $0.29(10)$  &  $-0.02(1)$ & $0.33(11)$
& $0.31(11)$  &  $0$ & $0.31(11)$ &  $0.02(1)$ & $0.29(10)$ & $0.27(10)$\\  
$1/N_{c}$~\cite{Buchmann:2002et}&$0.783$ & $0.783$  &  $0$ & $0.783$
& $0.869$  &  $0.108$ & $0.669$ &  $0.206$ & $0.561$ & $0.457$\\  
$1/N_{c}$~\cite{Flores-Mendieta:2015wir} & $1.048$  &  $1.101$ & $0.105$ & $0.891$
& $0.939$  &  $-0.031$ & $0.895$ &  $-0.098$ & $0.981$ & $1.042$\\  
$\chi$QM~\cite{Wagner:2000ii} &$0.77$ & $0.77$  &  $0$ & $0.77$
& $0.93$  &  $0.10$ & $0.74$ &  $0.20$ & $0.68$ & $0.78$\\ 
GBE CQM~\cite{Berger:2004yi}&$0.43$ & $0.43$  &  $0$ & $0.43$
& $0.42$  &  $0.37$ & $0.03$ &  $0.06$ & $0.33$ & $0.29$\\  
 \hline 
 \hline
\end{tabular}
\end{table}

In Table~\ref{tab:1}, we list the present results of the charge radii
of the baryon decuplet in comparison with those from various
theoretical approaches. The present results are in general larger than
those of lattice data, which was already expected when we compared 
the results of the $E0$ form factors with those of the lattice QCD 
previously. The results of the form factors from the lattice
calculation fall off more slowly than the present ones. The
covariant versions of chiral perturbation
theory~\cite{Geng:2009ys,Li:2016ezv} yield systematically smaller
values of the charge radii of the baryon decuplet, compared with the
present ones. On the other hand, the large $N_c$ expansions give
the results similar to the present ones~\cite{Buchmann:2002et,
  Flores-Mendieta:2015wir}. Wagner et al.~\cite{Wagner:2000ii} computed
the various observables of the baryon decuplet based on a potential
model including chiral residual interactions in addition to the
confining potentials. Interestingly, the results of
Ref.~\cite{Wagner:2000ii} are rather similar to the present ones.
Berger et al. used the Goldstone-boson-exchange (GBE) constituent
quark model (CQM). The GBE CQM is distinguished from that of
Ref.~\cite{Wagner:2000ii}. The GBE CQM is a relativistic one though it
also contains the chiral interactions for the Nambu-Goldstone
bosons. The results are quite underestimated in comparison with the
present ones but are comparable to those of
Refs.~\cite{Geng:2009ys,Li:2016ezv}.  

Comparing the results of the second row and those of the third one, we
see that the effects of flavor SU(3) symmetry breaking are marginal on
the electric monopole form factors of the baryon decuplet, as mentioned
already. Note that the charge radii of the neutral decuplet baryons
vanish when $m_s$ is set equal to zero, since the SU(3) symmetric parts
of the expression for the electric monopole form factors are proportional to
the corresponding charges. Thus, the main contributions arise from the
linear $m_s$ corrections. As shown explicitly in
Eq.~\eqref{eq:E0lmscorr}, there are two different $m_s$ contributions,
i.e. those from the collective operators and those from the
wavefunction corrections. Interestingly, the charge densities of these
two different contributions are oppositely polarized. That is, the
contributions from the wavefunction corrections provide positive
charge densities whereas the collective operator parts yield 
negative ones. It indicates that the charge radii of the neutral
decuplet baryons should be even smaller than that of the neutron. 

\setlength{\tabcolsep}{2pt}
\renewcommand{\arraystretch}{1.5}
\begin{table}
\caption{Magnetic dipole moments of the baryon decuplet in
  comparison with the results from lattice
QCD~\cite{Leinweber:1992hy,Lee:2005ds,Aubin:2008qp,Boinepalli:2009sq},  
the relativistic quark model~\cite{Schlumpf:1993rm},
 next-to-leading-order HB$\chi$PT~\cite{Butler:1993ej}, 
large $N_{c}$~\cite{Luty:1994ub}, QCD sum rules~\cite{Lee:1997jk},
the chiral quark model~\cite{Wagner:2000ii},
covariant $\chi$PT~\cite{Geng:2009ys}, $\chi$PT~\cite{Li:2016ezv} 
and the experimental
data~\cite{Wallace:1995pf,LopezCastro:2000cv,Kotulla:2002cg}.} 
\label{tab:2}
 \begin{tabular}{c | l l l l l l l l l l } 
  \hline 
    \hline 
 $ \mu_{B}  $& $\Delta^{++}$  &  $\Delta^{+}$ & $\Delta^{0}$ &
 $\Delta^{-}$ &  $\Sigma^{*+}$ &  $\Sigma^{*0}$ &  $\Sigma^{*-}$ &
 $\Xi^{*0}$ &  $\Xi^{*-}$ &  $\Omega^{-}$ \\
  \hline 
$m_{s} = 180$ &$3.65 $ & $ 1.72 $  &  $-0.21$ & $ -2.14$
& $1.91$  &  $-0.04$ & $-1.99$ & $0.13$ & $-1.84$ & $-1.69$\\  
$m_{s} = 0$ &$3.63$ & $1.82$  &  $0$ & $-1.82$
& $1.82$  &  $0$ & $-1.82$ &  $0$ & $-1.82$ & $-1.82$\\ 
With scaling &$4.96$ & $2.34$  &  $-0.29$ & $-2.91$
& $2.59$  &  $-0.06$ & $-2.70$ &  $0.17$ & $-2.50$ & $-2.30$\\ 
\cite{Wallace:1995pf,LopezCastro:2000cv,Kotulla:2002cg} & $6.14(51)$ 
& $2.7^{+1.0}_{-1.3}\pm1.5\pm3$&  $-$ & $-$
& $-$  &  $-$ & $-$ &  $-$ & $-$ & $-2.02(5)$\\
 \hline 
LQCD~\cite{Leinweber:1992hy} &$6.09(88)$ & $3.05 (44) $  &  
$0.00$ & $-3.05(44) $ & $3.16(40) $  &  $0.329(67) $ & 
$-2.50(29) $ &  $0.58(10)$ & $-2.08(24) $ & $-1.73(22) $\\  
LQCD~\cite{Lee:2005ds} &$5.24(18)$ & $0.97(8)$  &  $-0.04(0)$ & $-2.98(19)$
& $1.27(6)$  &  $0.33(5)$ & $-1.88(4)$ &  $0.16(4)$ & $-0.62(1)$ & $-$\\ 
LQCD~\cite{Boinepalli:2009sq} &$3.20(56)$ & $1.60(28)$  &  $0$ & $-1.60(28)$
& $1.76(18)$  &  $0.00(4)$ & $-1.75(13)$ &  $0.08(5)$ 
& $-1.76(8)$ & $-1.70(7)$\\ 
LQCD~\cite{Aubin:2008qp} &$3.70(12)$ & $2.40(6)$  &  $-$ & $-1.85(6)$
& $-$  &  $-$ & $-$ &  $-$ & $-$ & $-1.93(8)$\\ 
\hline

HB$\chi$PT~\cite{Butler:1993ej} &$4.0(4)$ & $2.1(2)$  
&  $-0.17(4)$ & $-2.25(25)$
& $2.0(2)$  &  $-0.07(2)$ & $-2.2(2)$ &  $0.10(4)$ 
& $-2.0(2)$ & $-1.94(22)$\\
$\chi$PT~\cite{Geng:2009ys} &$6.04(13)$ & $2.84(2)$  
&  $-0.36(9)$ & $-3.56(20)$
& $3.07(12)$  &  $0$ & $-3.07(12)$ &  $0.36(9)$ & $-2.56(6)$ & $-2.02$\\
$\chi$PT~\cite{Li:2016ezv} &$4.97(89)$ & $2.60(50)$  
&  $0.02(12)$ & $-2.48(32)$
& $1.76(38)$  &  $-0.02(3)$ & $-1.85(38)$ &  $-0.42(13)$ 
& $-1.90(47)$ & $-2.02(5)$\\ 
$1/N_{c}$~\cite{Luty:1994ub} &$5.9(4)$ & $2.9(2)$  &  $-$ & $-2.9(2)$
& $3.3(2)$  &  $0.3(1)$ & $-2.8(3)$ &  $0.65(20)$ & $-2.30(15)$ & $-1.94$\\ 
RQM~\cite{Schlumpf:1993rm} &$4.76$ & $2.38$  &  $0$ & $-2.38$
& $1.82$  &  $-0.27$ & $-2.36$ &  $-0.60$ & $-2.41$ & $-2.35$\\ 
$\chi$QM~\cite{Wagner:2000ii} &$6.93$ & $3.47$  &  $0$ & $-3.47$
& $4.12$  &  $0.53$ & $-3.06$ &  $1.10$ & $-2.61$ & $-2.13$\\ 
QCD-SR~\cite{Lee:1997jk} &$4.13(1.30)$ & $2.07(65)$  
& $0$ &  $-2.07(65)$ & $2.13(82)$
& $-0.32(15)$  &  $-1.66(73)$ & $-0.69(29)$ &  $-1.51(52)$ & $-1.49(45)$ \\ 
 \hline 
 \hline
\end{tabular}
\end{table}
Table~\ref{tab:2} lists the numerical results of the present
model. Before we discuss them, we want to mention two important
points related to the magnetic dipole moments within the
$\chi$QSM. Firstly, it is well known that the results of the magnetic
dipole moments of the baryon octet from any chiral solitonic
approaches are underestimated in comparison with the experimental
data. In order to improve them, Ref.~\cite{Ledwig:2008es} used the
nuclear magneton in terms of the classical soliton mass instead of the
physical nucleon mass. Thus, the results of the magnetic dipole moments were
obtained by an additional mass factor 1.36. With this factor included,
Ref.~\cite{Ledwig:2008es} reproduced nicely the experimental data on
the magnetic dipole moment of the $\Omega^-$ baryon. If we introduce this
additional factor in the present work, of course we are able to
reproduce almost the same results as in Ref.~\cite{Ledwig:2008es},
which we list in the fourth row of Table~\ref{tab:2}~\footnote{The
  differences of the numerical 
  values from those of   Ref.~\cite{Ledwig:2008es} come from the
  different values of the current quark mass.}. However, we present
also the results without this mass factor, since we want to
concentrate on the physics of the $\chi$QSM as such. Secondly, the
$\chi$QSM provides the SU(3) symmetric expressions of the magnetic
dipole moments, which are proportional to the corresponding charges of
the baryon decuplet, as shown in Eq.~\eqref{eq:M1leadingcon}. Thus,
all the magnetic moments of the neutral decuplet baryons vanish as
listed in the third row of Table~\ref{tab:2}, when the $m_s$
corrections turned off. In general, the effects of the $m_s$
corrections are again small on those of the 
positively charged decuplet baryons. The results are also compared
with those from various different theoretical approaches. 

\begin{table}
\caption{Magnetic radii of the baryon decuplet in comparison with
  those of the chiral quark model~\cite{Wagner:2000ii}, 
  $\chi$PT~\cite{Li:2016ezv}.}  
\label{tab:3}
 \begin{tabular}{ c | c c c c c c c c c c } 
  \hline 
    \hline 
$\langle r^{2}\rangle^{B}_{M}  $& $\Delta^{++}$  &  $\Delta^{+}$ &
$\Delta^{0}$ & $\Delta^{-}$  &  $\Sigma^{*+}$ &  $\Sigma^{*0}$ &  
$\Sigma^{*-}$ &  $\Xi^{*0}$ &  $\Xi^{*-}$ &  $\Omega^{-}$ \\ 
  \hline 
This work ($m_{s} = 180$)&$0.587 $ & $ 0.513 $  &  $1.786$ & $0.764$
& $0.599$  &  $3.356$ & $0.713$ & $0.784$ & $0.653$ & $0.582$\\  
This work ($m_{s} = 0$)&$0.582$ & $0.582$  &  $0.582$ & $0.582$
& $0.582$  & $0.582$ & $0.582$ &  $0.582$ & $0.582$ & $0.582$\\  
 \hline 
$\chi$PT~\cite{Li:2016ezv} &$0.61(15)$ & $0.64(14)$  & $0.07(12)$ &
 $0.55(19)$ & $0.59(16)$  &  $0$ & $0.59(16)$ &  $-0.07(12)$ &
 $0.64(14)$ & $0.70(12)$\\   
$\chi$QM~\cite{Wagner:2000ii}&$0.62$ & $0.62$  &  $0$ & $0.62$
& $0.67$  &  $0.82$ & $0.61$ &  $0.82$ & $0.58$ & $0.53$\\  
 \hline 
 \hline
\end{tabular}
\end{table}
In Table~\ref{tab:3}, we present the results of the magnetic radii of
the baryon decuplet. The magnetic radius of a baryon is defined as the
derivative of its magnetic dipole form factor 
\begin{align}
\langle r^2 \rangle_{M1}^B = \left. -\frac{6}{G_{M1}^B(0)}\frac{d
  G_{M1}^{B}(Q^2)}{dQ^2}\right |_{Q^2=0}.   
\label{eq:mag_radii}
\end{align} 
Thus, the SU(3) symmetric results of the baryon decuplet are all the
same. It means that the results of the magnetic radii exhibit
effectively the effects of the flavor SU(3) 
symmetry breaking. The $\Delta^-$ magnetic radius shows the strongest
enhancement due to the $m_s$ corrections except for the case of the
neutral baryons. However, the effects of the flavor SU(3) symmetry
breaking are in general small on the magnetic radii of the positively
charged decuplet baryons. Compared to the results of chiral
perturbation theory~\cite{Li:2016ezv} and the chiral quark-potential
model~\cite{Wagner:2000ii}, the present ones of the positively charged
baryons are in good agreement with them, whereas those of the neutral
baryons are rather different from each other. Note that the neutral
baryons have the same magnetic radii as those  
of the charged baryons, since the denominator and numerator SU(3)
$D$-functions are explicitly canceled out (see
Eq.~\eqref{eq:mag_radii}).

\setlength{\tabcolsep}{5pt}
\renewcommand{\arraystretch}{1.5}

\begin{table}
\caption{Electric quadrupole moments of the baryon decuplet in
  comparison with the quark model~\cite{Krivoruchenko:1991pm},
  HB$\chi$PT~\cite{Butler:1993ej}, the Skyrme model~\cite{Oh:1995hn},
  large $N_{c}$~\cite{Buchmann:2002et}, the chiral quark
  model~\cite{Wagner:2000ii}, the QCD sum
  rules~\cite{Azizi:2008tx,Aliev:2009pd}.} 
\label{tab:4}
 \begin{tabular}{ c | c c c c c c c c c c } 
  \hline 
    \hline 
 $Q_{B}$ & $\Delta^{++}$  &  $\Delta^{+}$ & $\Delta^{0}$ &
 $\Delta^{-}$  &  $\Sigma^{+}$ &  $\Sigma^{*0}$ &  $\Sigma^{*-}$ 
&  $\Xi^{*0}$ &  $\Xi^{*-}$ &  $\Omega^{-}$ \\
  \hline 
$m_{s} = 180$&$-0.102 $ & $-0.039$  &  $0.023$ & $0.085$
& $-0.070$  &  $0.003$ & $0.077$ & $-0.016$ & $0.069$ & $0.061$\\  
$m_{s} = 0$&$-0.109$ & $-0.054$  &  $0$ & $0.054$
& $-0.054$  &  $0$ & $0.054$ &  $0$ & $0.054$ & $0.054$\\  
\hline
HB$\chi$PT~\cite{Butler:1993ej}&$-0.08(5)$ & $-0.03(2)$  &  
$0.012(5)$ & $0.06(3)$
& $-0.07(3)$  &  $-0.013(7)$ & $0.04(2)$ &  $-0.035(2)$ & 
$0.02(1)$ & $0.009(5)$\\ 
1/$N_{c}$~\cite{Buchmann:2002et} &$-0.120$ & $-0.060$  &  $0$ & $0.060$
& $-0.069$  &  $0.014$ & $0.077$ &  $-0.023$ & $0.047$ & $0.027$\\ 
NQM~\cite{Krivoruchenko:1991pm}&$-0.093$ & $-0.046$  &  $0$ & $0.046$
& $-0.054$  &  $-0.007$ & $0.040$ &  $-0.013$ & $0.034$ & $0.028$\\  
Skyrme~\cite{Oh:1995hn} &$-0.088$ & $-0.029$  &  $0.029$ & $0.088$
& $-0.071$  &  $0$ & $0.071$ &  $-0.046$ & $0.046$ & $0$\\  
$\chi$QM~\cite{Wagner:2000ii}&$-0.252$ & $-0.126$  &  $0$ & $0.126$
& $-0.123$  &  $-0.021$ & $0.082$ &  $-0.030$ & $0.048$ & $0.026$\\  
QCD-SR~\cite{Azizi:2008tx,Aliev:2009pd} &$-0.028(8)$ & $-0.014(4)$  &
 $0$ & $0.014(4)$ & $-0.028(9)$  &  $0.0012(4)$ & $0.03(1)$ &  
$0.0025(8)$ & $0.045(15)$ & $0.12(4)$\\ 
 \hline 
 \hline
\end{tabular}
\end{table}
Table~\ref{tab:4} lists the results of the electric quadrupole moments
in comparison with those from other theoretical works. Since $Q_B$ reveals
how much the corresponding baryon is distorted from a spherical charge
distribution, it is of great interest to examine it carefully. As we
have already mentioned in the previous subsection, when $Q_B$ is
negative the corresponding baryon takes an oblate shape (cushion shape),
while when $Q_B>0$ it does a prolate one (rugby-ball shape). As shown
in Table~\ref{tab:4}, the positively charged baryons all have negative 
values of $Q_B$, whereas the negatively charged ones get the positive
values. This conclusion is shared by all other theoretical
predictions. However, when it comes to the case of the neutral
decuplet baryons, $Q_B$ are small but positive except for the
$\Xi^{*0}$. Note that heavy-baryon chiral perturbation
theory~\cite{Butler:1993ej}, the nonrelativistic quark model
(NQM)~\cite{Krivoruchenko:1991pm}, and the chiral quark-potential
model~\cite{Wagner:2000ii} give the negative values of $Q_{\Sigma^{*0}}$.      

\subsection{Chiral extrapolation and comparison with the lattice
  data\label{subsec:IV-D}}  
To compare the present results of the EM form factors of the
$\Delta^+$ and $\Omega^-$ baryons with those from lattice QCD, we made
a chiral extrapolation as done in Ref.~\cite{Goeke:2007fq}. We first
derived the profile functions of the soliton, which correspond to the
values of the \textit{unphysical} pion mass: $m_\pi=297$ MeV, 311 MeV,
353 MeV and 384 MeV. Using them, we computed the EM form factors of
the $\Delta^+$ and $\Omega^-$. 
  \begin{figure}[htp]
  \includegraphics[scale=0.20]{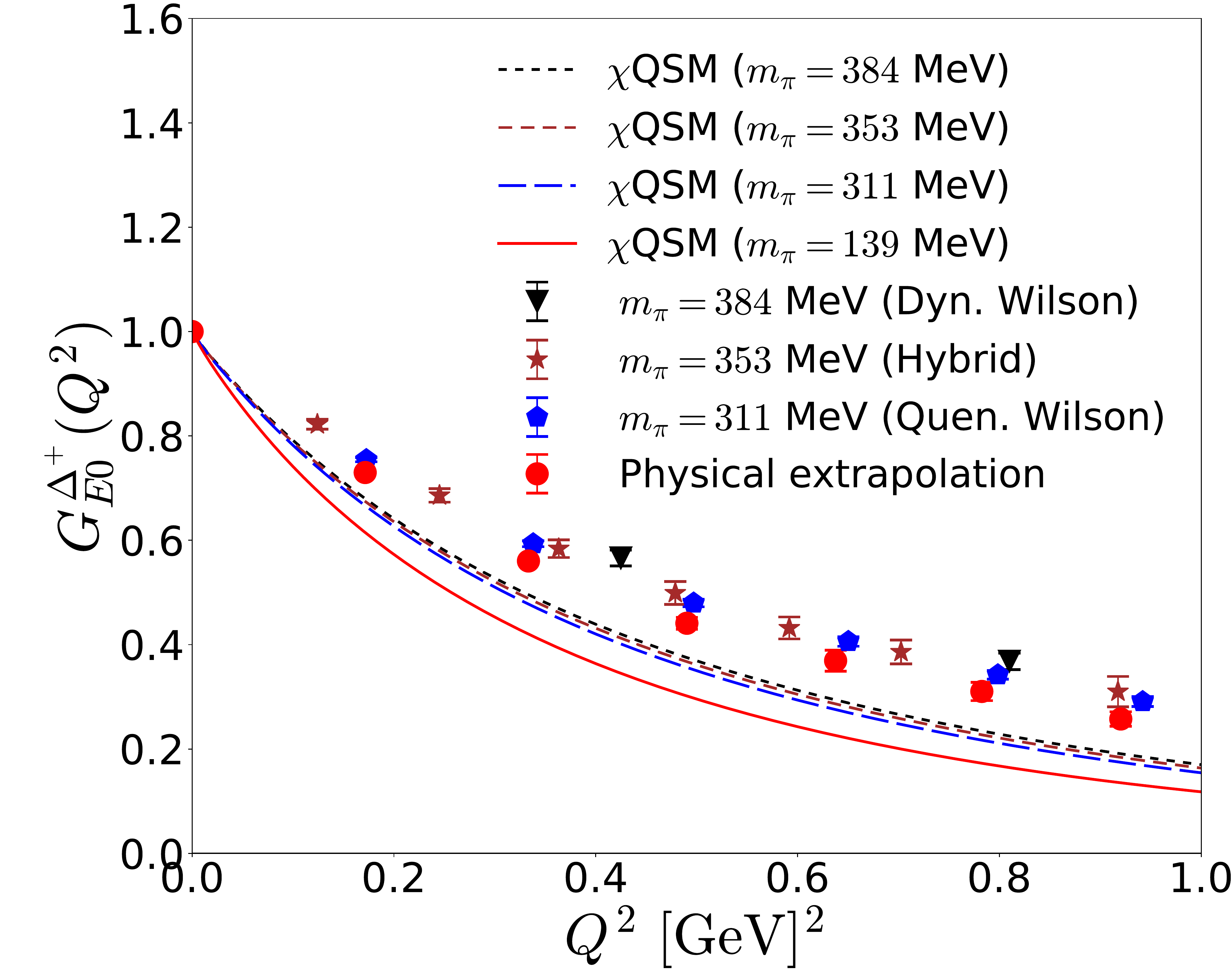}\hspace{0.5cm}
  \includegraphics[scale=0.20]{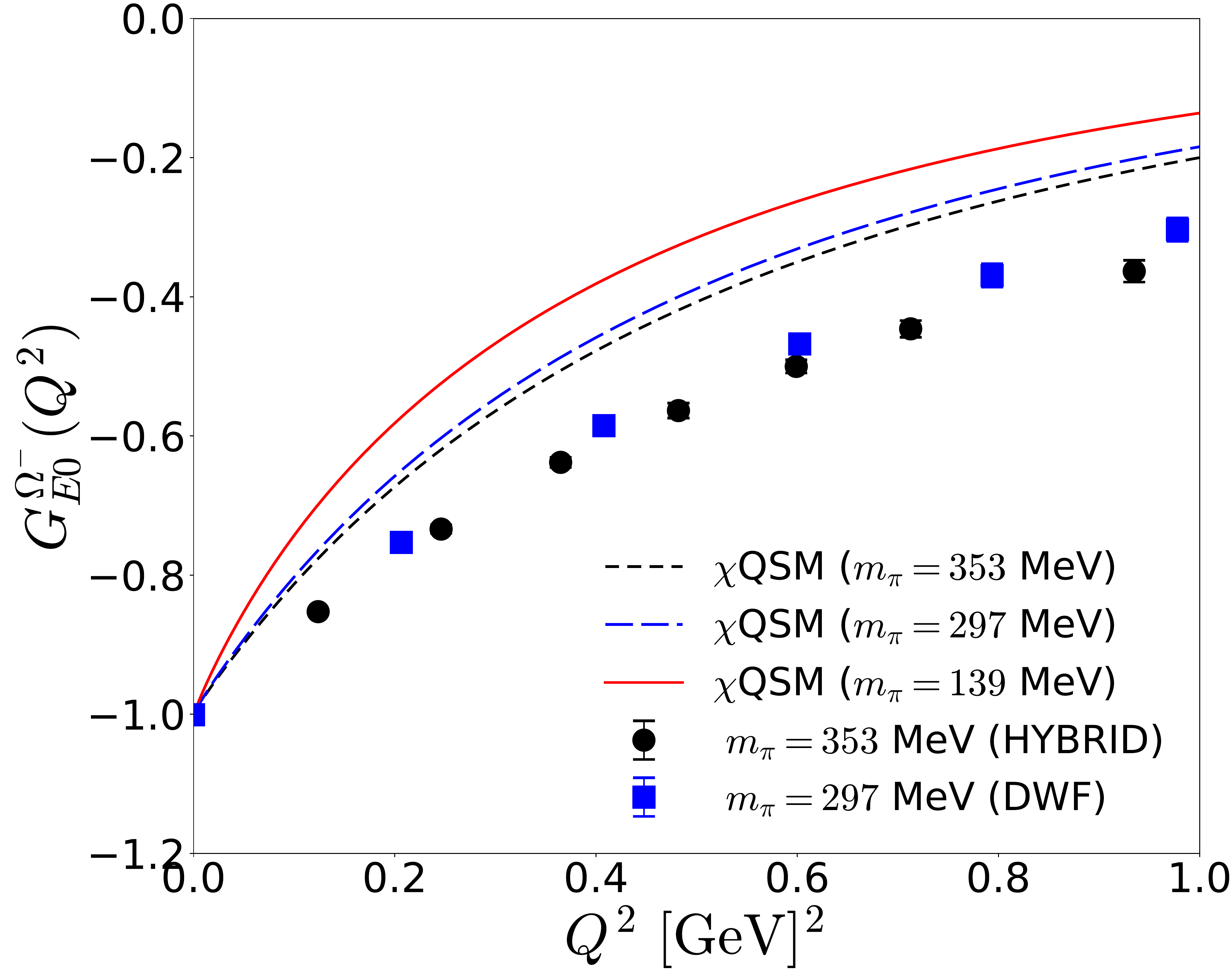}
  \includegraphics[scale=0.20]{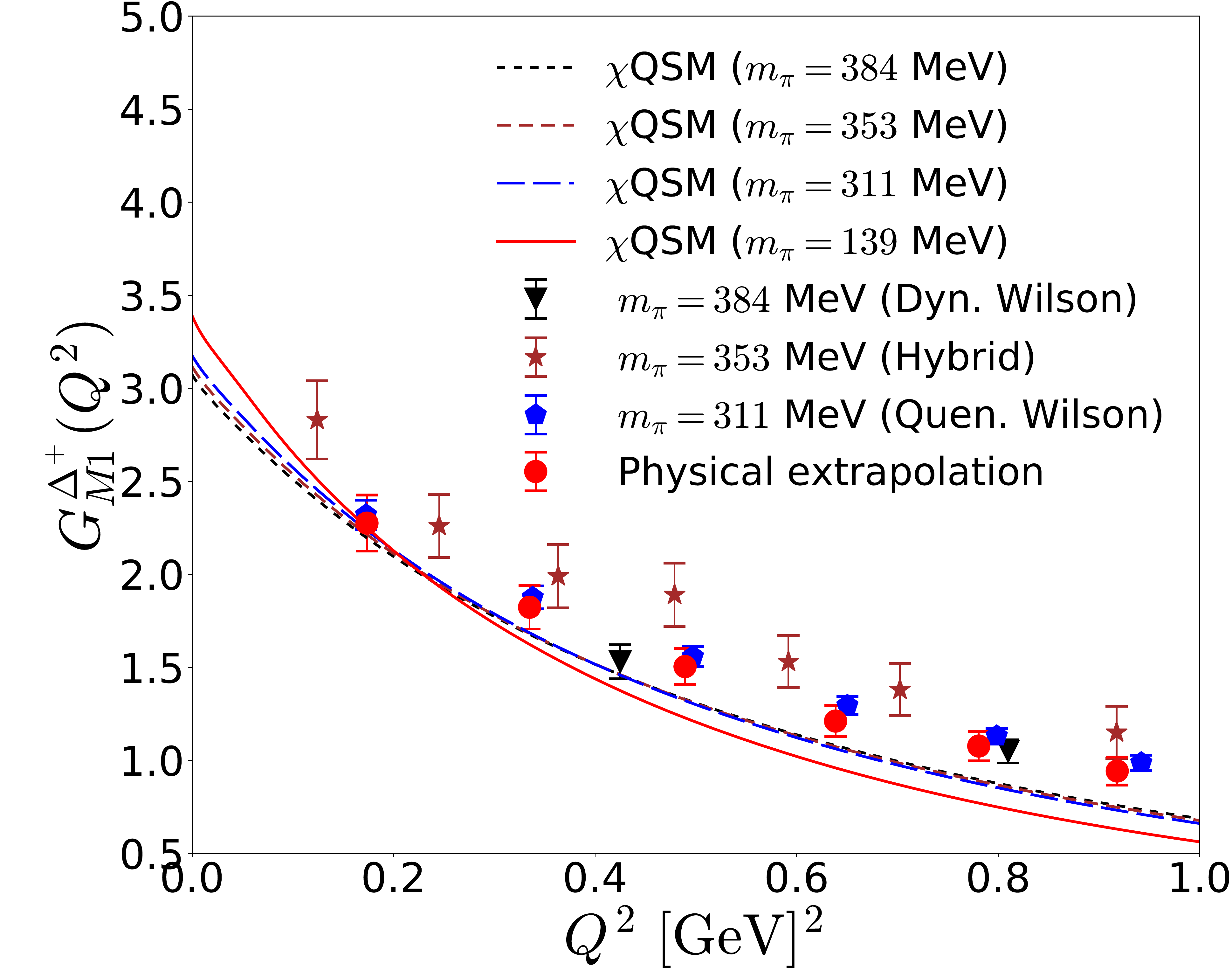}\hspace{0.5cm}
  \includegraphics[scale=0.20]{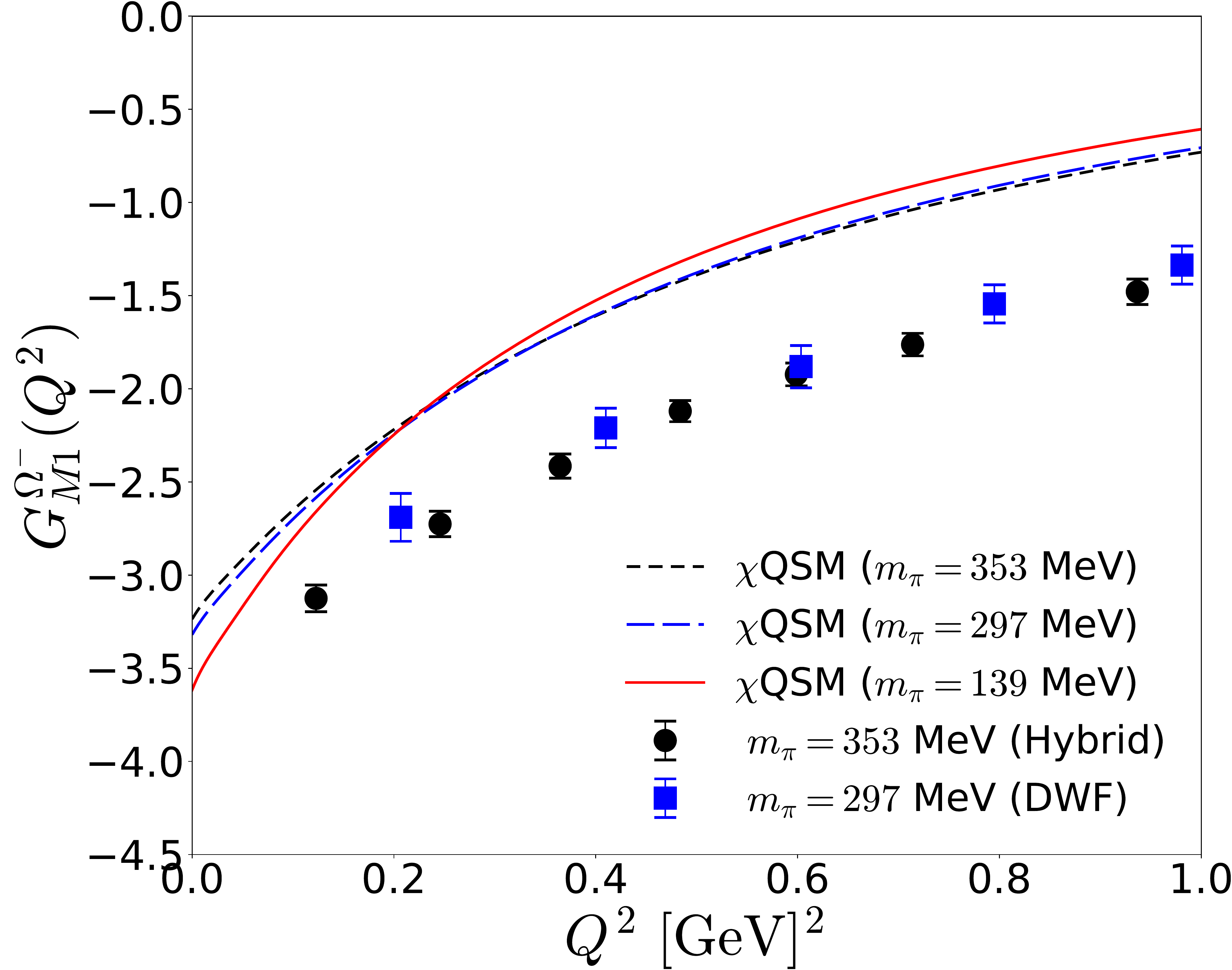}
  \includegraphics[scale=0.20]{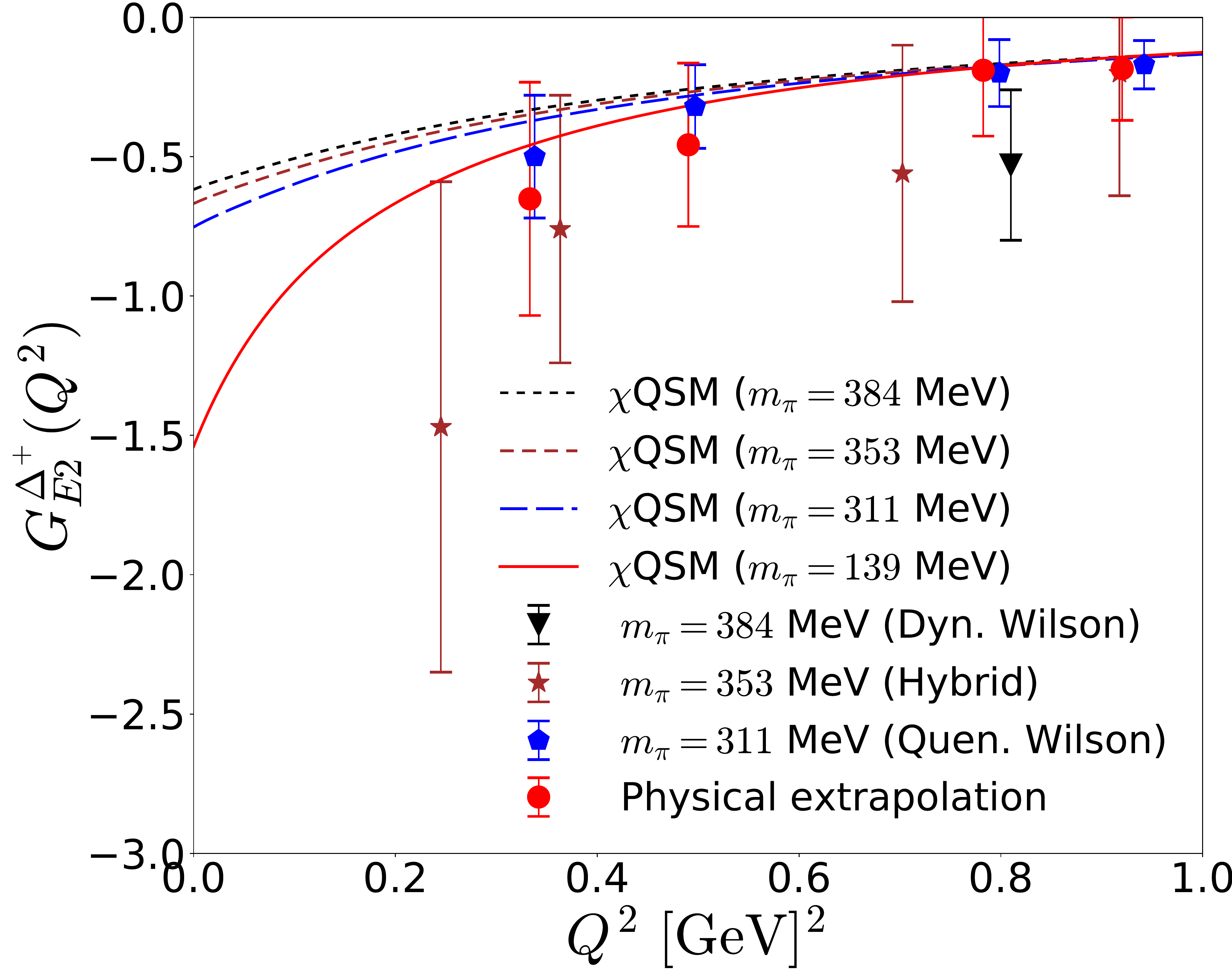}\hspace{0.5cm}
  \includegraphics[scale=0.20]{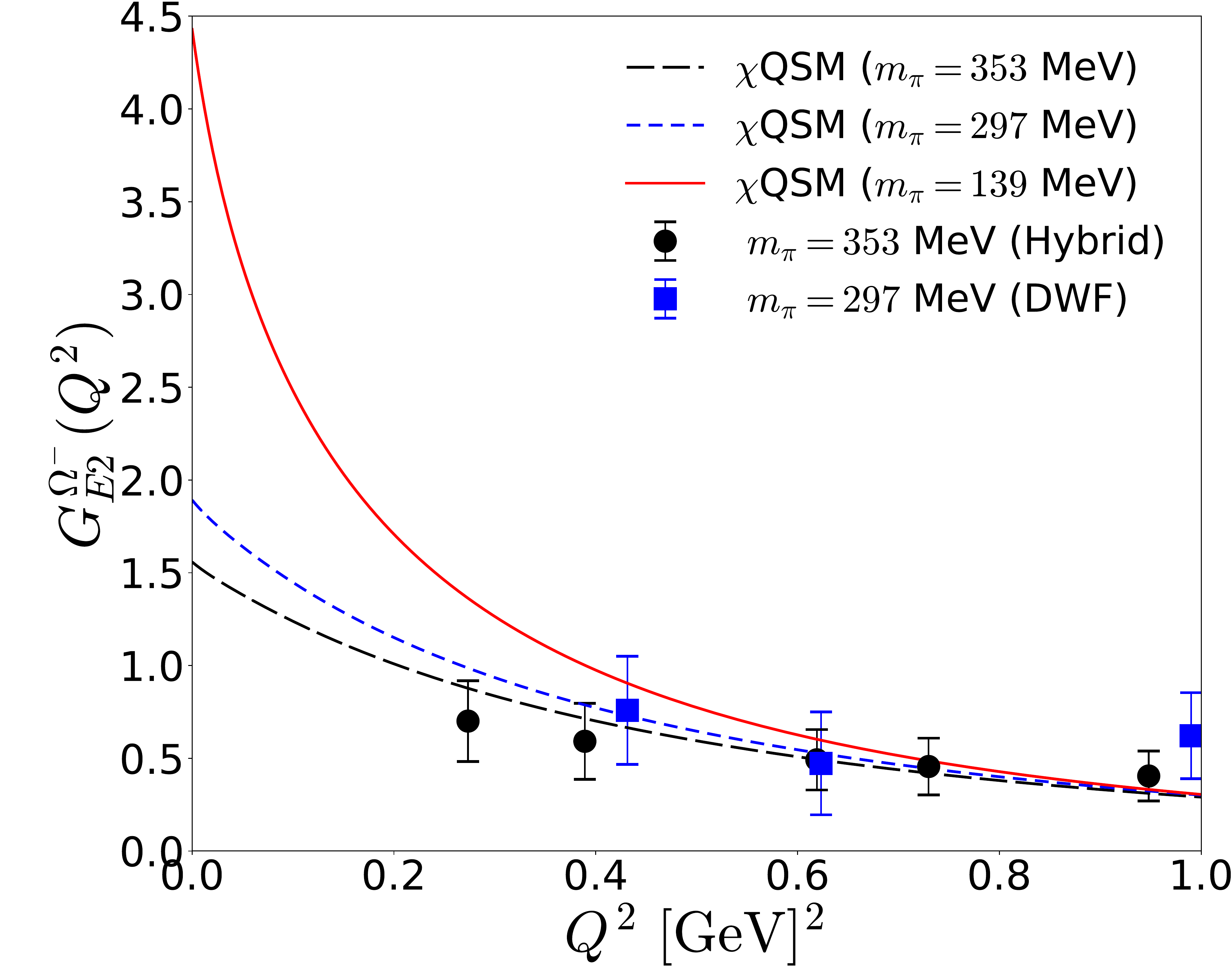}
  \caption{Numerical results of the EM form factors of the
    $\Delta^{+}$ and $\Omega^{-}$ baryons with the different pion
    masses in the calculations of lattice
    QCD~~\cite{Alexandrou:2007we, Alexandrou:2009hs,
      Alexandrou:2010jv} employed.}  
\label{fig:13}
\end{figure}

The results are drawn in Fig~\ref{fig:13}. In general, the larger 
values of the pion mass yield the EM form factors that fall off more
slowly than those with the physical pion mass ($m_\pi=139$
MeV). In the case of the electric form factors, the present results
get closer to the lattice ones with the value of $m_\pi$ increased. To
compare the present results with the lattice data more closely, we
have scaled the $M1$ form factors by introducing the corresponding
numerical results of the $\Delta^+$ and $\Omega^-$ magnetic moments
taken from a model-independent approach~\cite{Yang:2013qza}, which
reproduces the experimental data very well. Then it is clearly seen
that the $Q^2$ dependences of the $M1$ form factors become very
similar to the lattice data. When it comes to the $E2$ form factors,
the results are remarkable. The numerical results of the $E2$ form
factors are very sensitive to the values of the pion mass. In
particular, the present results of the $\Omega^-$ $E2$ form factors
are in good agreement with the lattice data when higher values of the
pion mass are used.

\section{Summary and conclusion}
\label{sec:5}
In the present work, we investigated the electromagnetic form factors
of the baryon decuplet together with other related observables, based
on the self-consistent SU(3) chiral quark-soliton model. We considered
the $1/N_c$ rotational corrections and the linear $m_s$ corrections
with the symmetry-conserving zero-mode quantization used. We first
computed the electric monopole form factors of the $\Delta^+$ and
$\Omega^-$ and compared the results with those from recent
calculations of lattice QCD. Taking into account the fact that any
results of hadronic form factors from lattice QCD fall off more slowly
than the experimental data, the results are comparable with the
lattice data. We presented the results of the electric monopole form
factors for all other members of the baryon decuplet. The
contributions of the sea-quark polarization are stronger to the
negatively charged decuplet baryons than to the positively charged
ones. However, the valence quark contributions are almost canceled by
the sea-quark ones. At $Q^2=0$, they are exactly canceled each other
as they should be. The results of the magnetic dipole form factors
show very similar $Q^2$ dependences in comparison with the lattice
data. While the sea-quark contributions are smaller than the valence
ones to the $M1$ form factors of both the positively and
negatively charged baryons, they are dominant over the valence-quark
contributions in the case of the neutral baryons except for the
$\Xi^{*0}$. We also computed the electric quadrupole form factors of
the $\Delta^+$ and $\Omega^-$ and the results are in agreement with
the lattice ones. The sea-quark contributions govern the form factor
in smaller $Q^2$ regions and then yield to the valence-quark 
contributions as $Q^2$ increases. This tendency is general also for
the $E2$ form factors of all other decuplet baryons. 
In the present framework, the magnetic octupole form factors vanish
because of the hedgehog symmetry. It implies that the $M3$ form
factors must be very small. 

We also examined the effects of flavor SU(3) symmetry breaking. In 
fact, there are two different sources of the linear $m_s$ corrections:
one from the effective chiral action and the other from the
wavefunction corrections that arise from the mixing of the baryon
decuplet wavefunctions with higher representations. Since all the form
factors of the baryon decuplet are proportional to the corresponding
charges when the $m_s$ corrections are switched off, the effects of
the flavor SU(3) symmetry breaking are in particular important on the
form factors of the neutral decuplet baryons. In fact, the linear
$m_s$ corrections are the leading-order contributions to them. 
While the $m_s$ corrections are marginal to most of the form factors,
their effects are nonnegligible on the electric quadrupole form
factors. We also presented the charge radii, the magnetic radii, the
magnetic dipole moments and the electric quadrupole moments of the
baryon decuplet, discussed them in comparison with those from other
theoretical works.  

Finally, we used the values of the unphysical pion mass to compare the
present results with the lattice data in a more quantitative
manner. Indeed, larger values of the pion mass bring the present
results closer to those of lattice QCD. In particular, the $Q^2$
dependence of the $\Delta^+$ and $\Omega^-$ $M1$ form factors are in
good agreement with the lattice data. We also showed that the $E2$
form factors are rather sensitive to the values of the pion mass. As a
result, the present result of the $\Omega^-$ $E2$ form factor is in
quantitative agreement with the lattice data, when larger values of
the pion mass are employed. 

It is also of great importance to study the electromagnetic transition
form factors of the baryon decuplet to the baryon octet. Though there
are previous works on the $\Delta\to N\gamma$ transitions within the
same theoretical framework, Other decay channels were not
investigated. However, it is indeed important to scrutinize the
radiative decays such as $\Sigma^*\to \Sigma \gamma$ and $\Xi^*\to \Xi
\gamma$, since they provide information not only on radiative decays
but also on the vector-meson couplings to the baryon octet and
decuplet. The corresponding investigation is under way. 
\begin{acknowledgments}
We are grateful to Gh.-S. Yang for valuable discussions.
The present work was supported by the Inha University Grant in 2018.  
\end{acknowledgments}

\appendix

\section{Matrix elements of the SU(3) Wigner ${D}$
  function \label{app:A}} 
In the following we list the results of the matrix elements of the
relevant collective operators for the EM form factors of the baryon
decuplet in Table~\ref{tab:E0_Leading}, \ref{tab:E0_Op},
\ref{tab:E0_WF27}, \ref{tab:E0_WF35}~\ref{tab:M1_Leading},
\ref{tab:M1_Op}, \ref{tab:M1_WF27}, \ref{tab:M1_WF35}. The
  collective operators for E2 form factors can be found in those for
  E0 and M1.  
   
   \begin{table}[htp]
  \caption{The matrix elements of the collective operators of the
    leading terms and the $1/N_c$ rotational corrections to the
    electric monopole form factors.}
  \label{tab:E0_Leading}
\begin{center}
\begin{tabular}{ c } 
 \hline 
  \hline  
$\langle \Delta |D^{(8)}_{88} | \Delta \rangle= \langle
  \Sigma^{*} |D^{(8)}_{88} | \Sigma^{*} \rangle = \langle
  \Xi^{*} |D^{(8)}_{88} | \Xi^{*} \rangle= \langle
  \Omega |D^{(8)}_{88} | \Omega \rangle =\frac{1}{8} Y $   \\ 
$\langle \Delta |D^{(8)}_{38} | \Delta \rangle= \langle
  \Sigma^{*} |D^{(8)}_{38} | \Sigma^{*} \rangle = \langle
  \Xi^{*} |D^{(8)}_{38} | \Xi^{*} \rangle= \langle
  \Omega |D^{(8)}_{38} | \Omega \rangle =\frac{\sqrt{3}}{12} T_{3}$ 
  \\ 
$\langle \Delta |D^{(8)}_{8i}J_{i} | \Delta \rangle= \langle
  \Sigma^{*} |D^{(8)}_{8i}J_{i} | \Sigma^{*} \rangle = \langle
  \Xi^{*} |D^{(8)}_{8i}J_{i} | \Xi^{*} \rangle= \langle
  \Omega |D^{(8)}_{8i}J_{i} | \Omega \rangle =-\frac{5\sqrt{3}}{16} Y$   \\ 
$\langle \Delta |D^{(8)}_{3i}J_{i} | \Delta \rangle= \langle
  \Sigma^{*} |D^{(8)}_{3i}J_{i} | \Sigma^{*} \rangle = \langle
  \Xi^{*} |D^{(8)}_{3i}J_{i} | \Xi^{*} \rangle= \langle
  \Omega |D^{(8)}_{3i}J_{i} | \Omega \rangle =-\frac{5}{8} T_{3}$   \\ 
$\langle \Delta |D^{(8)}_{8a}J_{a} | \Delta \rangle= \langle
  \Sigma^{*} |D^{(8)}_{8a}J_{a} | \Sigma^{*} \rangle = \langle
  \Xi^{*} |D^{(8)}_{8a}J_{a} | \Xi^{*} \rangle= \langle
  \Omega |D^{(8)}_{8a}J_{a} | \Omega \rangle =-\frac{\sqrt{3}}{8} Y $   \\ 
$\langle \Delta |D^{(8)}_{3a}J_{a} | \Delta \rangle= \langle
  \Sigma^{*} |D^{(8)}_{3a}J_{a} | \Sigma^{*} \rangle = \langle
  \Xi^{*} |D^{(8)}_{3a}J_{a} | \Xi^{*} \rangle= \langle
  \Omega |D^{(8)}_{3a}J_{a} | \Omega \rangle =-\frac{1}{4} T_{3} $   \\ 
 \hline 
 \hline
\end{tabular}
\end{center}
\end{table}

\begin{table}[htp]
  \caption{The matrix elements of the collective operators of the
    $m_s$ corrections to the electric monopole form factors.} 
  \label{tab:E0_Op}
\begin{center}
\begin{tabular}{ c | c c c c } 
 \hline 
  \hline 
 $ {\cal{R}}$ & \multicolumn{4}{c}{$\bm{{10}} \ (J_{3}=3/2)$}  \\  
B & $\Delta$ & $\Sigma^{*}$ & $\Xi^{*}$ & $\Omega$ \\  
 \hline
$\langle B_{{\cal{R}}} |D^{(8)}_{8i}D^{(8)}_{3i} | B_{{\cal{R}}}
  \rangle$  & $\frac{13\sqrt{3}}{252}T_{3}$&
 $\frac{\sqrt{3}}{21}T_{3}$
& $\frac{11\sqrt{3}}{252}T_{3}$ & $0$ \\  
$\langle B_{{\cal{R}}} |D^{(8)}_{8i}D^{(8)}_{8i} | B_{{\cal{R}}}
  \rangle$  & $\frac{17}{56}$ & $\frac{31}{84}$ & $\frac{25}{56}$ 
& $\frac{15}{28}$ \\  
$\langle B_{{\cal{R}}} |D^{(8)}_{8a}D^{(8)}_{3a} | B_{{\cal{R}}}
  \rangle$  & $-\frac{5\sqrt{3}}{126}T_{{3}}$& 
$-\frac{\sqrt{3}}{42}T_{{3}}$& $-\frac{\sqrt{3}}{126}T_{{3}}$ & $0$  \\   
$\langle B_{{\cal{R}}} |D^{(8)}_{8a}D^{(8)}_{8a} | B_{{\cal{R}}}
  \rangle$  & $\frac{15}{28}$ &$\frac{11}{21}$ &$\frac{13}{28}$ 
&$\frac{5}{14}$   \\  
$\langle B_{{\cal{R}}} |D^{(8)}_{88}D^{(8)}_{38} | B_{{\cal{R}}}
  \rangle$  & $-\frac{\sqrt{3}}{84}T_{3}$ &
 $-\frac{\sqrt{3}}{42}T_{3}$
& $-\frac{\sqrt{3}}{28}T_{3}$ & $0$  \\  
$\langle B_{{\cal{R}}} |D^{(8)}_{88}D^{(8)}_{88}| B_{{\cal{R}}}
  \rangle$  & $\frac{9}{56}$ & $\frac{3}{28}$ & $\frac{5}{56}$ 
& $\frac{3}{28}$  \\  
 \hline 
 \hline
\end{tabular}
\end{center}
\end{table}

\begin{table}[htp]
  \caption{The relevant transition matrix elements of the collective
    operators coming from the 27plet component of the baryon
    wavefunctions for the electric monopole form factors.}   
  \label{tab:E0_WF27}
\begin{center}
\begin{tabular}{ c | c  c c c } 
 \hline 
  \hline 
 $ {\cal{R}}$& \multicolumn{4}{c}{$\bm{{10}} \ (J_{3}=3/2)$}  \\  
B & $\Delta$ & $\Sigma^{*}$ & $\Xi^{*}$ & $\Omega$ \\  
\hline
$\langle B_ {\bm{27}} |D^{(8)}_{88} | B_{\cal{R}} \rangle$
& $\frac{\sqrt{30}}{16}$ & $\frac{1}{4}$ & $\frac{\sqrt{6}}{16}$& $0$\\
$\langle B_ {\bm{27}}  |D^{(8)}_{38} | B_{\cal{R}} \rangle$
& $-\frac{\sqrt{10}}{24}T_{3}$ & $-\frac{\sqrt{3}}{8}T_{3}$
& $-\frac{7\sqrt{2}}{24}T_{3}$ & $0$ \\  
$\langle B_ {\bm{27}}  |D^{(8)}_{8i}J_{i} | B_{\cal{R}}\rangle$
& $\frac{5\sqrt{10}}{32}$  & $\frac{5\sqrt{3}}{24}$  
& $\frac{5\sqrt{2}}{32}$ & $0$ \\  
$\langle B_ {\bm{27}}  |D^{(8)}_{3i}J_{i} | B_{\cal{R}}\rangle$
& $-\frac{5\sqrt{30}}{144}T_{3}$ & $-\frac{5}{16}T_{3}$
& $-\frac{35\sqrt{6}}{144}T_{3}$ & $0$ \\
$\langle B_ {\bm{27}}  |D^{(8)}_{8a}J_{a} | B_{\cal{R}}\rangle$
& $-\frac{\sqrt{10}}{16}$ & $-\frac{\sqrt{3}}{12}$ 
& $-\frac{\sqrt{2}}{16}$ & $0$ \\   
$\langle B_ {\bm{27}}  |D^{(8)}_{3a}J_{a} | B_{\cal{R}}\rangle$  
& $\frac{\sqrt{30}}{72}T_{3}$ & $\frac{1}{8}T_{3}$ &
$\frac{7\sqrt{6}}{72}T_{3}$ &$0$   \\   
 \hline 
 \hline
\end{tabular}
\end{center}
\end{table}

\begin{table}[htp]
  \caption{The relevant transition matrix elements of the collective
    operators coming from the 35plet component of the baryon wave 
    functions for the electric monopole form factors.}  
  \label{tab:E0_WF35}
\begin{center}
\begin{tabular}{ c | c  c c c  } 
 \hline 
  \hline 
 $ {\cal{R}}$& \multicolumn{4}{c}{$\bm{10} \  (J_{3}=3/2)$}   \\  
B & $\Delta$ & $\Sigma^{*}$ & $\Xi^{*}$ & $\Omega$ \\  
 \hline
$\langle B_ {\bm{35}} |D^{(8)}_{88} | B_{\cal{R}} \rangle$
& $\frac{5\sqrt{14}}{112}$ & $\frac{\sqrt{35}}{28}$
&  $\frac{3\sqrt{70}}{112}$ &  $\frac{\sqrt{35}}{28}$  \\  
$\langle B_ {\bm{35}}  |D^{(8)}_{38} | B_{\cal{R}} \rangle$
& $\frac{\sqrt{42}}{168}T_{3}$ &  $\frac{\sqrt{105}}{168}T_{3}$ &
$\frac{\sqrt{210}}{168}T_{3}$ & $0$  \\  
$\langle B_ {\bm{35}}  |D^{(8)}_{8i}J_{i} | B_{\cal{R}}\rangle$
& $-\frac{5\sqrt{42}}{224}$ & $-\frac{\sqrt{105}}{56}$ 
& $-\frac{3\sqrt{210}}{224}$ & $-\frac{\sqrt{105}}{56}$   \\  
$\langle B_ {\bm{35}}  |D^{(8)}_{3i}J_{i} | B_{\cal{R}}\rangle$ 
& $-\frac{\sqrt{14}}{112}T_{3}$ & $-\frac{\sqrt{35}}{112}T_{3}$ 
& $-\frac{\sqrt{70}}{112}T_{3}$ &$0$ \\  
$\langle B_ {\bm{35}}  |D^{(8)}_{8a}J_{a} | B_{\cal{R}}\rangle$ 
& $\frac{5\sqrt{42}}{112}$ &$\frac{\sqrt{105}}{28}$ 
& $\frac{3\sqrt{210}}{112}$ &$\frac{\sqrt{105}}{28}$   \\   
$\langle B_ {\bm{35}}  |D^{(8)}_{3a}J_{a} | B_{\cal{R}}\rangle$ 
& $\frac{\sqrt{14}}{56}T_{3}$ & $\frac{\sqrt{35}}{56}T_{3}$ &
$\frac{\sqrt{70}}{56}T_{3}$ &$0$   \\   
 \hline 
 \hline
\end{tabular}
\end{center}
\end{table}


   \begin{table}[htp]
  \caption{The matrix elements of the collective operators of the
    leading terms and the $1/N_c$ rotational corrections to the
    magnetic dipole form factors.} 
  \label{tab:M1_Leading}
\begin{center}
\begin{tabular}{ c } 
 \hline 
  \hline  
$\langle \Delta |D^{(8)}_{33} | \Delta \rangle= \langle
  \Sigma^{*} |D^{(8)}_{33} | \Sigma^{*} \rangle = \langle
  \Xi^{*} |D^{(8)}_{33} | \Xi^{*} \rangle= \langle
  \Omega |D^{(8)}_{33} | \Omega \rangle =-\frac{1}{4} T_{3} $   \\ 
$\langle \Delta |D^{(8)}_{83} | \Delta \rangle= \langle
  \Sigma^{*} |D^{(8)}_{83} | \Sigma^{*} \rangle = \langle
  \Xi^{*} |D^{(8)}_{83} | \Xi^{*} \rangle= \langle
  \Omega |D^{(8)}_{83} | \Omega \rangle =-\frac{\sqrt{3}}{8} Y$ 
  \\ 
$\langle \Delta |D^{(8)}_{38}J_{3} | \Delta \rangle= \langle
  \Sigma^{*} |D^{(8)}_{38}J_{3} | \Sigma^{*} \rangle = \langle
  \Xi^{*} |D^{(8)}_{38}J_{3} | \Xi^{*} \rangle= \langle
  \Omega |D^{(8)}_{38}J_{3} | \Omega \rangle =\frac{\sqrt{3}}{8} T_{3}$   \\ 
$\langle \Delta |D^{(8)}_{88}J_{3} | \Delta \rangle= \langle
  \Sigma^{*} |D^{(8)}_{88}J_{3} | \Sigma^{*} \rangle = \langle
  \Xi^{*} |D^{(8)}_{88}J_{3} | \Xi^{*} \rangle= \langle
  \Omega |D^{(8)}_{88}J_{3} | \Omega \rangle =-\frac{3}{16} Y$   \\ 
$\langle \Delta |d_{ab3}D^{(8)}_{3a}J_{b} | \Delta \rangle= \langle
  \Sigma^{*} |d_{ab3}D^{(8)}_{3a}J_{b} | \Sigma^{*} \rangle = \langle
  \Xi^{*} |d_{ab3}D^{(8)}_{3a}J_{b} | \Xi^{*} \rangle= \langle
  \Omega |d_{ab3}D^{(8)}_{3a}J_{b} | \Omega \rangle =\frac{1}{8} T_{3} $   \\ 
$\langle \Delta |d_{ab3}D^{(8)}_{8a}J_{b} | \Delta \rangle= \langle
  \Sigma^{*} |d_{ab3}D^{(8)}_{8a}J_{b} | \Sigma^{*} \rangle = \langle
  \Xi^{*} |d_{ab3}D^{(8)}_{8a}J_{b} | \Xi^{*} \rangle= \langle
  \Omega |d_{ab3}D^{(8)}_{8a}J_{b} | \Omega \rangle =\frac{\sqrt{3}}{16} Y $   \\ 
 \hline 
 \hline
\end{tabular}
\end{center}
\end{table}

\begin{table}[htp]
  \caption{The matrix elements of the collective operators of the
    $m_s$ corrections to the magnetic dipole form factors.} 
  \label{tab:M1_Op}
\begin{center}
\begin{tabular}{ c |  c c  c  c   } 
 \hline 
  \hline 
 $ {\cal{R}}$&  \multicolumn{4}{c}{$\bm{{10}} \ (J_{3}=3/2)$}  \\  
B & $\Delta$ & $\Sigma^{*}$ & $\Xi^{*}$ & $\Omega$ \\  
 \hline
$\langle B_{{\cal{R}}} |D^{(8)}_{88}D^{(8)}_{33} | B_{{\cal{R}}}\rangle$  
& $-\frac{5}{84}T_{3}$& $-\frac{1}{28}T_{3}$& $-\frac{1}{84}T_{3}$ & $0$ \\  
$\langle B_{{\cal{R}}} |D^{(8)}_{88}D^{(8)}_{83} | B_{{\cal{R}}}\rangle$  
& $\frac{\sqrt{3}}{56}$ & $\frac{\sqrt{3}}{84}$ & $-\frac{\sqrt{3}}{56}$
& $-\frac{\sqrt{3}}{14}$ \\  
$\langle B_{{\cal{R}}} |D^{(8)}_{83}D^{(8)}_{38} | B_{{\cal{R}}}\rangle$  
& $-\frac{5}{84}T_{3}$& $-\frac{1}{28}T_{3}$& $-\frac{1}{84}T_{3}$ & $0$ \\    
$\langle B_{{\cal{R}}} |D^{(8)}_{83}D^{(8)}_{88} | B_{{\cal{R}}}\rangle$  
& $\frac{\sqrt{3}}{56}$ & $\frac{\sqrt{3}}{84}$ & $-\frac{\sqrt{3}}{56}$
& $-\frac{\sqrt{3}}{14}$ \\  
$\langle B_{{\cal{R}}} |d_{ab3}D^{(8)}_{8a}D^{(8)}_{8b} | B_{{\cal{R}}}\rangle$  
&$\frac{5}{56}$ &$-\frac{1}{42}$ &$-\frac{5}{56}$ &$-\frac{3}{28}$   \\  
$\langle B_{{\cal{R}}} |d_{ab3}D^{(8)}_{3a}D^{(8)}_{8b}| B_{{\cal{R}}}\rangle$  
& $-\frac{11\sqrt{3}}{252}T_{3}$ & $-\frac{5\sqrt{3}}{84}T_{3}$
& $-\frac{19\sqrt{3}}{252}T_{3}$ & $0$  \\  
 \hline 
 \hline
\end{tabular}
\end{center}
\end{table}

\begin{table}[htp]
  \caption{The relevant transition matrix elements of the collective
    operators coming from the 27plet component of the baryon wave 
    functions for the magnetic dipole form factors.}  
  \label{tab:M1_WF27}
\begin{center}
\begin{tabular}{ c | c  c c c  } 
 \hline 
  \hline 
 $ {\cal{R}}$& \multicolumn{4}{c}{$\bm{10} \  (J_{3}=3/2)$}   \\  
B & $\Delta$ & $\Sigma^{*}$ & $\Xi^{*}$ & $\Omega$ \\  
 \hline
$\langle B_ {\bm{27}} |D^{(8)}_{33} | B_{\cal{R}} \rangle$
& $-\frac{1}{12} \sqrt{\frac{5}{6}} T_{3}$ & $-\frac{1}{8}T_{3}$
&  $-\frac{7}{12} \sqrt{\frac{1}{6}} T_{3}$ &  $0$  \\  
$\langle B_ {\bm{27}}  |D^{(8)}_{83} | B_{\cal{R}} \rangle$
& $\frac{1}{8} \sqrt{\frac{5}{2}} $ &  $\frac{1}{4} \sqrt{\frac{1}{3}} $
&$\frac{1}{8} \sqrt{\frac{1}{2}} $ & $0$  \\  
$\langle B_ {\bm{27}}  |D^{(8)}_{38}J_{3} | B_{\cal{R}}\rangle$  
& $-\frac{1}{8} \sqrt{\frac{5}{2}}T_{3} $ & $-\frac{3\sqrt{3}}{16} T_{3}$
& $-\frac{7}{8} \sqrt{\frac{1}{2}}T_{3} $ & $0$   \\  
$\langle B_ {\bm{27}}  |D^{(8)}_{88}J_{3} | B_{\cal{R}}\rangle$ 
& $\frac{3}{16} \sqrt{\frac{15}{2}} $ & $\frac{3}{8} $ 
& $\frac{3}{16} \sqrt{\frac{3}{2}} $ &$0$ \\  
$\langle B_ {\bm{27}}  |d_{ab3}D^{(8)}_{3a}J_{b} | B_{\cal{R}}\rangle$ 
& $-\frac{1}{24} \sqrt{\frac{5}{6}} T_{3} $ &$-\frac{1}{16} T_{3} $ 
& $-\frac{7}{24} \sqrt{\frac{1}{6}} T_{3} $ &$0$   \\   
$\langle B_ {\bm{27}}  |d_{ab3}D^{(8)}_{8a}J_{b} | B_{\cal{R}}\rangle$ 
& $\frac{1}{16} \sqrt{\frac{5}{2}} $ & $\frac{1}{8} \sqrt{\frac{1}{3}} $ 
&$\frac{1}{16} \sqrt{\frac{1}{2}} $ &$0$   \\   
 \hline 
 \hline
\end{tabular}
\end{center}
\end{table}

\begin{table}[htp]
  \caption{The relevant transition matrix elements of the collective
    operators coming from the 35plet component of the baryon wave 
    functions for the magnetic dipole form factors.}  
  \label{tab:M1_WF35}
\begin{center}
\begin{tabular}{ c | c  c c c  } 
 \hline 
  \hline 
 $ {\cal{R}}$& \multicolumn{4}{c}{$\bm{10} \  (J_{3}=3/2)$}   \\  
B & $\Delta$ & $\Sigma^{*}$ & $\Xi^{*}$ & $\Omega$ \\  
 \hline
$\langle B_ {\bm{35}} |D^{(8)}_{33} | B_{\cal{R}} \rangle$
& $-\frac{1}{20} \sqrt{\frac{1}{14}} T_{3}$ &
$-\frac{1}{8}\sqrt{\frac{1}{35}}T_{3}$&  $-\frac{1}{4}
 \sqrt{\frac{1}{70}} T_{3}$ &  $0$  \\    
$\langle B_ {\bm{35}}  |D^{(8)}_{83} | B_{\cal{R}} \rangle$
& $-\frac{1}{8}  \sqrt{\frac{3}{14}}$&$-\frac{1}{4}\sqrt{\frac{3}{35}}$
&$-\frac{3}{8}\sqrt{\frac{3}{70}}$&$-\frac{1}{4}\sqrt{\frac{3}{35}}$\\ 
$\langle B_ {\bm{35}}  |D^{(8)}_{38}J_{3} | B_{\cal{R}}\rangle$  
& $\frac{1}{8} \sqrt{\frac{3}{14}}T_{3} $ &
$\frac{1}{16}\sqrt{\frac{15}{7}} 
T_{3}$ & $\frac{1}{8}\sqrt{\frac{15}{14}} T_{3}$ & $0$   \\  
$\langle B_ {\bm{35}}  |D^{(8)}_{88}J_{3} | B_{\cal{R}}\rangle$ 
& $\frac{15}{16} \sqrt{\frac{1}{14}} $ &
$\frac{3}{8}\sqrt{\frac{5}{7}} $ 
& $\frac{9}{16} \sqrt{\frac{5}{14}} $ &$\frac{3}{8} \sqrt{\frac{5}{7}}$ \\  
$\langle B_ {\bm{35}}  |d_{ab3}D^{(8)}_{3a}J_{b} | B_{\cal{R}}\rangle$ 
& $-\frac{1}{8} \sqrt{\frac{1}{14}} T_{3} $
  &$-\frac{1}{16}\sqrt{\frac{5}{7}} T_{3} 
$ & $-\frac{1}{8} \sqrt{\frac{5}{14}} T_{3} $ &$0$   \\   
$\langle B_ {\bm{35}}  |d_{ab3}D^{(8)}_{8a}J_{b} | B_{\cal{R}}\rangle$ 
& $-\frac{5}{16} \sqrt{\frac{3}{14}} $ & $-\frac{1}{8}
 \sqrt{\frac{15}{7}} $ &$-\frac{3}{16} \sqrt{\frac{15}{14}} $ 
&$-\frac{1}{8}\sqrt{\frac{15}{7}}$   \\   
 \hline 
 \hline
\end{tabular}
\end{center}
\end{table}

\section{Densities of the electric quadrupole
  form factors} \label{app:B}
The densities of the electric quadrupole form factors are given as
\begin{align}
(-\sqrt{10})\frac{2}{N_{c}} {\cal{I} }_{1E2}(\bm{z})&= \sum_{n \ne
\mathrm{val} }\frac{1}{E_{n}-E_{\mathrm{val}}}{\langle\mathrm{val}
| \bm{\tau} | n\rangle} \cdot{\langle n |\bm{z} \rangle
\{ \sqrt{4\pi}Y_{2}  \otimes\tau_{1}  \}_{1}\langle \bm{z} |
\mathrm{val}\rangle} \cr 
&  + \frac{1}{2} \sum_{n,m} {\cal{R}}_{3}(E_n,E_m)   {\langle n |
  \bm{\tau} | m \rangle} \cdot{\langle m | \bm{z} \rangle \{
  \sqrt{4\pi} Y_{2}  \otimes \tau_{1}  \}_{1} \langle \bm{z} | n
  \rangle}  ,\cr 
(-\sqrt{10})\frac{2}{N_{c}} {\cal{K} }_{1E2}(\bm{z})&=  \sum_{n \ne
 \mathrm{val} } \frac{1}{E_{n}-E_{\mathrm{val}}} {\langle\mathrm{val}
 |  \gamma^{0} \bm{\tau} | n \rangle} \cdot {\langle n | \bm{z} \rangle
 \{ \sqrt{4\pi} Y_{2}  \otimes \tau_{1}  \}_{1} \langle \bm{z} |
 \mathrm{val} \rangle} \cr  
& + \frac{1}{2} \sum_{n,m} {\cal{R}}_{5}(E_n,E_m)     {\langle n |
  \gamma^{0} \bm{\tau} | m \rangle} \cdot {\langle m | \bm{z} \rangle
  \{ \sqrt{4\pi} Y_{2}  \otimes \tau_{1}  \}_{1} \langle \bm{z} | n
  \rangle}, 
\end{align}
where the regularization functions are defined as 
\begin{align}
&{\cal{R}}_{3}(E_{n},E_{m}) = \frac{1}{2 \sqrt{\pi}} \int^{\infty}_{0}
  \phi(u) \frac{du}{\sqrt{u}} \left[ \frac{ e^{-u E_{m}^{2}}- e^{-u
  E_{n}^{2}}}{u(E^{2}_{n} - E^{2}_{m})} -\frac{E_{m} e^{-u
  E_{m}^{2}}+E_{n} e^{-u E_{n}^{2}}}{E_{n} + E_{m}}  \right ], \cr 
&{\cal{R}}_{5}(E_{n},E_{m}) =
  \frac{\mathrm{sign}(E_{n})-\mathrm{sign}(E_{m})}{2(E_{n}-E_{m})}.
\end{align}
Here, $|\mathrm{val}\rangle$ and $|n\rangle$ denotes the state of the 
valence and sea quarks with the corresponding eigenenergies
$E_{\mathrm{val}}$ and $E_n$ of the single-quark Hamiltonian $h(U_c)$,
respectively.  

\section{Comparison with the nucleon EM form factors} 
\label{app:c}
The results on the nucleon form factors were computed with the same
set of parameters within the same framework.  
  \begin{figure}[htp]
  \includegraphics[scale=0.225]{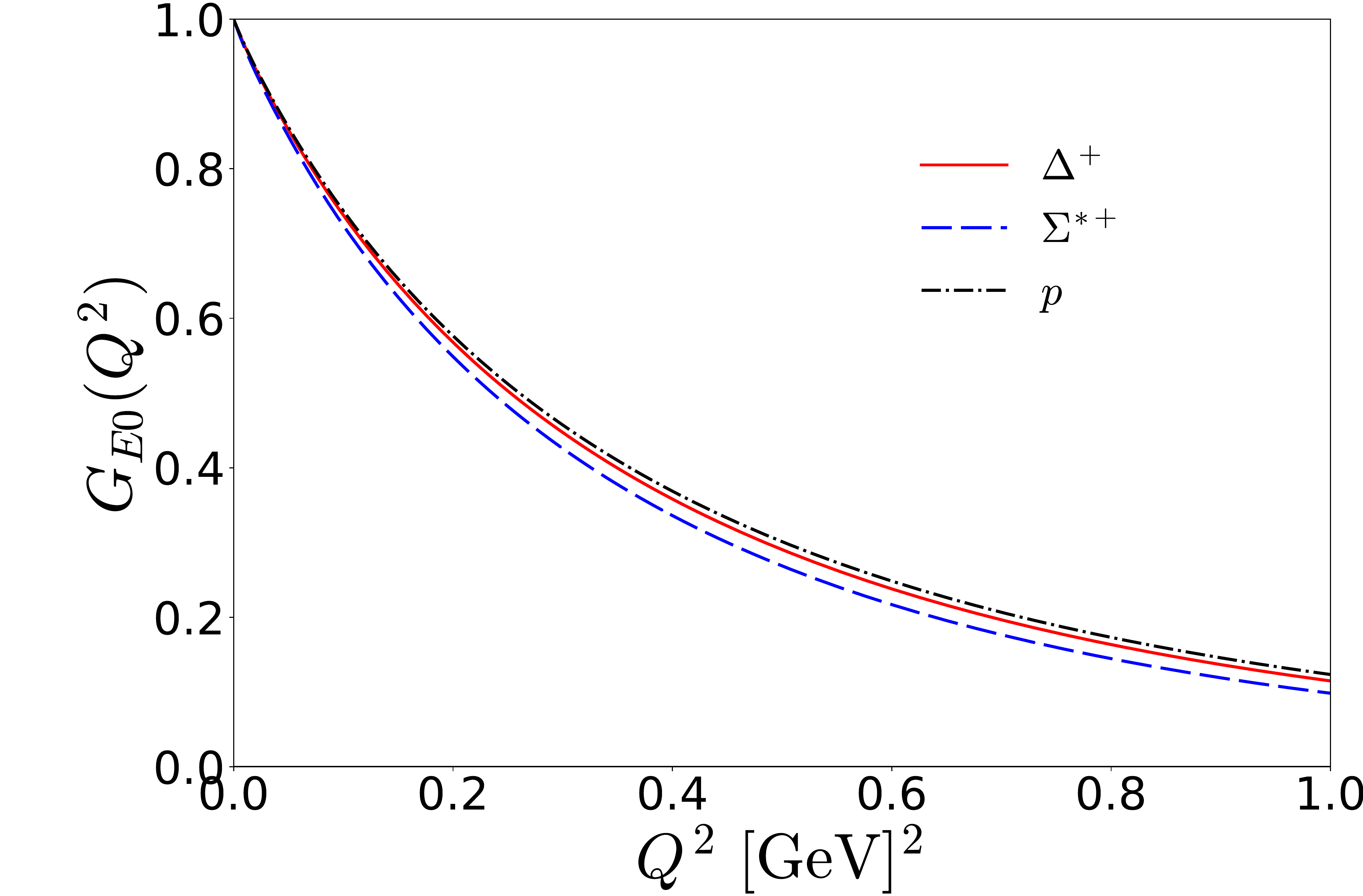}
  \includegraphics[scale=0.225]{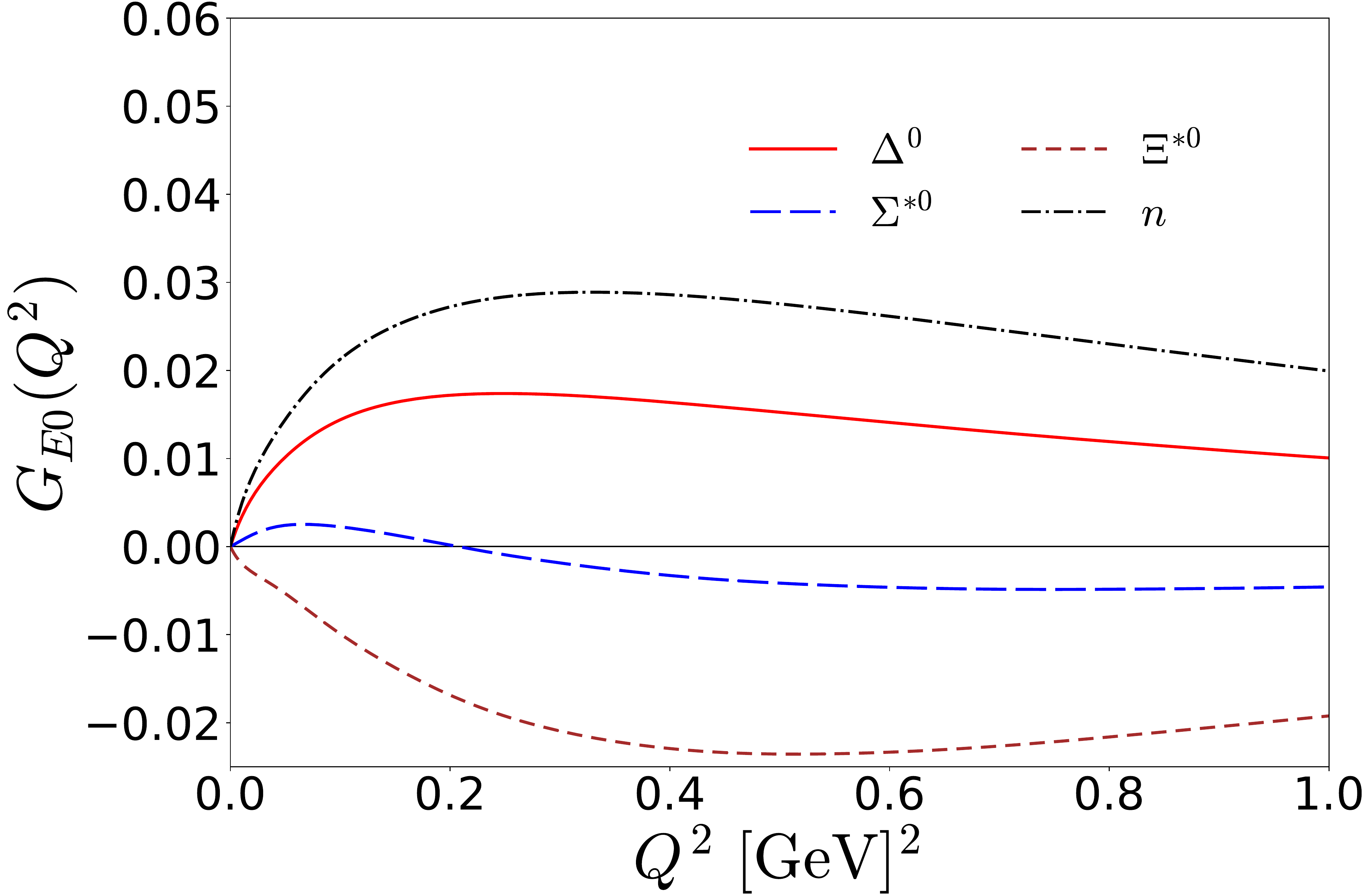}
    \caption{Electric monopole form factors of the baryon decuplet
     in comparison with those of the nucleon.}
\label{fig:14}
\end{figure}
In the left panel of Fig.~\ref{fig:14}, we compare the results of the
$E0$ form factors of the positively singly charged baryon decuplet
with those of the proton electric form factor, whereas in
the right panel of Fig.~\ref{fig:14}, we do those of the neutral
baryon decuplet with those of the neutron electric form factor. The
charged baryon form factors behave in a way very similar to the proton
one. On the other hand, the neutral $E0$ form factors are in general
different from that of the neutron. For example, that of $\Xi^{*0}$
becomes even negative. 
 
  \begin{figure}[htp]
  \includegraphics[scale=0.225]{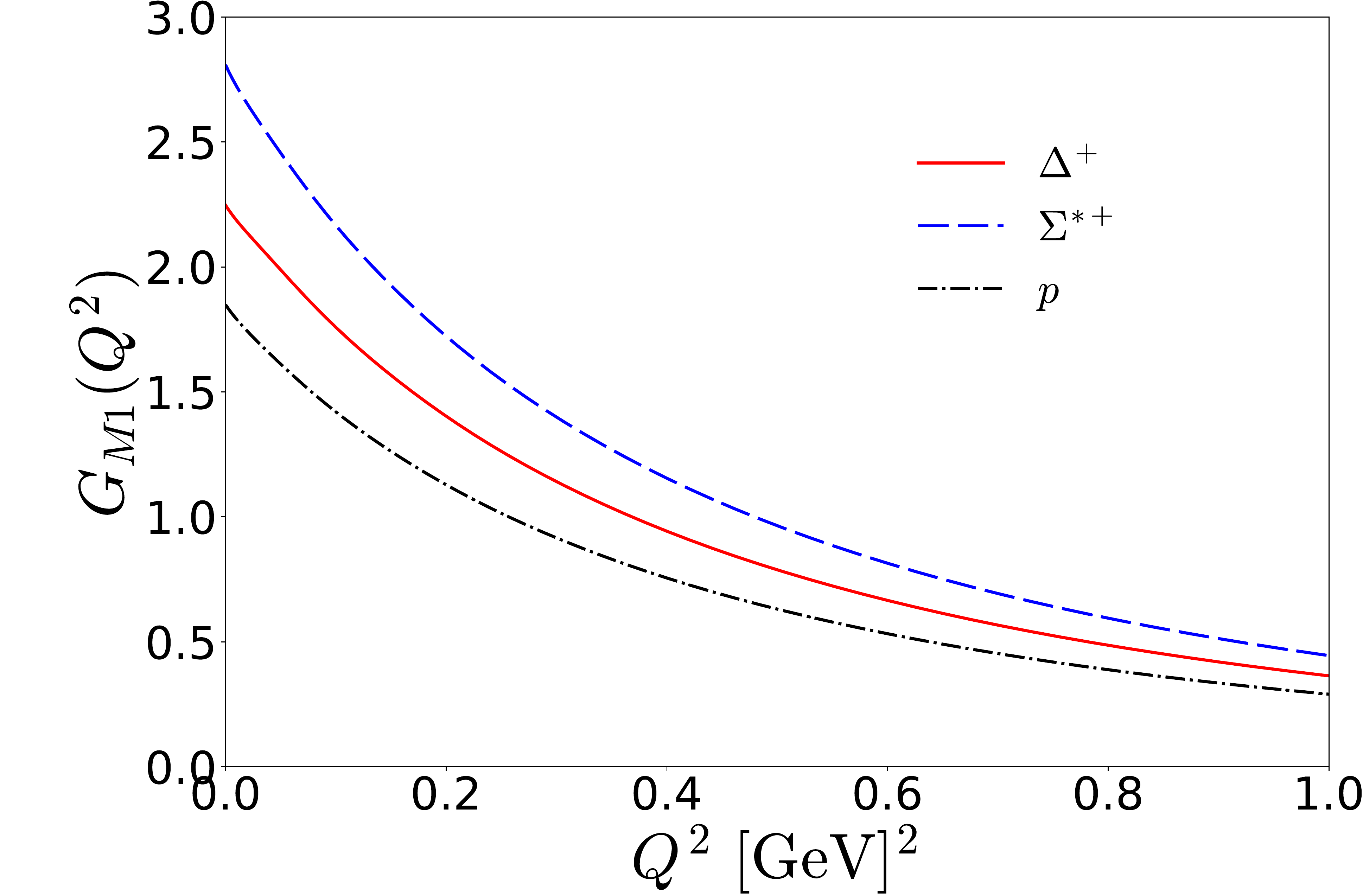}
  \includegraphics[scale=0.225]{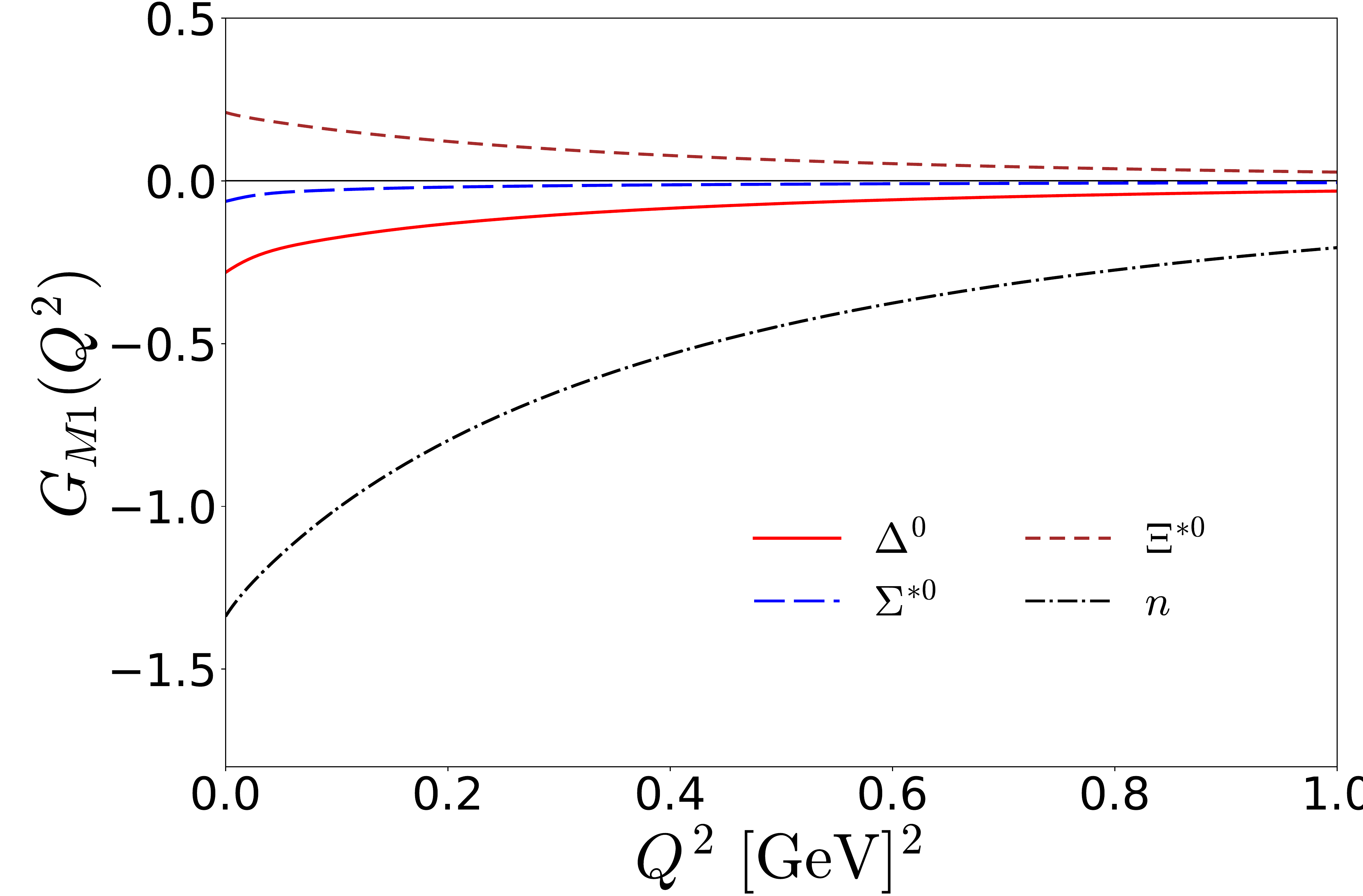}
    \caption{Magnetic dipole form factors of the baryon decuplet
     in comparison with those of the nucleon.}
\label{fig:15}
\end{figure}
The left panel of Fig.~\ref{fig:15} presents the results of the $M1$
form factors of the positively singly charged baryon decuplet in
comparison with those of the proton electric form factor, whereas in 
the right panel of Fig.~\ref{fig:15}, the right panel shows those of
the neutral baryon decuplet being compared with those of the neutron
electric form factor. The general tendency of the results on the $M1$
form factors of the charged baryon decuplet are similar to the proton
one. However, those of the neutral baryons are rather different from
thse of the neutron. The magnitudes of the magnetic dipole
moments of the neutral decuplet baryons are very small, compared to
that of the neutron. 


\end{document}